\documentclass[prd,onecolumn,superscriptaddress, nofootinbib, preprintnumbers]{revtex4}
\usepackage{amsmath, graphicx, float, bm, epsfig}
\pdfoutput=1 

\usepackage[normalem]{ulem}
\usepackage[colorlinks=true,urlcolor=brown,linkcolor=blue,citecolor=magenta]{hyperref}
\usepackage{color}
\usepackage[dvipsnames]{xcolor}
\usepackage{multirow}
\usepackage{cleveref} 

\newcommand{\beq}{\begin{equation}}
\newcommand{\eeq}{\end{equation}}
\newcommand{\bea}{\begin{eqnarray}}
\newcommand{\eea}{\end{eqnarray}}
\newcommand{\nn}{\nonumber\\}
\newcommand{\dd}{\text{d}}
\newcommand{\dec}{\text{dec}}
\newcommand{\rec}{\text{rec}}

\newcommand{\eff}{\text{eff}}
\newcommand{\zdec}{z_\text{dec}}
\newcommand{\zrec}{z_\text{rec}}

\newcommand{\Neff}{N_\text{eff}}
\newcommand{\Geff}{G_\text{eff}}

\newcommand{\meff}{\sum m}
\newcommand{\mk}{\mathbf{k}}
\newcommand{\mq}{\mathbf{q}}
\newcommand{\ml}{\mathbf{l}}
\newcommand{\avgrate}{\left< \Gamma \right>}

\begin{document}

\title{Confronting interacting dark radiation scenarios with cosmological data}

\author{Thejs Brinckmann}
\email{thejs.brinckmann@gmail.com}
\affiliation{Dipartimento di Fisica e Scienze della Terra, Universit\'a degli Studi di Ferrara, via Giuseppe Saragat 1, 44122 Ferrara, Italy}
\affiliation{Istituto Nazionale di Fisica Nucleare (INFN), Sezione di Ferrara, Via Giuseppe Saragat 1, 44122 Ferrara, Italy}
\affiliation{C.~N.~Yang Institute for Theoretical Physics and Department of Physics \& Astronomy, \\ Stony Brook University, Stony Brook, NY 11794, USA}
\author{Jae Hyeok Chang}
\email{jaechang@umd.edu}
\affiliation{Maryland Center for Fundamental Physics, University of Maryland, College Park, MD 20742, USA}
\affiliation{Department of Physics and Astronomy, Johns Hopkins University, Baltimore, MD 21218, USA}
\author{Peizhi Du}
\email{peizhi.du@rutgers.edu}
\affiliation{New High Energy Theory Center, Rutgers University, Piscataway, NJ 08854, USA}
\affiliation{C.~N.~Yang Institute for Theoretical Physics and Department of Physics \& Astronomy, \\ Stony Brook University, Stony Brook, NY 11794, USA}
\author{Marilena LoVerde}
\email{mloverde@uw.edu}
\affiliation{Physics Department, University of Washington, Seattle, WA 98195-1560, USA}
\affiliation{C.~N.~Yang Institute for Theoretical Physics and Department of Physics \& Astronomy, \\ Stony Brook University, Stony Brook, NY 11794, USA}

\begin{abstract}
Dark radiation (DR) is generally predicted in new physics scenarios that address fundamental puzzles of the Standard Model or tensions in the cosmological data. Cosmological data has the sensitivity to constrain not only the energy density of DR, but also whether it is interacting. In this paper, we present a systematic study of five types of interacting DR (free-streaming, fluid, decoupling, instantaneous decoupling, and recoupling DR) and their impact on cosmological observables. We modify the Boltzmann hierarchy to describe all these types of interacting DR under the relaxation time approximation. We, for the first time, robustly calculate the collision terms for recoupling scalar DR and provide a better estimation of the recoupling transition redshift. We demonstrate the distinct features of each type of DR on the CMB and matter power spectra. We perform MCMC scans using the Planck 2018 data and BAO data. Assuming no new physics in the SM neutrino sector, we find no statistically significant constraints on the couplings of DR, although there is a slight preference for the fluid-like limit of all the cases. In the case of instantaneous decoupling DR, this limit corresponds to a late transition redshift around recombination. The $\Delta N_{\rm eff}$ constraint varies marginally depending on the type of DR.
\end{abstract} 
\maketitle

\section{Introduction}

       The Standard Model (SM) of particle physics is a great success because it unifies all elementary particles we have detected and passes extensive experimental tests. The SM, however, is not complete, because of several fundamental puzzles, such as the existence of cold dark matter (CDM), the Higgs hierarchy problem, and non-zero neutrino masses. Physics beyond the SM (BSM) is needed to resolve these puzzles. Many BSM models that are proposed to address these fundamental puzzles predict some light particles that behave like radiation during the epochs probed by the Cosmic Microwave Background (CMB). Since these light particles typically have negligible interactions with the SM, they are usually called dark radiation (DR). For example, some well-motivated models addressing the Higgs hierarchy problem tend to generate dark radiation, e.g., Twin Higgs models~\cite{Chacko:2005pe,Chacko:2016hvu,Craig:2016lyx,Csaki:2017spo} and N-naturalness models~\cite{Arkani-Hamed:2016rle}. On the cosmological side, the so-called  $\Lambda$CDM model describes the evolution of our universe extremely well and explains the cosmological data across many scales \cite{Planck:2018vyg}. Despite this success, there are tensions between CMB measurements and late universe measurements. For example, the Hubble and $S_8$ tensions, referring to disagreement between the predicted values of today's expansion rate of the Universe and matter clustering from CMB data in comparison with locally determined quantities (see, e.g., \cite{Verde:2019ivm,DiValentino:2020zio,Schoneberg:2021qvd} for reviews on the Hubble tension and Ref.~\cite{Amon:2022ycy} for a study of the $S_8$ tension), have attracted significant attention. Many BSM models that attempt to alleviate one or both of these two tensions involve dark radiation~(see, e.g.,~\cite{Wilkinson:2014ksa,Archidiacono:2014nda,Buen-Abad:2015ova,Lesgourgues:2015wza,Archidiacono:2015oma,Archidiacono:2016kkh,Chacko:2016kgg,Ko:2016uft,Ko:2016fcd,Ko:2017uyb,Archidiacono:2017slj,Buen-Abad:2017gxg,PhysRevD.97.043513,PhysRevD.97.075039,Buen-Abad:2018mas,Stadler:2019dii,Archidiacono:2019wdp,Archidiacono:2020yey,Heimersheim:2020aoc,Choi:2020pyy,Becker:2020hzj,Mosbech:2020ahp,Ghosh:2021axu,Cyr-Racine:2021oal,Hooper:2021rjc,Aloni:2021eaq,Corona:2021qxl,Bansal:2021dfh,Hooper:2022byl,Joseph:2022jsf,Buen-Abad:2022kgf}). Constraining the properties of dark radiation is therefore crucial to test BSM models that address puzzles of the SM, with a potential link to tensions in cosmological data.
       
      Since dark radiation has extremely weak couplings with the SM particles, it is challenging to constrain it directly from terrestrial and astrophysical measurements. Cosmological data, however, is sensitive to components that only have gravitational interactions with the SM sector. For example, CMB datasets can determine the dark matter density as well as the density of any decoupled radiation, including dark radiation. One well-known example of such decoupled radiation is SM neutrinos, which decouple from the SM bath when the temperature of the universe is around the MeV scale, much earlier than the time of CMB decoupling. Therefore, SM neutrinos only have gravitational effects on the CMB.  The current constraints from Planck collaboration on the energy density of decoupled radiation, usually denoted as $N_{\rm eff}$, assumes such radiation is free-streaming. This is because SM neutrino interactions become negligible by the epochs probed by the CMB so the particles propagate freely at the speed of light. The effect of radiation on the CMB power spectrum is to increase damping of the higher $\ell$ modes, and free-streaming radiation additionally introduces a phase shift in the CMB peaks, due to the fact that free-streaming radiation propagating super-sonically pushes the SM sound horizon towards a larger scale~\cite{Bashinsky:2003tk,Hou:2011ec}. Current CMB data has the precision to resolve this damping and phase shift, so it is able to distinguish whether the decoupled radiation is free-streaming or  interacting.  
            
      Going beyond the SM, neutrinos or dark radiation can have non-trivial self-interactions that prevent them from free-streaming. Non-standard neutrino interactions can generate strong self-interactions among neutrinos. One example is the Majoron model, where the Majoron is the Goldstone boson from lepton number breaking and is related to the generation of neutrino masses~\cite{Gelmini:1980re,Chikashige_1981,Georgi:1981pg}.  The Majoron will mediate self-interactions among SM neutrinos. If the coupling is strong, neutrinos behave as a coupled fluid and the constraints on  $N_{\rm eff}$ vary compared to the free-streaming case due to the distinct signatures on the CMB power spectrum~\cite{Cyr-Racine:2013jua, Baumann:2015rya, Brust:2017nmv,Choi:2018gho, Blinov:2020hmc}. More generally, these same type of models can also generate decoupling and recoupling features in neutrinos depending on the mass of the Majoron~\cite{Chacko:2003dt,Hannestad:2004qu,PhysRevD.72.103514,Bell:2005dr,Sawyer:2006ju,Friedland:2007vv,Basboll:2008fx,Jeong:2013eza,Archidiacono:2013dua,Forastieri:2015paa,Forastieri:2017oma,Forastieri:2019cuf,Das:2020xke,RoyChoudhury:2020dmd,Brinckmann:2020bcn,Escudero:2020dfa,Taule:2022jrz,Dvorkin:2022jyg,Abazajian:2022ofy}. Recently, these models are studied extensively in the literature because they were also proposed to solve the Hubble tension~\cite{Kreisch:2019yzn,Escudero:2019gvw,Escudero:2021rfi}. However, it has been shown that the decoupling neutrino models do not solve the Hubble tension \cite{Das:2020xke,RoyChoudhury:2020dmd,Brinckmann:2020bcn}. Separately, these models are generally in conflict with other terrestrial and astrophysical constraints because they tend to modify SM neutrino interactions significantly~\cite{Blinov:2019gcj,Brdar:2020nbj,PhysRevD.102.051701,Lyu:2020lps}. 
      
     Much more freedom is allowed if we study interacting dark radiation and assume no new physics in the neutrino sector. Practically, we fix the $N_{\rm eff}$ of neutrinos to its SM prediction and study constraints on additional contributions to DR ($\Delta N_{\rm eff}$). In this paper, we present a systematic study of five types of DR, categorized by the behavior of their self-interactions. We first discuss the standard free-streaming and fluid DR. We then study decoupling DR, which are light dark fermions mediated via a heavy scalar mediator. We also study instantaneous decoupling DR, that can arise from models with dark recombination or dark sector confinement. 
For the recoupling case, we choose to study a self-interacting scalar DR model that can originate from models with axion-like particles. There is another popular scenario for recoupling DR: the Majoron-like model with a light mediator. In this scenario, the light mediator is unavoidably generated in the process of recoupling, leading to additional signatures that we do not consider here. 
     
     To  describe the effects of these different kinds of DR on the perturbations in the CMB we modify the Boltzmann equations to account for these interactions. Refs.~\cite{Oldengott:2014qra,Oldengott:2017fhy} provide the general framework to calculate the collision terms due to interactions using the relaxation time approximation. Under this approximation, the collision term is proportional to the product of thermally averaged interaction rate $\langle\Gamma\rangle$  and some numerical coefficient $\alpha$, we dub the relaxation time coefficient, that needs to be calculated for different type of interactions. Refs.~\cite{Oldengott:2014qra,Oldengott:2017fhy} calculated the coefficient $\alpha$ for the decoupling case and so far no calculation is done for the recoupling case. In this work, we, for the first time, calculate the relaxation time coefficient for the recoupling case with scalar DR. Only after this can one provide a robust map between the particle physics model parameters and the cosmological constraints. Moreover, the general expectation of the relaxation time coefficient is order unity, which is the case for decoupling DR. However, we found the relaxation time coefficient for the recoupling case is nearly one order of magnitude smaller than unity. This means the naive estimation of the time of the transition will also be modified by including this coefficient. In this paper, we provide a new  definition of the transition time (redshift) by setting $\alpha \langle\Gamma\rangle=H$ instead of usual $\langle\Gamma\rangle=H$. We demonstrate that this new definition gives a better estimate of the transition redshift according to numerical calculations of the DR behavior. 
     
    We modify the public Boltzmann solver CLASS \cite{Blas:2011rf,Lesgourgues:2011re,Lesgourgues:2011rh} to account for each type of interacting DR, and show the changes to the CMB temperature, polarization and lensing power spectra, together with matter power spectrum. We discuss different features of these observables for each type of DR and also give an intuitive physical understanding of how they arise. After performing MCMC scans using current CMB and BAO data, we place a updated constraints on BSM models with dark radiation. We find no statistically significant bounds on the coupling constants of DR, although we find a slight preference for a late transition redshift for instantaneous decoupling DR at around recombination, and for the fluid-like limit of all the cases. The data exhibits interesting features at some specific times/redshifts in the early universe, but more constraining data is required to derive statistically significant bounds. The constraints on  $\Delta N_{\rm eff}$ also differ marginally under different assumptions. 
 
  The rest of this paper is organized as the follows.  In Sec.~\ref{sec:models}, we discuss the qualitative effect of five kinds of DR on the CMB, including free-streaming, fluid, decoupling, instantaneous decoupling and recoupling DR. We also present example particle physics models of these types of DR. In Sec.~\ref{sec:Boltzmanneq}, we derive the Boltzmann hierarchy that can account for all these types of DR, under the relaxation time approximation. We calculate the relaxation time coefficient for decoupling and recoupling cases (with more details in Appendix~\ref{app:relaxation_time_coeff}) and present a better way to estimate the redshift of the transition. In Sec.~\ref{sec:observables}, we present the CMB temperature, polarization and lensing spectra, together with the linear theory matter power spectrum from CLASS. In Sec.~\ref{sec:method}, we show the results of MCMC scans based on  Planck 2018 data and BAO data. We conclude in Sec.~\ref{sec:discussion}. In Appendix~\ref{app:scalar_DR}, we discuss a simple model for scalar DR that has the feature of recoupling DR. In Appendix~\ref{app:full_MCMC}, we show more details of the MCMC scans.

\section{Dark radiation models}
\label{sec:models}
For the purpose of this work, we classify DR according to its gravitational effects on cosmological perturbations. The energy density of DR contributes to the parameter $N_\eff$, which is defined by
\beq
\rho_r = \rho_\gamma +\rho_\nu + \rho_\textrm{DR} \equiv \rho_\gamma \left[1+ \frac{7}{8}\left( \frac{4}{11} \right)^{4/3} N_\eff \right],
\eeq
where $\rho_r$ denotes the energy density of total radiation in the early universe and $\rho_\gamma$ is the energy density of photons, $\rho_\nu$ the energy density of neutrinos, and $\rho_{{\rm DR}}$ the energy density of DR. We fix the SM neutrino contribution to $N_\eff$ to the $\Lambda$CDM value of 3.046\footnote{Note that more recent calculations prefer 3.044~\cite{Froustey:2020mcq,Bennett:2020zkv,Akita:2020szl}.} and dub the contribution from DR as $\Delta N_\eff$. That is, we define 
\beq
N_\eff = 3.046 + \Delta N_\eff\,.
\eeq
Since SM neutrinos decouple from the photon bath much earlier than CMB decoupling, neutrinos can be treated as free-streaming for CMB analysis. The primary constraints on free-streaming neutrinos from CMB power spectrum arise from the phase shift of the peaks and suppression of high $\ell$ modes \cite{PhysRevD.87.083008}. If the DR is also free-streaming, it has similar effects on CMB as neutrinos, and the size of the effects is controlled by $\Delta N_\eff$.

More generically, DR may have non-negligible self-interactions or interactions with other particles that prevent DR from free-streaming. For example, DR  can behave as a tightly coupled fluid if interactions are always strong compared to the Hubble rate. One key difference between free-streaming and tightly coupled fluid is that tightly coupled fluid has vanishing quadrupole and higher multipoles of its density perturbation, while these moments are nonzero and have non-trivial time evolution if DR is free-streaming (see Sec.~\ref{sec:Boltzmanneq} for more detail). Because of this, tightly coupled DR generates different signatures compared to the free-streaming case with the same $\Delta N_\eff$ \cite{Baumann:2015rya}. 

 In many theoretical models, DR may have additional  features that are not captured in the above two cases. For example, the rate of interactions that keep DR in equilibrium are typically temperature-dependent and can therefore drop below the Hubble rate at an early time or a later time. Depending on the type of interactions, DR may be tightly coupled in the early universe and starts to free-stream later (the decoupling case), or the other way around (the recoupling case). The effects of these types of DR on observables should therefore be between those of free-streaming and tightly coupled cases with the same $\Delta N_\eff$. Moreover, due to different properties before and after the transition,  decoupling/recoupling DR leaves distinct features on the spectrum that depend on the transition time. These features are discussed in detail in Sec.~\ref{sec:observables}.

In what follows, we will identify DR by the behavior during epochs probed by CMB and large-scale structure (LSS) data.\footnote{While big bang nucleosynthesis probes cosmic history at an earlier time than the CMB, relic abundances are sensitive to the total amount of radiation rather than the perturbations to the radiation. BBN is therefore not generically sensitive to interactions in DR.} We classify DR into five groups according to their interactions and for each discuss examples of particle physics models that can lead to these types of DR. Note that our analysis in this paper is not restricted to the specific model we present. It can be generalized to other models where the interaction rate has the same temperature dependence.

\begin{enumerate}
\item {\bf Free-streaming:} This type of DR does not have any self-interactions or the interaction strength is negligible compared to the Hubble rate during epochs probed by CMB and LSS data. This kind of DR behaves identically to massless SM neutrinos. Free-streaming DR is therefore a natural consequence of BSM models involving light sterile neutrinos. Gravitational waves are another example of free-streaming DR. The effect of free-streaming DR on cosmological observables can be described by $\Lambda$CDM model with additional $\Delta \Neff$. 

\item {\bf Fluid:} We refer to DR as a fluid or fluid-like if it has strong self-interactions (the interaction rate is always large compared to the Hubble parameter) during the whole cosmological time of interest. For example, non-Abelian gauge fields in the dark sector can behave like fluid DR~\cite{Buen-Abad:2015ova}. Moreover, fluid DR is the strong coupling limit of the decoupling and recoupling DR mentioned below. Due to these strong interactions, the anisotropic stress and all higher multiples of fluid DR are dynamically set to zero. 

\item {\bf Decoupling:} If the self-interactions among DR are mediated by a heavy mediator, the thermal averaged interaction rate $\Gamma$ may scale as $T^3$ or $T^5$, depending on the particle nature of DR and the mediator. 
In the early universe, this interaction rate can be larger than the Hubble rate, which scales as $T^2/M_\textrm{pl}$, so DR interacts strongly and is tightly coupled. However, if $\Gamma$ drops below the Hubble rate at a later time, this interaction decouples and DR starts to free-stream. We dub this case decoupling.

Many particle physics models predict decoupling DR. To be concrete, we consider the example of DR that is an effectively massless Majorana fermion $\chi$ that couples to a massive mediator $\phi$. The Lagrangian can be written as
\bea\label{eq:Majoron}
-\mathcal L \supset \frac{1}{2}m_\phi^2 \phi^2 + \frac{1}{2} g_\phi \phi \bar\chi \chi,
\eea
where $m_\phi$ is the mass of $\phi$ and $g_\phi$ is the Yukawa coupling between $\phi$ and $\chi$.
This Lagrangian is similar to that of Majoron models, where the scalar $\phi$ couples to SM neutrinos instead of DR $\chi$. 
At $T \ll m_\phi$, $\phi$ can be integrated out and we obtain an effective Lagrangian with 4-fermi interactions
\bea\label{eq:Heavy_mediator}
-\mathcal L \supset \frac{1}{8} \Geff  \bar\chi \chi \bar\chi \chi,
\eea
where $\Geff\equiv g_\phi^2/m_\phi^2$. This term generates self-interactions among DR and the thermally averaged interaction rate scales as $G_\textrm{eff}^2 T^5$, which can lead to decoupling.

\item {\bf Instantaneous Decoupling:} This is a special type of decoupling DR, which undergoes much faster transition than the case discussed above. In this case, we can simply treat the DR as a fluid before a certain redshift $\zdec$, which characterizes the time of decoupling, and free-streaming after $\zdec$. Such DR can arise in models that contain a complex dark sector with a bound state that is neutral to DR. DR may decouple from the bath of the dark sector when dark sector particles form bound states, either a dark atom similar to recombination in the SM sector or dark baryons similarly to the QCD phase transition (see, for example, Atomic Dark Matter models~\cite{Kaplan:2009de,Cyr-Racine:2012tfp,Bansal:2022qbi} and Twin Higgs models~\cite{Chacko:2018vss}). In these cases, the interaction rate includes a factor $\exp(-B/T)$, where $B$ is the binding energy of the dark atom or the confinement scale of dark baryons.\footnote{$T$ here is the temperature of dark radiation, which may be different from the photon temperature.} This exponential factor changes decoupling qualitatively, making the duration of the decoupling transition much shorter than the standard decoupling described above. We assume dark atoms or dark baryons form only a small fraction of dark matter so we can ignore any changes to cold dark matter perturbations due to interactions between dark atoms and DR. That is, our instantaneous case only changes the behavior of $\Delta N_\eff$ and no other cosmological parameters. 

\item {\bf Recoupling:} Recoupling radiation free-streams in the early universe and becomes tightly coupled at a later time. In this paper, we model recoupling DR as a light scalar ($\phi$) with $\phi^4$ interaction
\bea\label{eq:fourphi}
-\mathcal L \supset \frac{\lambda_\phi}{4!} \phi^4,
\eea
where $\lambda_\phi$ is the dimensionless coupling constant. We present one explicit realization of such DR in models of axion-like particles (ALPs) in Appendix~\ref{app:scalar_DR}. The interaction in Eq.~(\ref{eq:fourphi}) leads to a thermal averaged rate that scales as $\lambda_\phi^2 T$, which decreases more slowy than the Hubble rate. Therefore, this type of DR has the desired recoupling feature: the interaction rate can be smaller than the Hubble rate initially but become stronger later.

We note that the recoupling DR can also arise in models with fermionic DR that interact with a light mediator. For example, using the same Lagrangian in Eq.~(\ref{eq:Majoron}) but setting $m_\phi\to 0$, the rate of self-interactions among DR ($\chi$) can be estimated to be $\sim g_\phi^4 T$, which has the right $T$ dependence for recoupling. However, the light mediator $\phi$ will be produced via $\chi\chi\to \phi\phi$, and will come into equilibrium with $\chi$ after recoupling \cite{Escudero:2020dfa}. Therefore, the abundance and perturbations of $\phi$, which are not covered in our analysis in this paper, can potentially have important impact on cosmological observables.

\end{enumerate}

The qualitative effects of decoupling or recoupling DR on cosmological observables can be viewed from the time dependence of their interactions.  
The opacity of DR provides an intuitive way to visualize the time dependence of different types of interactions. We define the opacity for decoupling and recoupling DR as
\bea
O_\dec(t) &=& 1-\exp\left[-\int_t^{t_0} \langle \Gamma\rangle\dd t \right], \label{eq:opacitydec}\\
O_\rec(t) &=& 1-\exp\left[-\int_0^{t} \langle \Gamma\rangle \dd t \right], \label{eq:opacityrec}
\eea
where  $\langle \Gamma\rangle$ is the thermal averaged interaction rate and $t_0$ is the time today (the exact definition of $\Gamma$ for the models above will be given below in Eq.~(\ref{eq:thermalavgrateapprox})). The opacity indicates the fraction of particles that have been scattered from the time $t$ to the time today $t_0$ for decoupling cases, or from the start of the universe to the time $t$ for recoupling cases. Zero opacity means DR is free-streaming, while fluid DR corresponds to opacity being unity. For decoupling and recoupling cases, the opacity will start from 1 and evolve to 0 (decoupling) or vice versa (recoupling).

\begin{figure}[t]
	\centering
	\includegraphics[width = .7 \columnwidth]{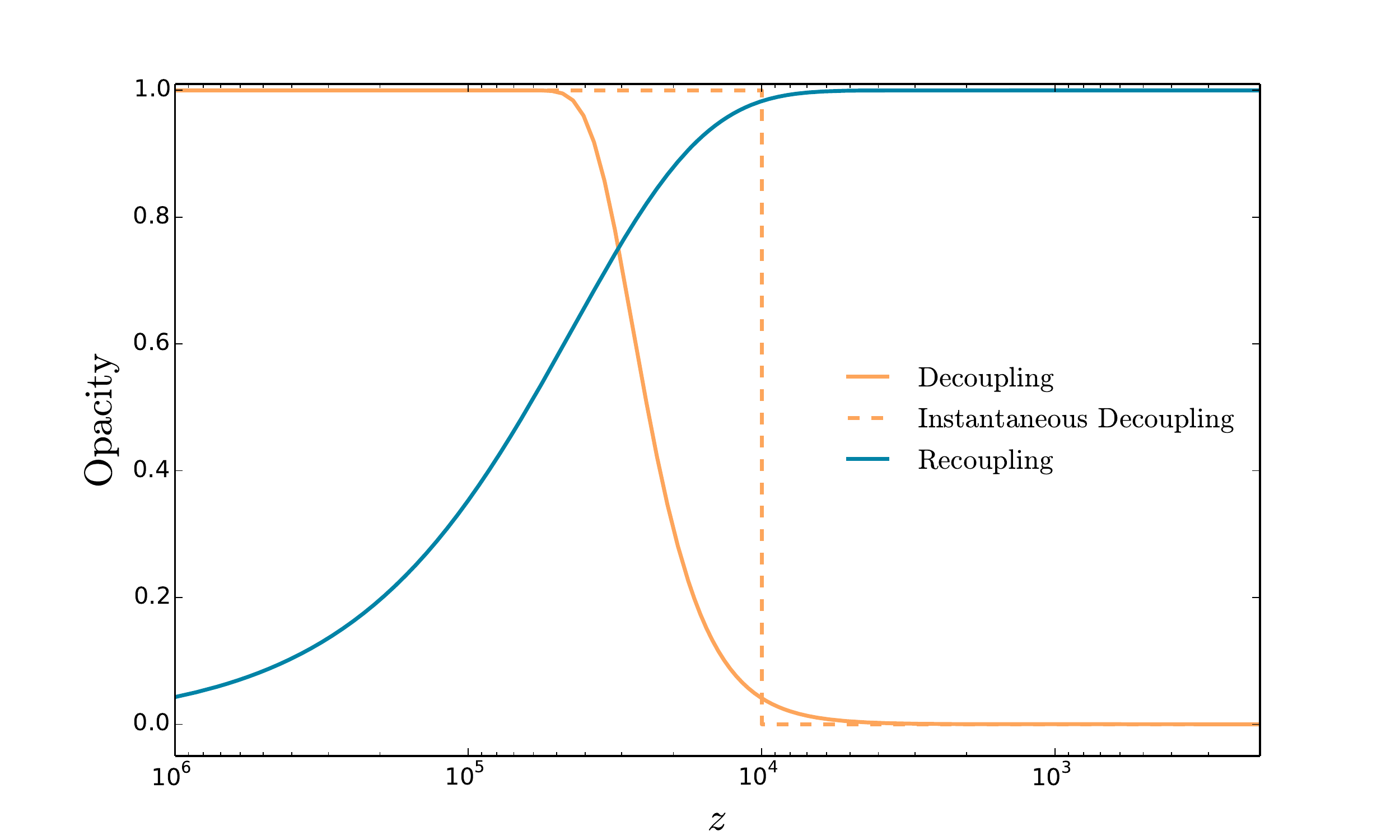}
 	\caption{The opacity for decoupling DR (orange), instantaneous decoupling DR (orange dashed), and recoupling DR (blue) as a function of redshift $z$. Couplings $G_{\rm eff}$ and $\lambda_\phi$ are chosen such that $z_\textrm{dec,rec} \sim10^4$. In this paper, we define the decoupling/recoupling transitions in terms of the perturbations, rather than the background opacity, which leads to the visual offset of the transitions above. Our precise definition of the decoupling/recoupling redshift will be given in Sec.~\ref{sec:Boltzmanneq} (see Eq.~(\ref{eq:z_dec_rec})), once we discuss the perturbations to dark radiation.}
	\label{fig:opacity}
\end{figure}

The opacity ($ O_{\dec,\rec}$) for three difference cases (decoupling, instantaneous decoupling and recoupling DR) is shown as a function of redshift in Fig.~\ref{fig:opacity}. We have chosen a step function to model $O_{\rm dec}(z)$ for the instantaneous decoupling case where the transition occurs at $z = 10^4$, and selected values of the coupling constants ($G_{\rm eff}$, $\lambda_\phi$ for decoupling and recoupling cases) such that they lead to the same transition redshift $z_\textrm{dec,rec}=10^4$ (our precise definition of $z_\textrm{dec,rec}=10^4$ will be given later in Eq.~(\ref{eq:z_dec_rec})). While the instantaneous decoupling transition is much faster than standard decoupling, it is also clear from Fig.~\ref{fig:opacity} that the transition of the recoupling case is slower than the decoupling case. This can be understood from the fact that decoupling rate has a stronger $T$ dependence ($\propto T^5$) than that of recoupling ($\propto T$). As we shall see, the time dependences visible in the opacity in Fig. \ref{fig:opacity} will lead to distinct signatures on observables.

\section{Boltzmann equations for self-interacting dark radiation}
\label{sec:Boltzmanneq}
In this section, we present the Boltzmann equations relevant for self-interacting dark radiation. We use the synchronous gauge and write perturbations around a flat Friedmann-Robertson-Walker metric written as 
\bea
ds^2=a^2(\tau)(-d\tau^2+(\delta_{ij}+H_{ij})dx^idx^j),
\eea
where $\tau$ is the conformal time and $a(\tau)$ is the scale factor. The metric perturbation $H_{ij}$ has the following expression in Fourier space,
\bea
H_{ij}(\mathbf{k},\tau)=\hat { k}_i\hat  { k}_j h(\mathbf k,\tau) +\left(\hat {k}_i\hat{k}_j -\frac{1}{3}\delta_{ij}\right)6\eta(\mathbf k,\tau),
\eea
where $h $ and $\eta$ denotes the trace and traceless longitudinal part of $H_{ij}$ respectively. Here $\mathbf{k}\equiv k \hat {\mathbf k} $ is Fourier conjugate to comoving position $\mathbf x$. We consider the phase space distribution of DR $f(\mathbf{q},\mathbf{k},\tau)$, which can be decomposed as 
\bea\label{eq:f_expansion}
f(\mathbf{q},\mathbf{k},\tau)\equiv \bar f(q,\tau) (1+\Psi(\mathbf{q},\mathbf{k},\tau)),
\eea
where $\bar f(q,\tau)$ is the averaged phase space distribution and $\mathbf{q}\equiv q \hat{\mathbf {q}}$ is the comoving momentum of DR, which is related to the physical momentum $\mathbf p$ as $\mathbf{q}=a \mathbf{p}$. In this work, we demand that all kinds of DR in our analysis have nearly thermal distributions, which means the averaged phase space distribution is set to the equilibrium distribution, 
\beq
\bar f(q,\tau)\approx \bar{f}^{\rm eq}(q)=N e^{-q/T_{D,0}},
\eeq 
where $T_{D,0}$ is the temperature of DR today. Here, we use the Maxwell-Boltzmann distribution as an approximation, and the normalization factor $N$ is chosen to match the \textit{energy} density between the Maxwell-Boltzmann  and Bose-Einstein/Fermi-Dirac distribution: $\int dq\, q^3 \bar{f}^{\rm eq}(q)=\int dq\, q^3 f^{\rm BE/FD}(q)$. Therefore, we will get  $N= \pi^4/90$ for bosons and $N=7 \pi^4/720$ for fermions.

Since the equilibrium distribution is independent of time, all non-trivial time evolution due to self-interactions is encoded in the perturbation $\Psi(\mathbf{q},\mathbf{k},\tau)$.
 The nearly thermal condition can be simply achieved in decoupling DR, because strong self-interactions keep DR in equilibrium before decoupling and the later evolution only generates perturbations. For the recoupling case, however, self-interactions cannot keep DR in equilibrium initially. To satisfy this condition, we assume DR is a thermal relic (like SM neutrinos) that inherits a thermal distribution from other processes. We present an example of a scalar DR model in Appendix~\ref{app:scalar_DR} that predicts recoupling DR with thermal initial conditions.  

As mentioned before, the non-trivial time evolution of $f(\mathbf{q},\mathbf{k},\tau)$ appears in the perturbation $\Psi(\mathbf{q},\mathbf{k},\tau)$ defined in Eq.~(\ref{eq:f_expansion}). The Boltzmann equation for $\Psi $ is given as:
\bea\label{eq:Boltzmann_eq_general}
\dot{\Psi}+i k P_1( \hat {\mathbf k} \cdot \hat {\mathbf q})\Psi+\frac{d \ln \bar f}{d \ln q}\left[-\frac{\dot h}{6}-\frac{P_2( \hat {\mathbf k} \cdot \hat {\mathbf q})}{3}(\dot h+6\dot \eta)\right]=C[f],
\eea
where $P_\ell$ are the Legendre polynomials and $\, \dot{} \,$ denotes the derivative respect to $\tau$. $C[f]$ is the collision term that accounts for self-interactions. The full expression of $C[f]$ was derived in \cite{Oldengott:2014qra,Oldengott:2017fhy} and is summarized in Appendix~\ref{app:relaxation_time_coeff}. 
Since we deal with purely relativistic particles, it is convenient to use $F$ instead of $\Psi$ in the analysis,
\bea
F(\mathbf k, \hat{\mathbf q},\tau)\equiv\frac{\int dq q^3\bar  f\Psi}{\int dq q^3\bar  f}\equiv \sum_{\ell=0}^{\infty}(-i)^\ell (2\ell+1) F_{\ell}(k,\tau)P_\ell( \hat {\mathbf k} \cdot \hat {\mathbf q}) .
\eea

 The explicit expression for $C[f]$ is complicated, we then use the relaxation time approximation \cite{Hannestad:2000gt} to simplify the formula. The basic assumption is that the leading order contribution in $C[f]$ is linear in $F$: $C[f]\propto -F$. It has been checked in \cite{Oldengott:2017fhy} that this approximation agrees well with the exact results for the decoupling case. To our knowledge, the explicit check for the  recoupling case has not been performed yet. However, for both cases the key approximations are the same: i) keeping only linear order in perturbations and ii) applying the separable ansatz: $\Psi_\ell(k,q,\tau)\approx -\frac{1}{4}\frac{d \ln \bar f}{d\ln q} F_\ell(k, \tau)$~\cite{Oldengott:2017fhy}, which can be easily derived for radiation using the separate universe argument: the only $\mathbf x$ dependence of $\Psi$ is in temperature $T(\mathbf x)$. Therefore, we expect the approximation works for the recoupling case as well.
 
Under the relaxation time approximation, the full Boltzmann hierarchy for $ F_{ \ell}$ can be simplified as
  \bea\label{eq:Boltzmann_hierarchy_decoupling}
&&\dot{F}_{0}=-kF_{1}-\frac{2}{3}\dot h,\nn
&&\dot{F}_{ 1}=\frac{k}{4}F_{1}-\frac{k}{2}F_{ 2},\nn
&&\dot{F}_{2}=\frac{2k}{5}F_{ 1}-\frac{3k}{5}F_{3}+\frac{4}{15}\dot h+\frac{8}{5}\dot \eta-\alpha_2a \avgrate {F}_{ 2},\nn
&&\dot{F}_{ \ell}=\frac{k}{2\ell+1}[\ell F_{\ell-1}-(\ell +1)F_{\ell+1}]-\alpha_\ell a \avgrate {F}_{\ell}, ~~~~\ell\geq3,
\eea
where $\avgrate$ is the thermally averaged rate for self-interactions and $\alpha_\ell$ is the relaxation time coefficient. The above formula is general and can be applied to all kinds of self-interacting DR.  As we discuss below,  different types of interactions will lead to different values of $\avgrate$ and $\alpha_\ell$.

The thermal averaged rate for a $2\to 2$ self-interaction process ($p_1+p_2\to p_3+p_4$) is simplified  under assumption of $\bar{f}(E) \ll 1$ \cite{Brinckmann:2020bcn}:
\begin{eqnarray}
\langle \Gamma \rangle &\equiv& \frac{1}{\bar n}\int d\Pi_1 d\Pi_2 d\Pi_3 d\Pi_4 \bar{f}(E_1) \bar{f}(E_2) (1 \pm \bar{f}(E_3))(1 \pm \bar{f}(E_4)) \langle |\mathcal{M}|^2 \rangle (2\pi)^4\delta^{(4)}(p_1+p_2-p_3-p_4)\nonumber\\
&\approx& \frac{g^2}{\bar n}\int \frac{d^3 \mathbf p_1}{(2\pi)^3} \frac{d^3 \mathbf p_2 }{(2\pi)^3} \bar{f}(E_1) \bar{f}(E_2)  \sigma_{2 \rightarrow 2}v_{\rm rel} \label{eq:thermalavgrateapprox}
\end{eqnarray}
where $p_i=(E_i,\mathbf{p}_i)$ is the 4-momentum of each particle and $d\Pi_i\equiv\frac{g d^3 \mathbf{p}_i}{(2\pi)^3 2E_i}$ with spin degeneracy $g $, $\bar n\equiv g \int d^3 \mathbf{p}/(2\pi)^3 \bar f(E)$ is the equilibrium number density and we use $n= g N T^3/\pi^2$ for massless particles. The quantity $\sigma_{2 \rightarrow 2}$ is the cross section for $2\to2$ self-interaction and $v_{\rm rel}=s/(2E_1E_2)$ is the relative velocity of initial particles. The matrix element squared $\langle |\mathcal{M}|^2 \rangle$ is averaged over \textit{initial} and \textit{final spins} of all particles, and includes a factor  $1/(N_i ! N_f!)$ that removes the double counting of identical particles, as well as a factor of $N_i$, because the number of initial identical particles changes by $N_i$ units per interaction in the Boltzmann equations. Here $N_i (N_f)$ denotes the number of identical particles in the initial (final) states. In our case, we always consider $2\leftrightarrow 2$ process of all identical particles. Therefore, this factor becomes $N_i/(N_i ! N_f!)=1/2$. We will show the analytical expression of $\langle \Gamma \rangle $ for the different cases below.

The calculations of the relaxation time coefficients $\alpha_\ell$ are shown in Appendix~\ref{app:relaxation_time_coeff}, where we follow calculations in \cite{Oldengott:2014qra,Oldengott:2017fhy}.\footnote{Note we define $\alpha_\ell$ with respect to the thermal averaged rate, not the rate from the dimensional analysis used in~\cite{Oldengott:2014qra,Oldengott:2017fhy}.} We note that for all models, $\alpha_{0}=\alpha_1=0$ due to energy and momentum conservation.

We summarize the thermal averaged rate and the relaxation time coefficients for our DR scenarios below.
 
\begin{enumerate}
\item\textbf{Free-streaming:}
Free-streaming DR has no self-interactions, therefore we should set  $\avgrate=0$ in Eq.~(\ref{eq:Boltzmann_hierarchy_decoupling}). This result coincides with the Boltzmann hierarchy for free-streaming neutrinos.
\item\textbf{Fluid:}
Fluid DR has very strong self-interactions and is always tightly coupled. This corresponds to the limit $\avgrate \to \infty$  in Eq.~(\ref{eq:Boltzmann_hierarchy_decoupling}). In this limit, the term $-\alpha_\ell a \avgrate F_{ \ell}$ dominates in the time evolution, and the solution is $F_{ \ell}\propto \textrm{exp}(-\alpha_\ell \int a \avgrate d\tau)$ for $\ell \geq 2$. This indicates that all $\ell \geq 2$ modes will be dynamically set to zero for the whole thermal history. Therefore, we can set $F_{\ell}(k,\tau)=0$ for $\ell \geq 2$ for all $\tau$. This means that the anisotropic stress and all higher multiples of DR are always zero.
 
\item\textbf{Decoupling:}
As mentioned in Sec.~\ref{sec:models}, we consider a Majorana fermion $\chi$ with a heavy scalar mediator $\phi$ as a model for decoupling DR. The Lagrangian is given in Eq.~(\ref{eq:Heavy_mediator}) with $G_{\rm eff}\equiv g_\phi^2/m_\phi^2$ being the effective Fermi constant. We assume the mediator is heavy for the interest of cosmological evolution and thus its abundance can be neglected. The only effect of the heavy mediator is that it generates self-interactions among dark radiation. This case will lead to, 
\beq\label{eq:ave_rate_decoupling}
a\avgrate= \frac{7\pi G_{\rm eff}^2T_{D,0}^5}{576 a^4}\,,
\eeq
in Eq.~(\ref{eq:Boltzmann_hierarchy_decoupling}), where $T_{D,0}$ is the temperature of the DR today. Note that
$a\avgrate\propto a^{-4}$ blows up for very small $a$ (early times). Physically, this limit corresponds to the radiation behaving as a fluid, but to avoid numerical issues, we set a cut-off on $a\avgrate$: $a\avgrate=\textrm {Min}[ 7\pi G_{\rm eff}^2T_{D,0}^5/(576 a^4),\Lambda]$, with $\Lambda$  chosen be $100/\textrm{Mpc}$. We have checked numerically that the observables are not very sensitive to the precise value of $\Lambda$ chosen, for $\Lambda \ge 100/\textrm{Mpc}$. The relaxation time coefficients are $\alpha_2=1.39, \alpha_3=1.48, \alpha_4=1.57, \alpha_5=1.62$ in this model (see details in Appendix~\ref{app:relaxation_time_coeff}). Since the $ \alpha_\ell$ are very similar in value and $\alpha_{\ell>2}$ have weaker effects on observables, we simply set $\alpha_\ell=\alpha_2\,(2\leq\ell\leq\ell_{\rm max})$, where $\ell_{\rm max}=17$ is the standard cut-off on the number of moments in CLASS. We verify that this approximation, compared to setting $\alpha_2=1.39, \alpha_3=1.48,  \alpha_4=1.57, \alpha_{\ell(5\leq\ell\leq\ell_{\rm max})}=1.62$,  makes a negligible effect on  the CMB and matter power spectra we consider.

\item\textbf{Instantaneous Decoupling:}
Instantaneous decoupling dark radiation behaves like a coupled fluid at early times, and rapidly transitions to free-streaming radiation at some redshift $z_{\rm dec}$. This case mimics the transition of photons in the SM sector that behave as an interacting fluid due to interactions with free electrons, but quickly transition to free-streaming after hydrogen formation. As mentioned before, the fluid regime can be modeled as $F_{  \ell}=0\,(\ell\geq 2)$, while the free-streaming case corresponds to setting $\avgrate=0$ in Eq.~(\ref{eq:Boltzmann_hierarchy_decoupling}). To account for both regions, we set the initial condition for $F_\ell$ as $F_{\ell\geq2}=0$ and  modify the Boltzmann equation in Eq.~(\ref{eq:Boltzmann_hierarchy_decoupling}) to be
\bea\label{eq:Instantaneous_decoupling}
&&\dot{F}_{  2}=\left(\frac{2k}{5}F_{  1}-\frac{3k}{5}F_{ 3}+\frac{4}{15}\dot h+\frac{8}{5}\dot \eta\right)g(z, z_{\rm dec}),\nn
&&\dot{F}_{  \ell}=\left(\frac{k}{2\ell+1}[\ell F_{ \ell-1}-(\ell +1)F_{ \ell+1}]\right)g(z, z_{\rm dec}), ~~~~\ell\geq3,
\eea
where $g(z, z_{\rm dec})$ is a function of $z$ and $z_{\rm dec}$, which has the feature that $g=0$ for $z>z_{\rm dec}$ and $g=1$ for $z<z_{\rm dec}$. 
We choose a smooth function for $g(z,z_{\rm dec})$:
\bea
\label{eq:gdec}
 g(z,z_{\rm dec})=\frac{1}{2}\left[\tanh\left(\frac{z_{\rm dec}-z}{\Delta z_{\rm dec}}\right)+1\right],
\eea
with $\Delta z_{\rm dec}=0.01z_{\rm dec}$. Typically, the decoupling width $\Delta \zdec$ for instantaneous decoupling cases is a few percent of $\zdec$, but the observables we show are not particularly sensitive to $\Delta z_{\rm dec}$ for values of $\Delta z_{\rm dec}/z_{\rm dec} \lesssim 0.1$, as we discuss in Sec. \ref{sec:observables}.  We have checked that the results from Eq.~(\ref{eq:Instantaneous_decoupling}) are almost identical to those from the general Eq.~(\ref{eq:Boltzmann_hierarchy_decoupling}) with a large  $\langle\Gamma\rangle$ at $z>z_{\rm dec}$ and $\langle\Gamma\rangle=0$ for $z<z_{\rm dec}$. We note that the above choice of $g(z,z_{\rm dec})$ can also model decoupling DR with heavy mediators, with the choice of $\Delta z_{\rm dec}=0.4z_{\rm dec}$ \cite{Brinckmann:2020bcn}.

\item \textbf{Recoupling:}
As mentioned in Sec.~\ref{sec:models}, we consider a light scalar with a $\phi^4$ self-interaction as a model for recoupling DR. This type of recoupling DR (with Lagrangian shown in Eq.~(\ref{eq:fourphi})) can be described by Eq.~(\ref{eq:Boltzmann_hierarchy_decoupling}) with, 
\beq\label{eq:ave_rate_recoupling}
a\avgrate = \frac{\pi \lambda_\phi^2 T_{D,0}}{23040}\,,
\eeq
and $\alpha_2=0.188, \alpha_3=0.294, \alpha_4=0.356, \alpha_5=0.395$ (see Appendix~\ref{app:relaxation_time_coeff} for details). Since $\Gamma\propto T$ and $H\propto T^2$ in radiation dominated era, the self-interaction rate will be significant compared to the Hubble rate at late times, leading to recoupling. This means dark radiation is free-streaming at early times, but behaves as a fluid at late times. Similar to the case of decoupling DR, we make the approximation: $\alpha_\ell=\alpha_2\,(2\leq\ell\leq\ell_{\rm max})$ for recoupling DR, which we have checked (compared to setting $\alpha_2=0.188, \alpha_3=0.294,  \alpha_4=0.356, \alpha_{\ell(5\leq\ell\leq\ell_{\rm max})}=0.395$) does not affect physical observables. 
\end{enumerate}
While all expressions above allow for a general $T_{D,0}$, for simplicity in all numerical calculations from here on we will set $T_{D,0}=T_{\nu,0}$, where $T_{\nu,0}$ is the SM neutrino temperature today. 

We choose adiabatic initial conditions at superhorizon scales for DR perturbations assuming that DR is free-streaming~\cite{Ma:1994dv, Bashinsky:2003tk}. This initial condition provides the correct free-streaming limit when the rate $\langle \Gamma \rangle$ goes to zero. If the rate $\langle \Gamma \rangle$ is large around the initial time, i.e., the fluid limit,  the higher moments $F_{\ell\geq2}$ of DR are dynamically set to zero rapidly and remain zero as long as $\langle \Gamma \rangle$ is large (for the instantaneous decoupling case we simply set $F_{\ell\geq2}=0$ as discussed before).  As a result, other perturbations (e.g. $F_{1}$ and metric perturbation $\eta$ in the synchronous gauge) will also deviate from free-streaming values and quickly approach the fluid limit of the corresponding perturbations. As long as the mode evolution begins sufficiently early ($k\tau \ll 1$),  this is equivalent to setting initial conditions that treat DR as fluid directly. We have checked that these two types of initial conditions agree in the fluid limit.

\begin{figure}[t!]
	\centering
	$\begin{array}{ccc}
	\includegraphics[width=5.8cm]{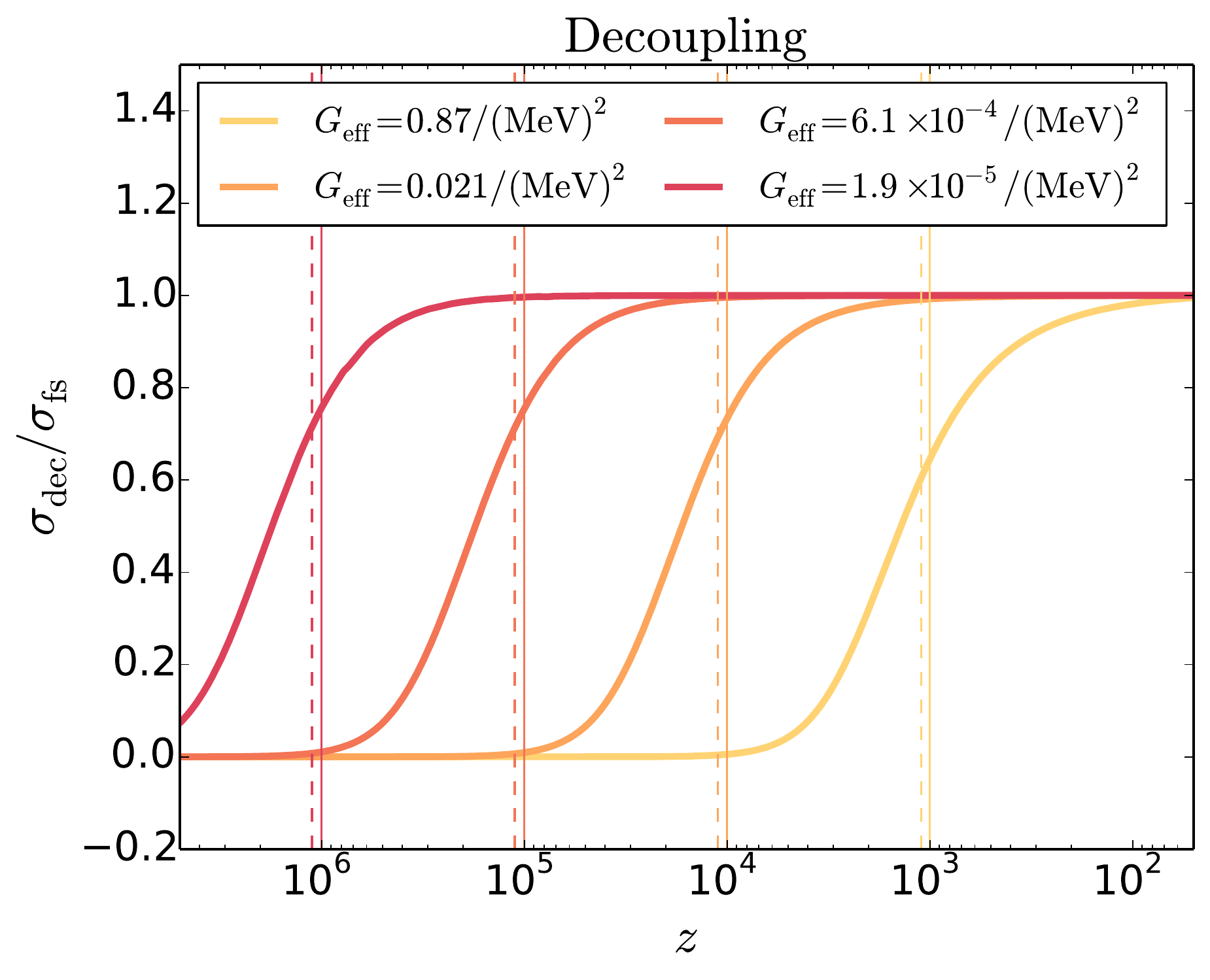} &
	\includegraphics[width=5.8cm]{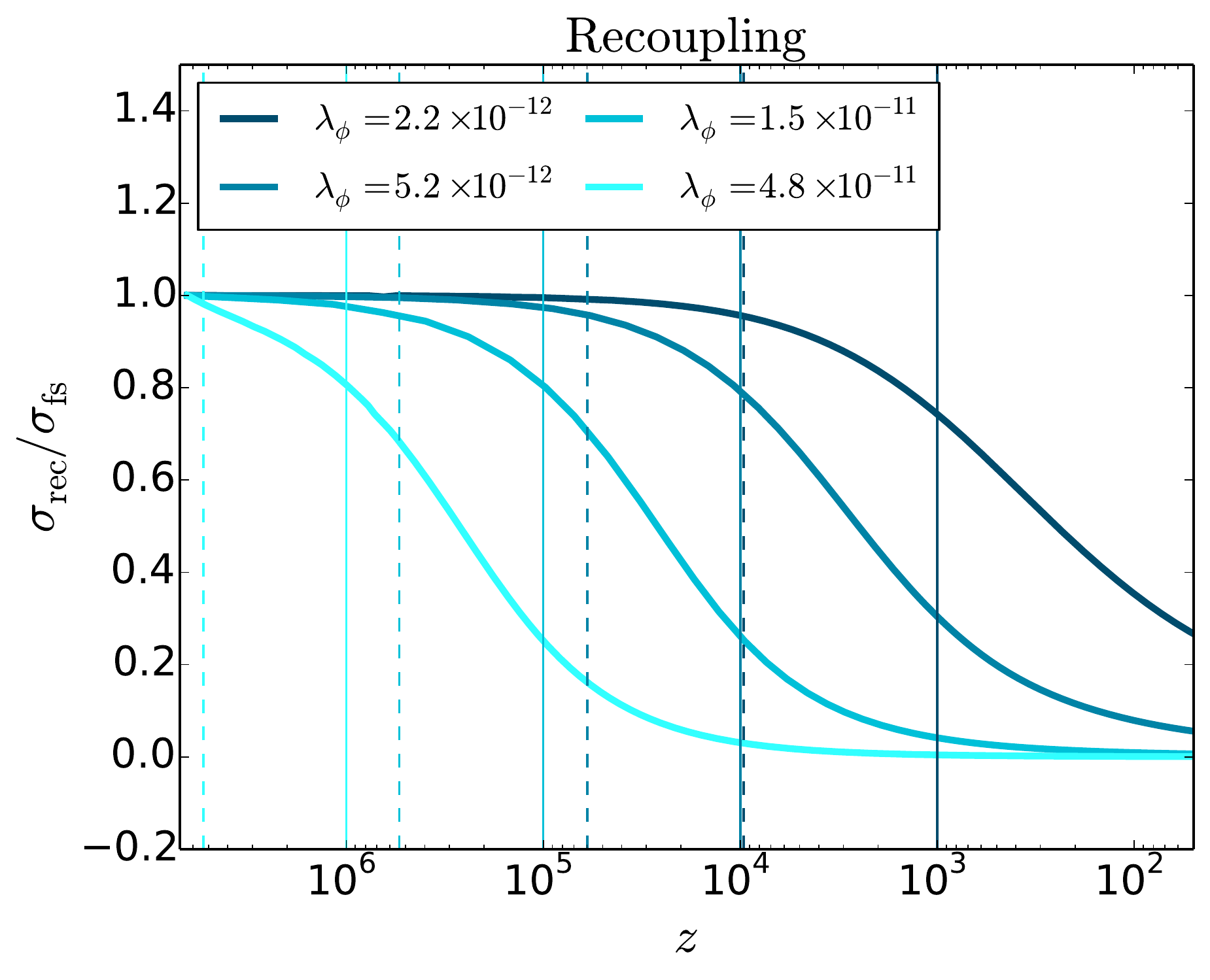}  &
	\includegraphics[width=5.8cm]{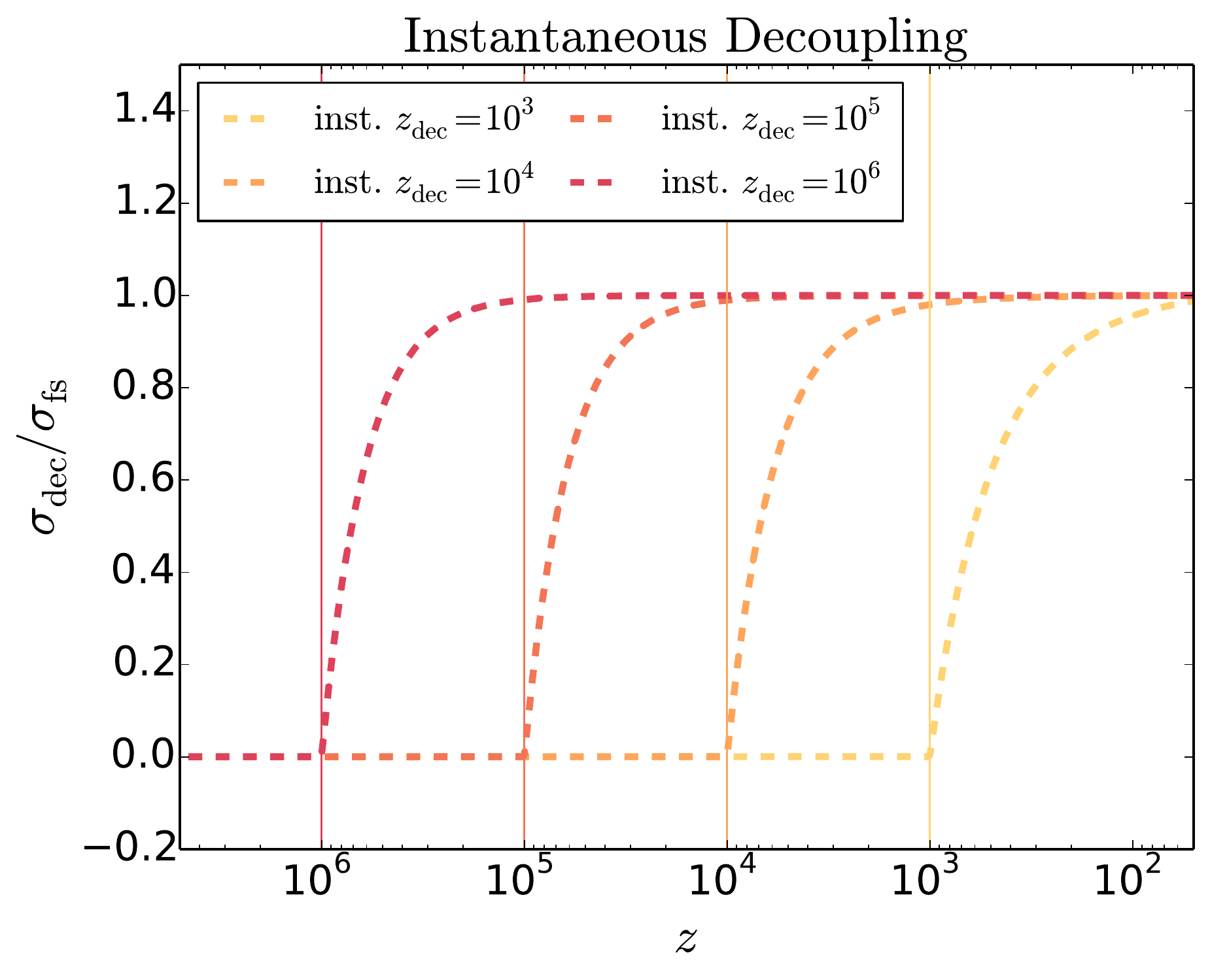}\\
	\mbox{(a)}&\mbox{(b)}&\mbox{(c)}
	\end{array}$
	\caption{The ratio of anisotropic stress of decoupling $(a)$, recoupling $(b)$ and instantaneous decoupling $(c)$ DR to that of free-streaming DR ($\sigma_{\rm dec,rec}/\sigma_{\rm fs}$) as a function of redshift $z$. In all panels we choose $k=10^{-3}\textrm{/Mpc}$ and a cosmology with $\Delta N_{\rm eff}=0.5$ in interacting DR compared with free-streaming DR.  Panel $(a)$ shows multiple values of $G_{\rm eff}$ for the decoupling case and panel $(b)$ multiple values of $\lambda_\phi$ for the recoupling case. Vertical lines show the estimate of $z_{\rm dec}$ $(a)$ and $z_{\rm rec}$ $(b)$ from our definition $\alpha_2 \langle\Gamma\rangle=H$ (solid lines, in both panels chosen to be $z_{\rm dec/rec} = 10^3,\, 10^4,\, 10^5,\, 10^6$) and the usual definition $\langle\Gamma\rangle=H$ (dashed lines). The two estimates are similar for decoupling DR, but for the recoupling case the standard definition ($\langle\Gamma\rangle=H$) produces estimates of the recoupling redshift that are significantly larger (note that the dashed line for $\lambda_\phi = 2.2\times 10^{-12}$ is nearly overlapping with the solid line for $\lambda_\phi = 5.2 \times 10^{-12}$). Panel $(c)$ shows the same ratio for the instantaneous decoupling case, for which the two estimates of $z_{\rm dec}$ are identical.} 
	\label{fig:sigma_ratio}
\end{figure}

To develop insight into the physical effects of DR, it is helpful to study the time dependence of self-interactions and identify the time, or redshift, of de/re-coupling transitions. The opacity given in Eqs.~(\ref{eq:opacitydec}) and ~(\ref{eq:opacityrec}), and shown in Fig.~\ref{fig:opacity}, is one way of doing this and characterizes the average scattering of DR particles. The observable effects of interacting DR, however, are due to changes to the behavior of DR perturbations. Therefore, we will define the transition redshift based on the effects on DR perturbations

 As shown in the general Boltzmann hierarchy, Eq.~(\ref{eq:Boltzmann_hierarchy_decoupling}), the time dependence of self-interactions comes from $\alpha_\ell a \langle\Gamma\rangle$. Since the dominant effect comes from $\ell=2$ moment, we define the $z_{\rm dec/rec}$ as
\bea\label{eq:z_dec_rec}
\alpha_2 \langle\Gamma\rangle(z)=H(z)\Big|_{z=z_{\rm dec/rec}}.
\eea
We note that this is different than the usual definition $\langle\Gamma\rangle=H$ by a relaxation time coefficient $\alpha_2$. These two definitions are similar if $\alpha_2$ is close to unity, as in the decoupling DR case. For recoupling DR, however, $\alpha_2$ is nearly one order of magnitude smaller than unity and the rate has only weak $z$ dependence ($ \langle\Gamma\rangle\propto z$), resulting a big shift of the value of $z_{\rm dec/rec}$. More generally,  $\alpha_2$ might be different than one for other types of interactions, and thus a dedicated calculation of $\alpha_2$ is needed to determine $z_{\rm dec/rec}$. 

To illustrate the time dependence of self-interactions on perturbations and the estimate of $z_{\rm dec/rec}$, we present in Fig.~\ref{fig:sigma_ratio} the ratio of the anisotropic stresses from interacting DR and free-streaming DR, $\sigma_{\rm dec,rec}/\sigma_{\rm fs}$, where $\sigma = 2F_2$. These are shown for different choices of $G_{\rm eff}$ for the decoupling case and $\lambda_\phi$ for the recoupling case. Plotted are the ratios of the $k = 10^{-3}/\textrm{Mpc}$ modes, which are super-horizon for the times shown and clearly illustrate the time-dependent changes to the behavior of DR. Fluid-like DR should have $\sigma =0$, whereas free-streaming DR will have non-zero $\sigma$. Normalizing by the anisotropic stress of the free-streaming case removes the additional cosmological time dependence. 

The ratio in Fig.~\ref{fig:sigma_ratio} has the expected behavior, evolving from $0$ to $1$ for the decoupling cases and $1$ to $0$ for the recoupling cases. The vertical grid lines show two definitions of $z_{\rm dec/rec}$: our definition  $\alpha_2 \langle\Gamma\rangle=H$ (solid) and the usual definition $\langle\Gamma\rangle=H$ (dashed). In the decoupling panel $(a)$ in Fig.~\ref{fig:sigma_ratio}, the two definitions of $z_{\rm dec}$ provide similar results because $\alpha_2$ here is close to one and $\langle\Gamma\rangle\propto z^5$. For the recoupling case in panel $(b)$, however, the two definitions of $z_{\rm rec}$ differ by nearly one order of magnitude. Our definition gives a better estimate of the transition time for $\sigma_{\rm rec}/\sigma_{\rm fs}$. For comparison we also plot the ratio of $\sigma_{\rm dec}/\sigma_{\rm fs}$ for the instantaneous decoupling case in panel $(c)$. For instantaneous decoupling, the interaction rate has a sharp transition around the parameter $z_{\rm dec}$ in Eq.~(\ref{eq:Instantaneous_decoupling}) with a large  $\langle\Gamma\rangle$ at $z>z_{\rm dec}$ and $\langle\Gamma\rangle=0$ for $z<z_{\rm dec}$. Both definitions ($\alpha_2 \langle\Gamma\rangle=H$ or $ \langle\Gamma\rangle=H$) therefore provide the identical results: the parameter $z_{\rm dec}$ naturally denotes the decoupling redshift.

\section{Impact on Observables}
\label{sec:observables}

In this section, we discuss how the interacting dark radiation scenarios described in Sec.~\ref{sec:models} affect the CMB temperature and polarization power spectra, and the matter power spectrum. While the physical effects of fluid-like, decoupling, and recoupling radiation on CMB power spectra are discussed in, e.g., \cite{Bashinsky:2003tk, Cyr-Racine:2013jua, Baumann:2015rya, Choi:2018gho, Archidiacono:2013dua} and on matter power spectra in \cite{Kreisch:2019yzn}, we will review them here for completeness. Our focus in this section is to extend the analysis of the effects of interacting DR on CMB and matter power spectra to describe the different changes made by different types of interactions. The main new results are a discussion of the differences between interacting scenarios with radiation that decouples instantaneously versus via a slower transition described by decoupling from four-Fermi interaction, or radiation that starts out as free-streaming and recouples at a later time. As we shall see, each of the five examples raised in Sec.~\ref{sec:models} --  free-streaming, fluid, decoupling, instantaneous decoupling, and recoupling -- leave a distinct imprint on the CMB and matter power spectra. 

The CMB and matter power spectra are sensitive to interactions in relativistic dark-sector particles because interactions change the behavior of linear perturbations in the stress-energy tensor of the DR, and the DR is gravitationally coupled to the photon and matter perturbations we observe in the CMB and large-scale structure. We first review physics of the two extreme cases, free-streaming and fluid DR, before moving on to discuss radiation that transitions between the two extremes. 

Radiation with frequent interactions can be characterized as fluid-like. Perturbations in a relativistic fluid are completely characterized by the equation of state $w = P/\rho$ and sound speed $c_s = \delta p/\delta\rho$. For a tightly coupled relativistic fluid, these quantities take simple values, $w = c_s^2 = c^2/3$. The only non-zero quantities in the fluid stress-energy tensor are the energy density and pressure, the fluid anisotropic stress and all higher moments vanish, and fluid perturbations propagate at $c_s = c/\sqrt{3}$. On the other hand, perturbations in relativistic free-streaming particles will have the same equation of state $w= 1/3$, and adiabatic sound speed $c_s = \dot{P}/\dot{\rho}$, but perturbation fronts propagate at $c$. Free-streaming particles also have significant anisotropic stress and the full Boltzmann hierarchy, rather than just the continuity equation, is required to describe the evolution of perturbations.  Decoupling or recoupling DR transitions between these two regimes. As shown in Figs.~\ref{fig:opacity} and \ref{fig:sigma_ratio}, the duration of the transition varies with the model. The process of decoupling from self-interactions is relatively slow in comparison with instantaneous DR decoupling and the process of recoupling is even slower.

\begin{figure}
	\centering
	$\begin{array}{ccc}
	\includegraphics[width = 5.8cm]{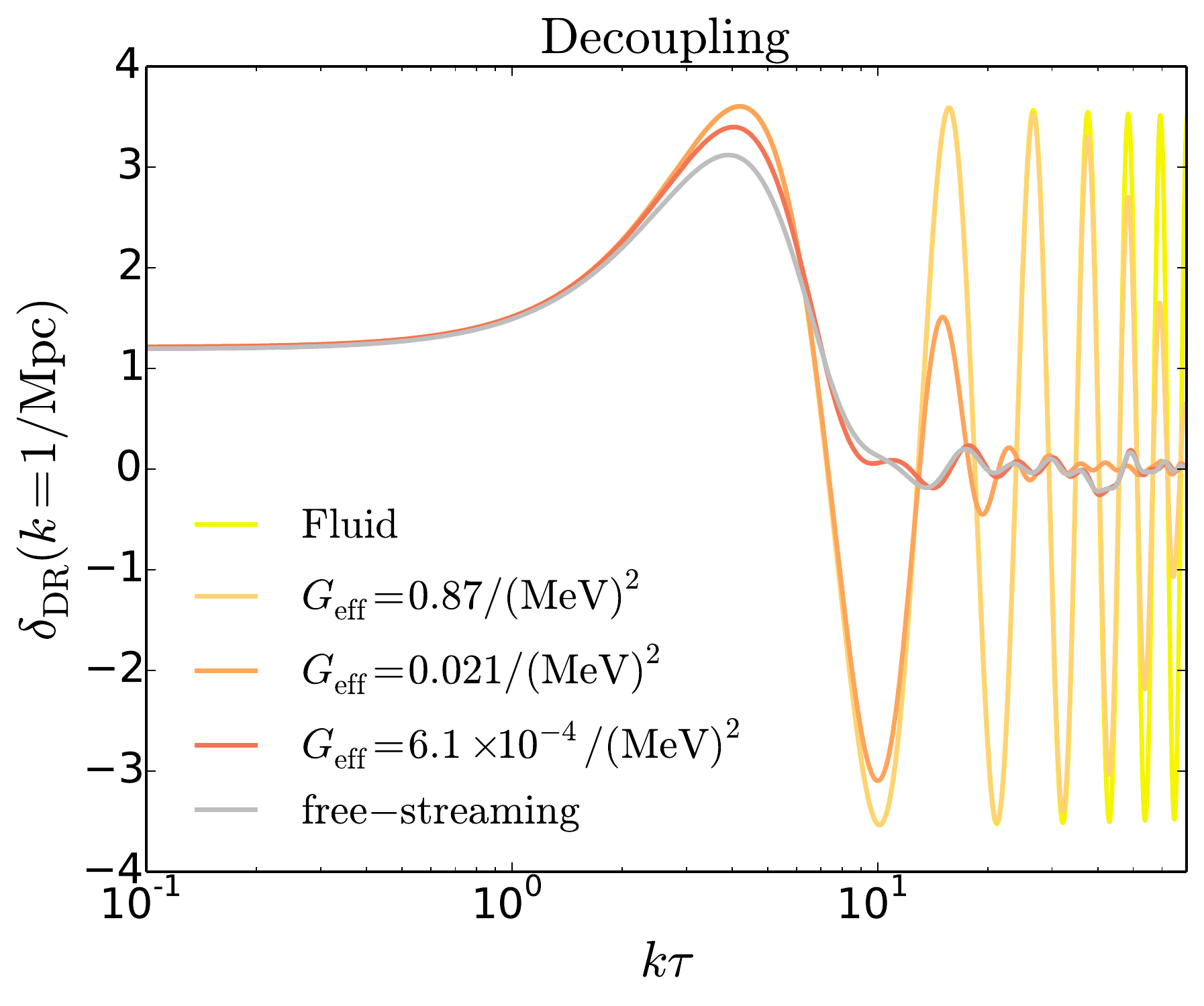}&\includegraphics[width = 5.8cm]{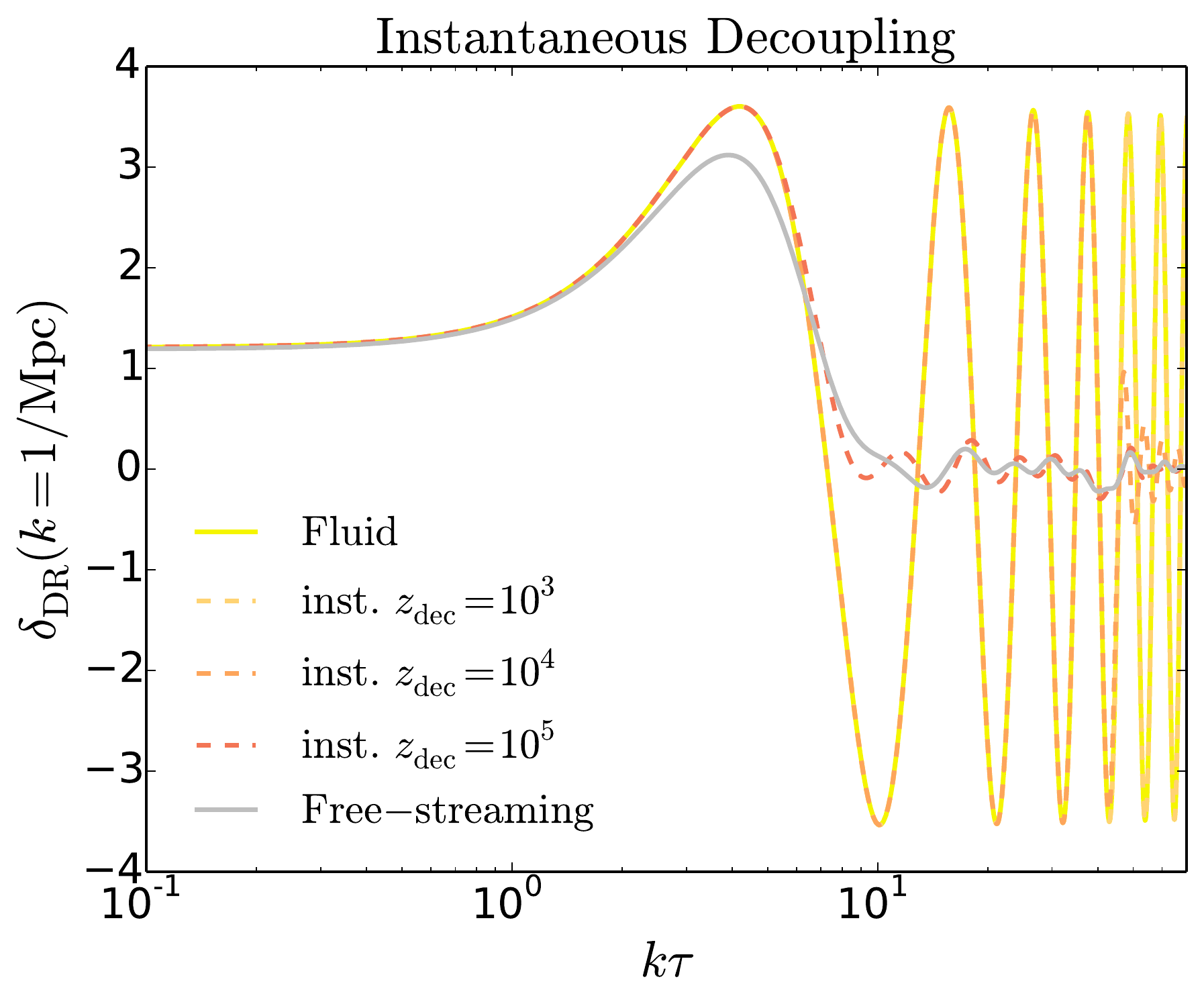} &\includegraphics[width = 5.8cm]{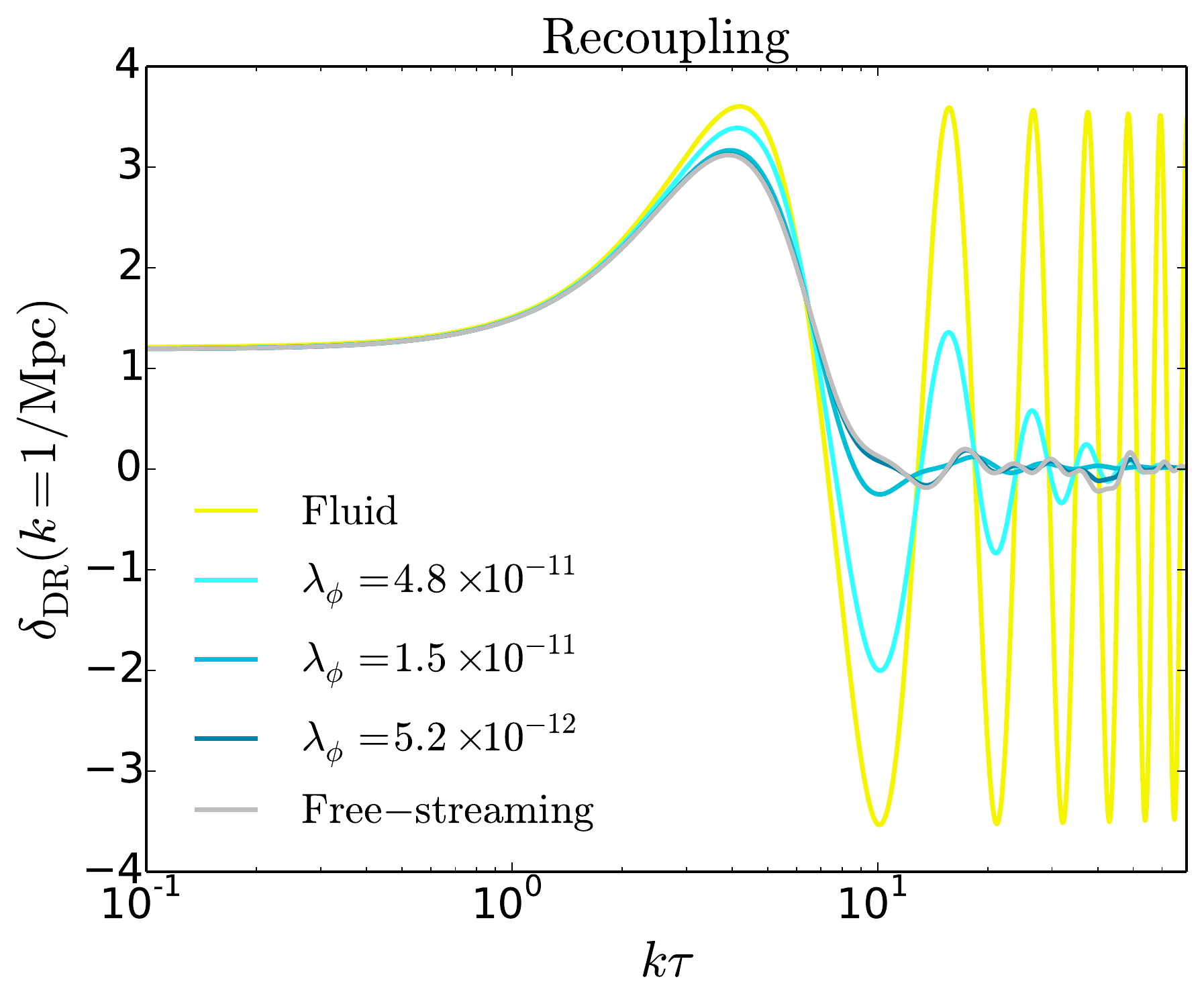}\\
		\mbox{(a)} & \mbox{(b)} & \mbox{(c)} 
	\end{array}$
 	\caption{The evolution of perturbations in the Newtonian gauge dark radiation energy density, $\delta_{\rm DR}$, plotted as a function of $k\tau$, with $k = 1/$Mpc. The left plot compares free-streaming and fluid-like radiation perturbations to decoupling models with values of $\Geff = 0.87/$(MeV)$^2$, $0.021/$(MeV)$^2$, and $6.1\times 10^{-4}/$(MeV)$^2$, corresponding to decoupling at $z_{\rm dec} \sim 10^3, 10^4, 10^5$. The center plot illustrates perturbations that decouple instantaneously at $z_{\rm dec} \sim 10^3, 10^4, 10^5$. The right plot compares free-streaming and fluid-like radiation perturbations to recoupling models with values of $\lambda_\phi = 5.2\times10^{-12},\, 1.5\times 10^{-11},\,$ and $4.8\times 10^{-11}$, corresponding to recoupling at $z_{\rm rec} \sim 10^4, 10^5, 10^6$. The recoupling curve with $\lambda_\phi = 5.2\times 10^{-12}$ is nearly indistinguishable from the free-streaming curve.}
	\label{fig:Deltarad}
	\end{figure}

Differences in the duration of the transition between fluid-like and free-streaming change the evolution of perturbations in the DR energy density. Figure~\ref{fig:Deltarad} illustrates these differences for perturbations to the DR energy density,  $\delta_{\rm DR} = \delta\rho_{\rm DR}/\rho_{\rm DR}$,  with wavenumber $k  = 1/{\rm Mpc}$.\footnote{For figures in this section we use a larger cut-off $\Lambda=10^8/\textrm{Mpc}$ in order to achieve good accuracy at very high $k\tau$ (see discussion below Eq.~(\ref{eq:ave_rate_decoupling})).} The horizon-crossing transition occurs at $c_s k\tau \sim 2\pi$, which for this mode is around $z\sim 10^5$, deep in the radiation-dominated era. For each DR example, we show a range of de/re-coupling times corresponding to transitions occurring prior to horizon crossing, during horizon-crossing, and after horizon crossing. If the decoupling or recoupling transition is complete prior to horizon crossing, then the DR perturbations match the free-streaming and fluid-like cases exactly and we do not show those curves. As expected, the DR perturbations that decouple instantaneously show a nearly immediate transition between following fluid and free-streaming curves. On the other hand, the decoupling radiation with slower transition begins to deviate from the fluid-like curve at an earlier time and takes longer time to approach the free-streaming solution. Finally, the recoupling DR transition is sufficiently slow that even for $z_{{\rm rec}} \sim 10^6$ ($\lambda_\phi = 4.8\times 10^{-11}$), the evolution does not fully match the fluid solution.

The DR perturbations couple gravitationally to the photon and matter perturbations that generate CMB and matter power spectra. Fluid DR will undergo acoustic oscillations along with the photon-baryon fluid (see Fig. \ref{fig:Deltarad}), enhancing the amplitude of acoustic oscillations in photon-baryon fluid, and therefore CMB temperature and polarization power spectra. Free-streaming particles do not participate in acoustic oscillations, and thereby suppress the amplitude of oscillations in CMB power spectra. Additionally, since perturbations in a relativistic fluid to propagate at a speed $c_s < c$, while perturbations in free-streaming radiation propagate at $c$, there is a slight difference in horizon crossing times. This subtle difference, just barely visible as a shift to the right for the interacting cases in Fig.~\ref{fig:Deltarad}, induces a phase shift in the acoustic peaks in CMB power spectra. This phase shift is a particularly distinct signature that is hard to mimic via changes in other cosmological parameters \cite{Bashinsky:2003tk, Baumann:2015rya}. 

For DR that transitions between fluid-like and free-streaming during the radiation dominated era, the observables acquire a scale-dependence, with the behavior of each mode depending on the behavior of the DR around horizon crossing. For instance, photon perturbations that cross the horizon while the DR is fluid-like ($k \gg (c_s \tau_{\rm dec})^{-1}$ or $k \ll (c_s \tau_{\rm rec})^{-1}$) will have larger amplitude oscillations and no phase shift, while those that cross while the species is free-streaming  ($k \ll (c_s \tau_{\rm dec})^{-1}$ or $k \gg (c_s \tau_{\rm rec})^{-1}$) will have reduced oscillation amplitude, and a phase shift with respect to the fluid-like case \cite{Cyr-Racine:2013jua, Choi:2018gho, Kreisch:2019yzn}.  Modes that cross the horizon during the transition have intermediate behavior. This is visible in Fig.~\ref{fig:Deltarad}. Consider the decoupling case in panel (a), the curve with $G_{\rm eff} \ge 0.021$ ($z_{\rm dec} \lesssim 10^4$) matches the fluid curve through horizon crossing and only later starts to depart. The curve with $G_{\rm eff} = 6.1\times 10^{-4}$ ($z_{\rm dec} \sim 10^5$) matches the fluid case as it begins to oscillate, then gradually switches to follow the free-streaming curve before the first oscillation is complete. Similar features are visible in panel (b), though the transition between fluid and free-streaming behavior is much more abrupt. The recoupling scenarios are shown in panel (c). Recoupling is so gradual that none of the curves follow the fluid one exactly, though for the earliest recoupling case,  $\lambda = 4.8 \times 10^{-11}$ corresponding to $z_{\rm rec}\sim 10^6$, the DR is sufficiently coupled to undergo acoustic oscillations.

\begin{figure}[t!]
\centering
	$\begin{array}{cc}
	\includegraphics[width = 7cm]{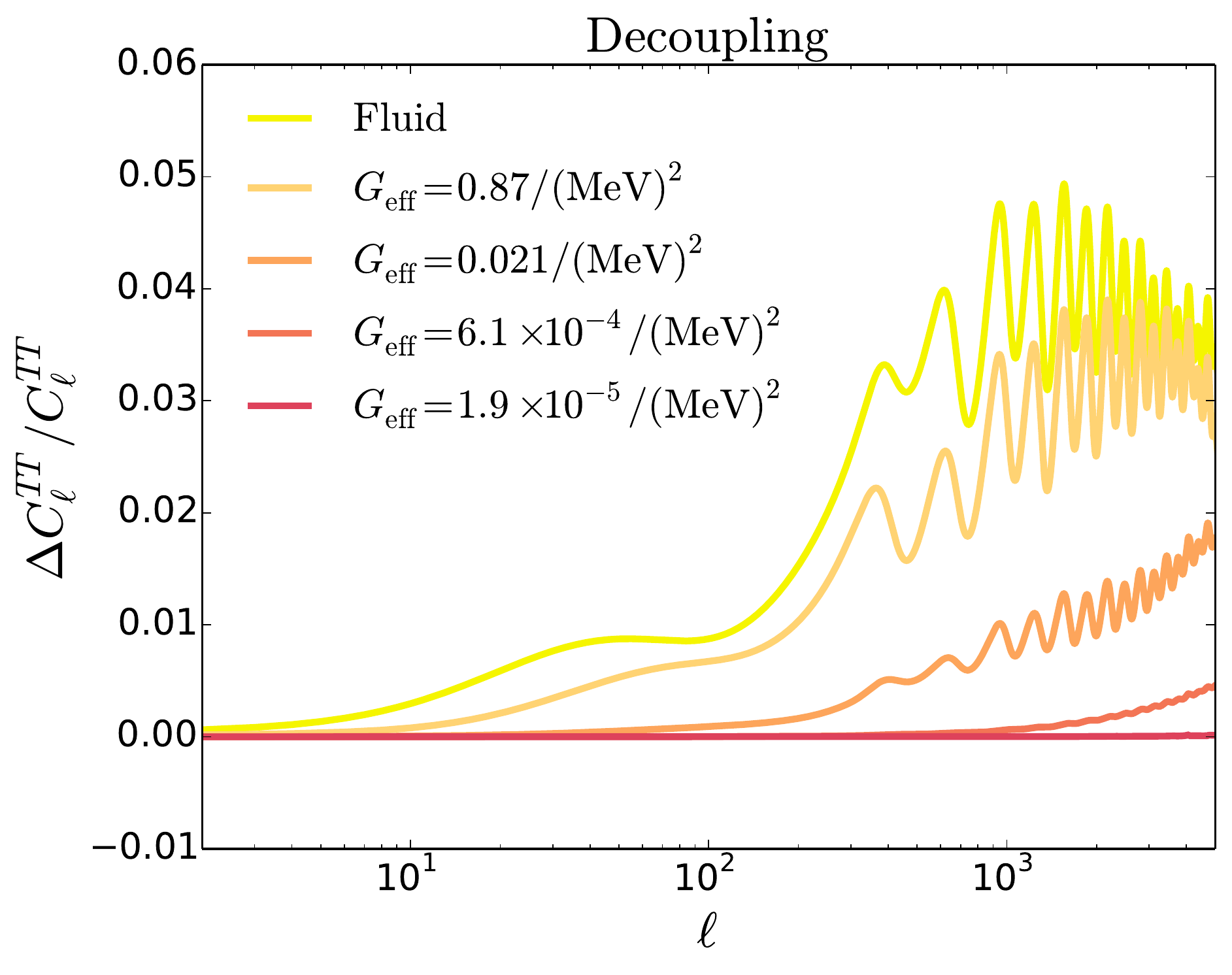} &\includegraphics[width = 7cm]{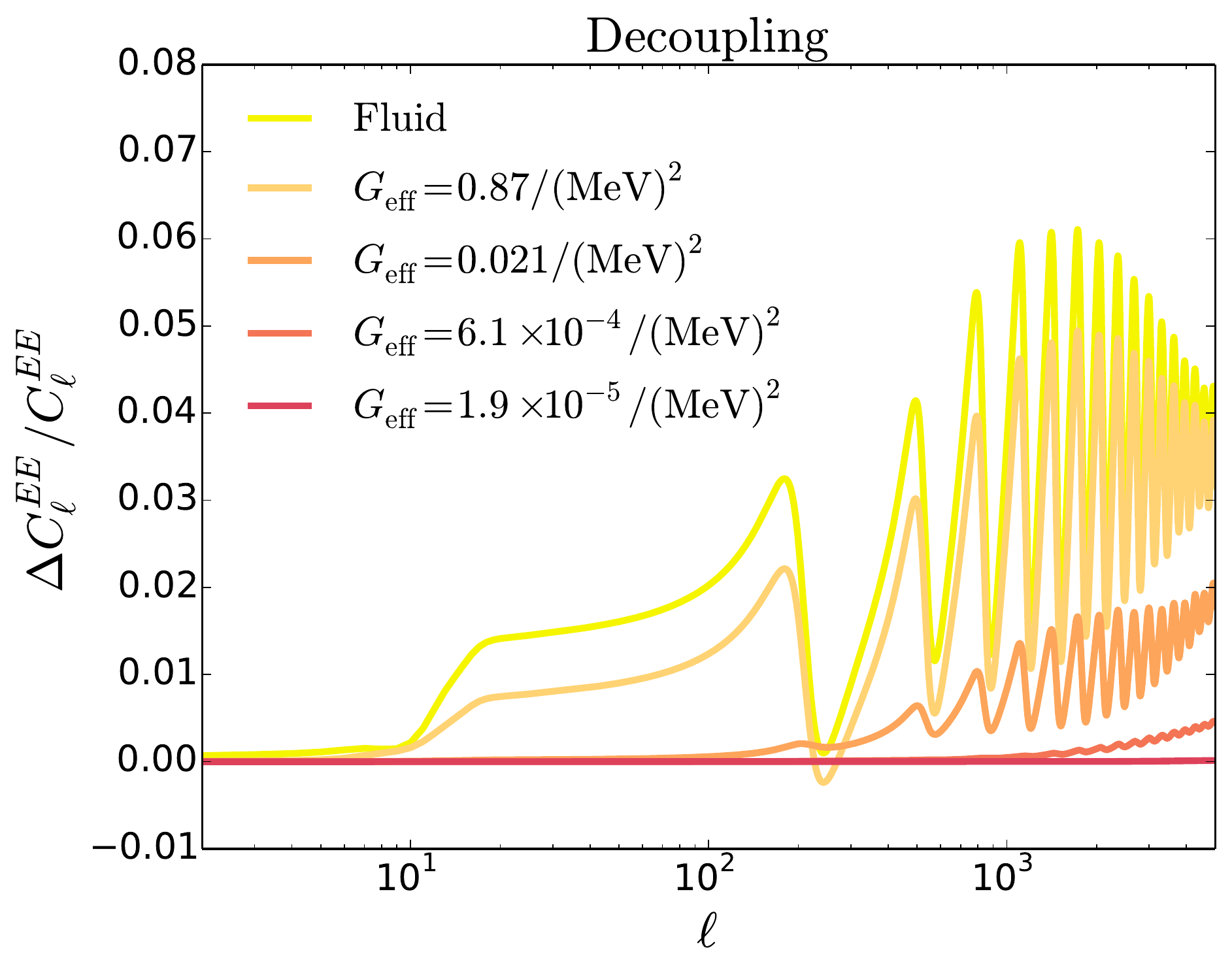}\\
	\mbox{(a)} & \mbox{(b)}\\
	\includegraphics[width = 7cm]{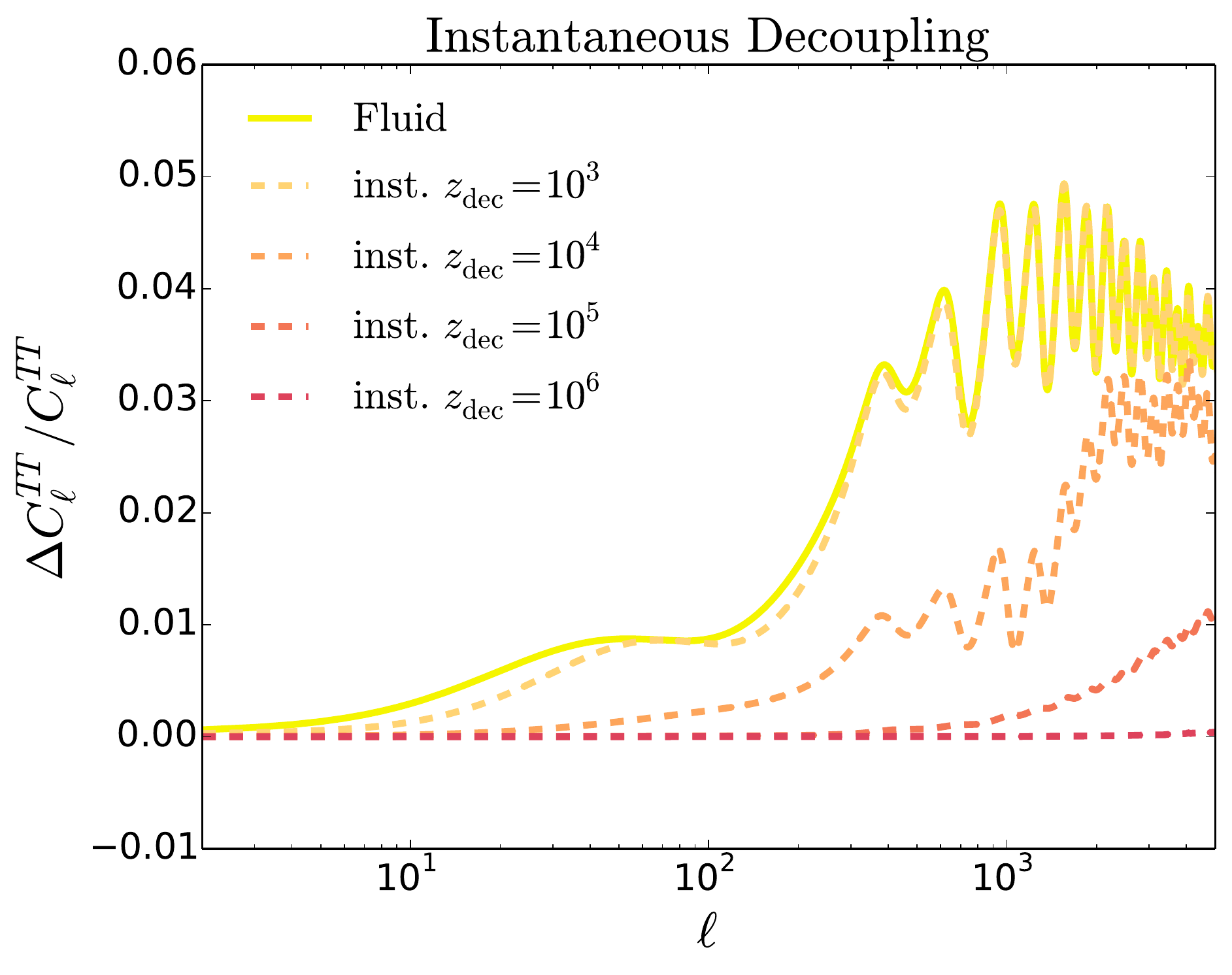} &\includegraphics[width = 7cm]{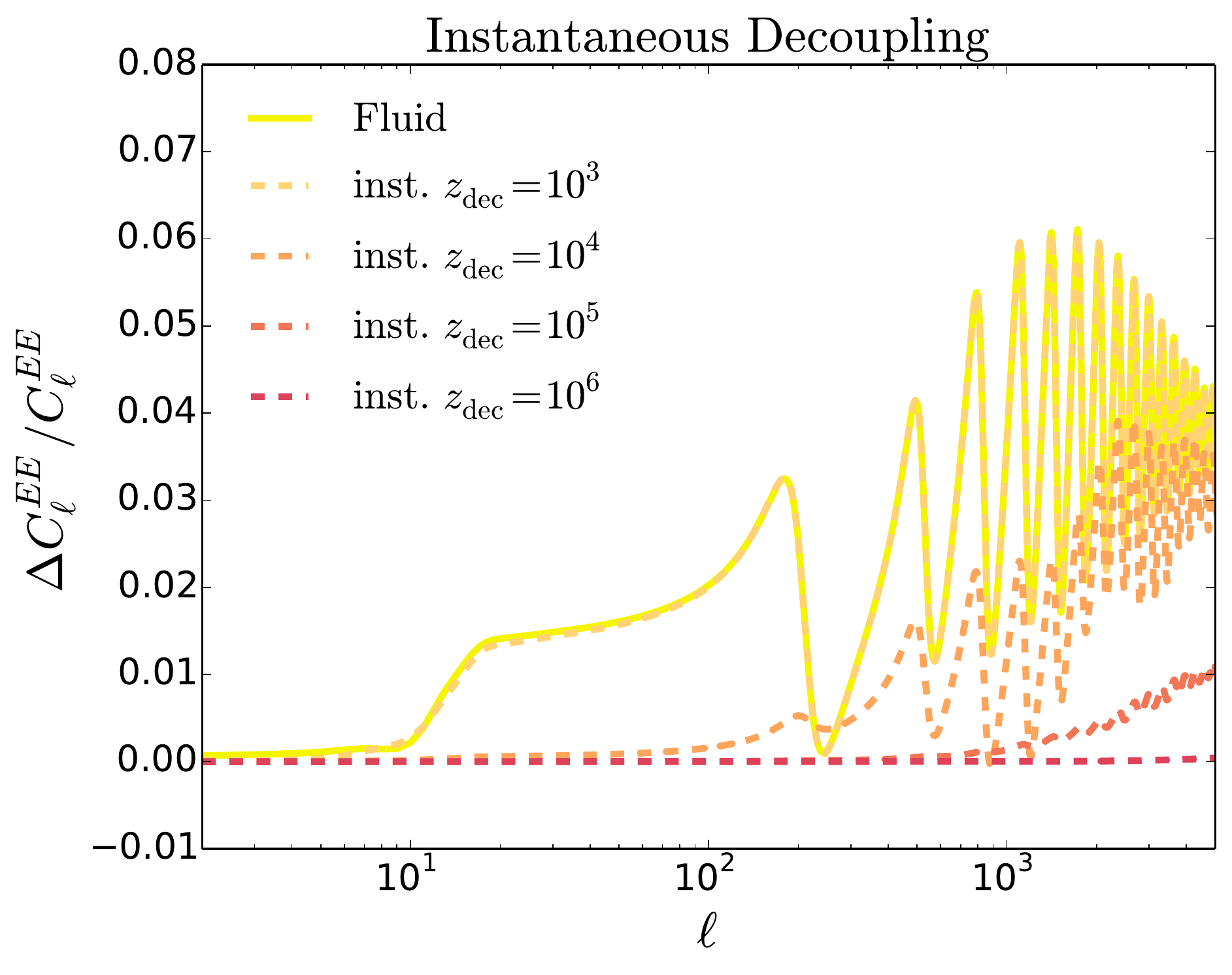}\\
	\mbox{(c)} & \mbox{(d)}\\
	\includegraphics[width = 7cm]{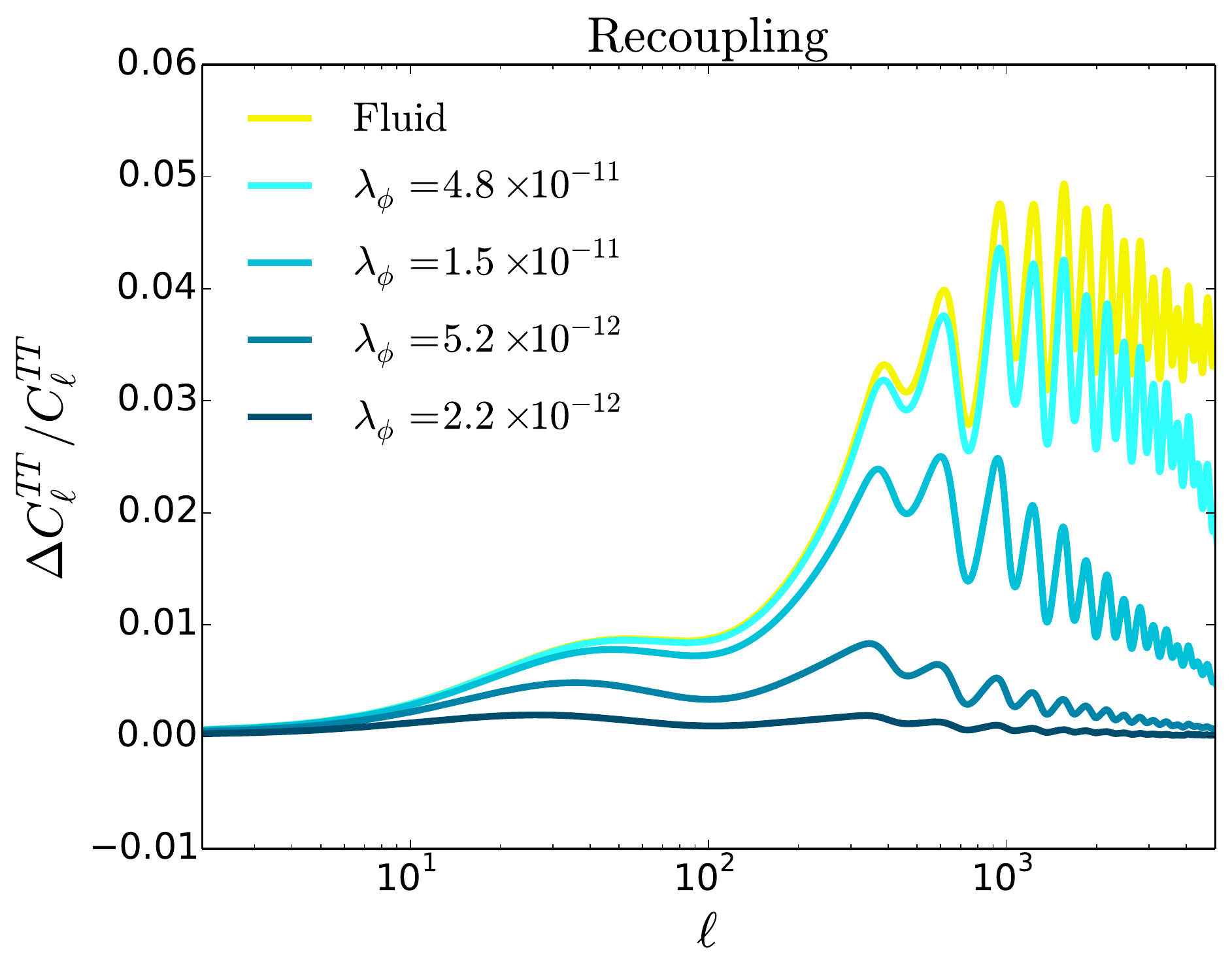} &\includegraphics[width = 7cm]{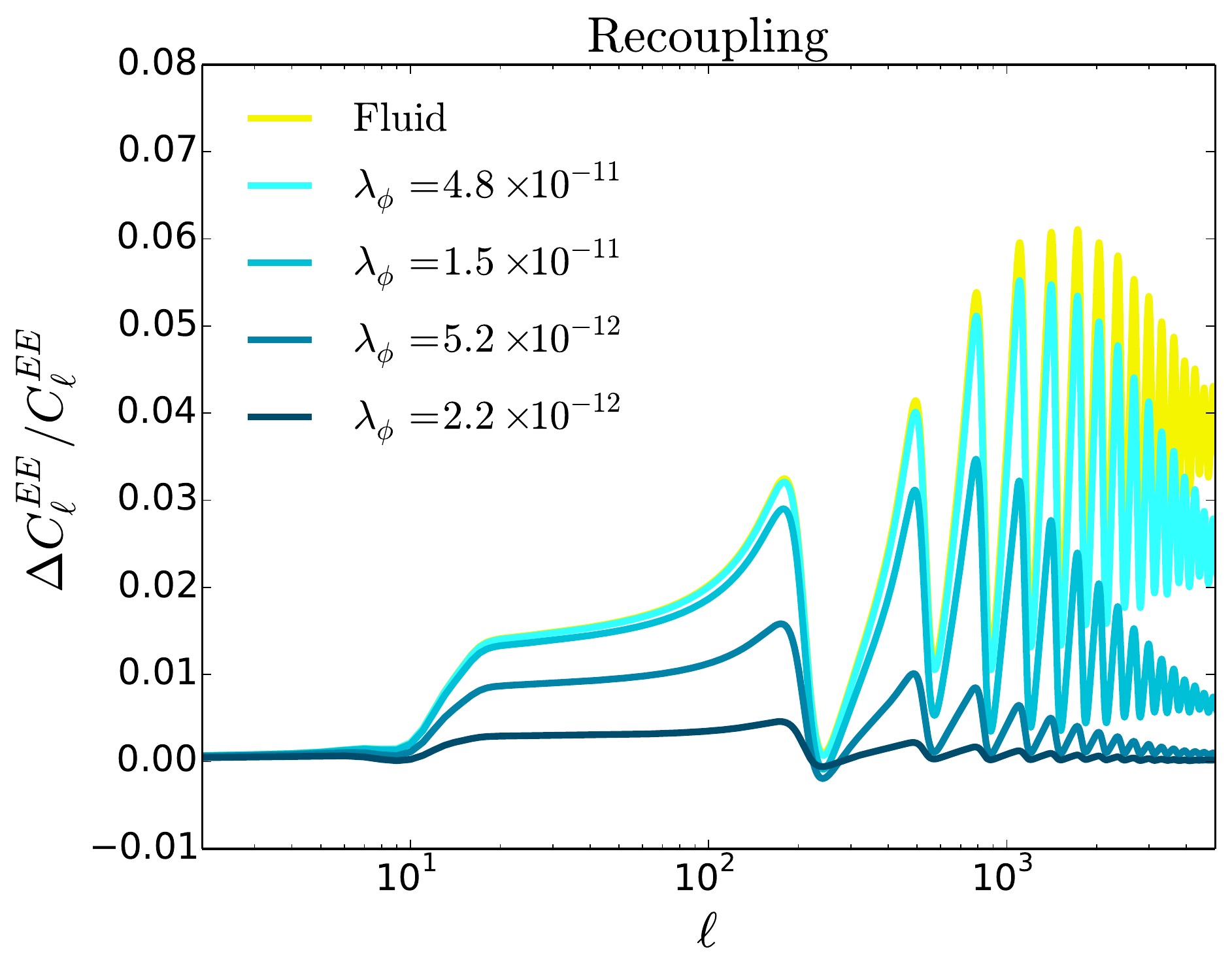}\\	
	\mbox{(e)} & \mbox{(f)}
\end{array}$
 	\caption{Fractional changes in the primary CMB power spectra when an additional interacting component of DR is added with $\Delta \Neff =0.5$ compared to power spectra for a cosmology with the total $\Neff$ entirely composed of free-streaming radiation. Panels (a) and (b) show the fractional change in $C^{TT}_{\ell}$ and $C^{EE}_\ell$ for the decoupling model and for a fluid-like species that never decouples. Panels (c) and (d) show the fractional change in $C^{TT}_{\ell}$ and $C^{EE}_\ell$ for the instantaenous decoupling model (dashed lines) along with the fluid case for comparison. The values of $\zdec$ in (c) and (d) are chosen to match the decoupling redshifts for each value of $\Geff$ used in (a) and (b). Panels (e) and (f) show the fractional change in $C^{TT}_\ell$ and  $C^{EE}_\ell$ for recoupling species and a fluid-like species.}
	\label{fig:DeltaCls}
	\vspace{-0.2cm}
\end{figure}

The impact of DR on primary CMB power spectra is plotted in Fig.~\ref{fig:DeltaCls}. Each panel illustrates the changes to the CMB power spectra between cosmologies with common background evolution with same total energy density in radiation, $\Neff = 3.046 +\Delta N_\eff$, with $\Delta N_\eff = 0.5$, but differ in the behavior of perturbations of the DR contributing to $\Delta N_\eff$. The different curves compare $\Delta \Neff =0.5$ in interacting DR, relative to $\Delta \Neff =0.5$ in free-streaming radiation. As expected, the amplitude of CMB spectra is enhanced at for all interacting DR and the oscillatory features demonstrate that the location of the peaks is shifted slightly. The enhancement is largest for DR that is interacting for longer, either decoupling at a later time, or recoupling at an earlier time. After matter-radiation equality, the fractional contribution of DR to the energy density drops as $1/a$ and whether the DR is interacting or not is increasingly less relevant. For instance, recoupling transition at $z\sim 1000$, the effect of the power spectra is very small. For decoupling at $z \gtrsim 10^6$ ($G_{\rm eff} \lesssim 10^{-5}$/MeV$^2$) the power spectra in this $\ell$-range are indistinguishable from the purely free-streaming case. 

In Fig.~\ref{fig:DeltaCls}, the primary difference between the decoupling and instantaneous decoupling scenarios is the amplitude of the change to the power spectra: since the decoupling model begins the transition to free-streaming earlier (see Fig.~\ref{fig:opacity} and Fig.~\ref{fig:sigma_ratio}), the effect on CMB spectra is smaller. As noted in \cite{Brinckmann:2020bcn}, there is some degeneracy between the width of the decoupling transition and the amplitude of the effect. There are, however, additional features that are distinct between the two scenarios. The CMB spectra at multipole $\ell$ have peak sensitivity to wavenumber $k \sim \pi \ell/(2d_{\rm LSS})$, where $d_{\rm LSS}$ is the distance to the surface of last-scattering.  For instantaneous decoupling at $z_{\rm dec} = 10^4$ , the decoupling transition is visible as a relatively sharp change in the CMB spectra around $\ell_{\rm dec} \approx (2/\pi) k_{\rm dec} d_{\rm LSS} \approx 2200$, where $k_{\rm dec} = 2\pi/(c_s\tau_{\rm dec})$. For the earlier decoupling redshifts this feature as at too high of $\ell$ to be detectable in the foreseeable future. For the standard decoupling model the more gradual transition washes this feature out. 

\begin{figure}
	\centering
	$\begin{array}{ccc}
	\includegraphics[width = 5.8cm]{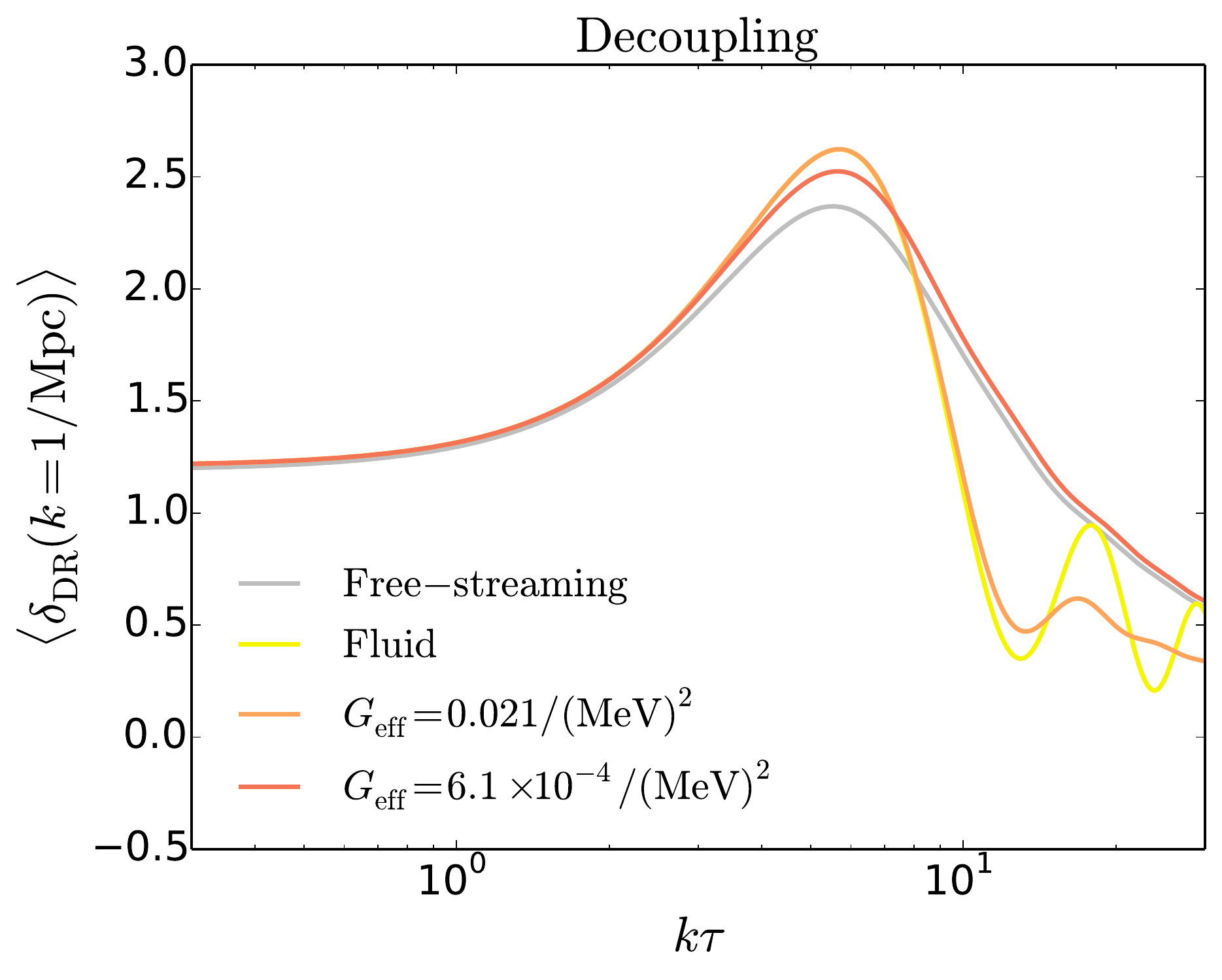}&\includegraphics[width = 5.8cm]{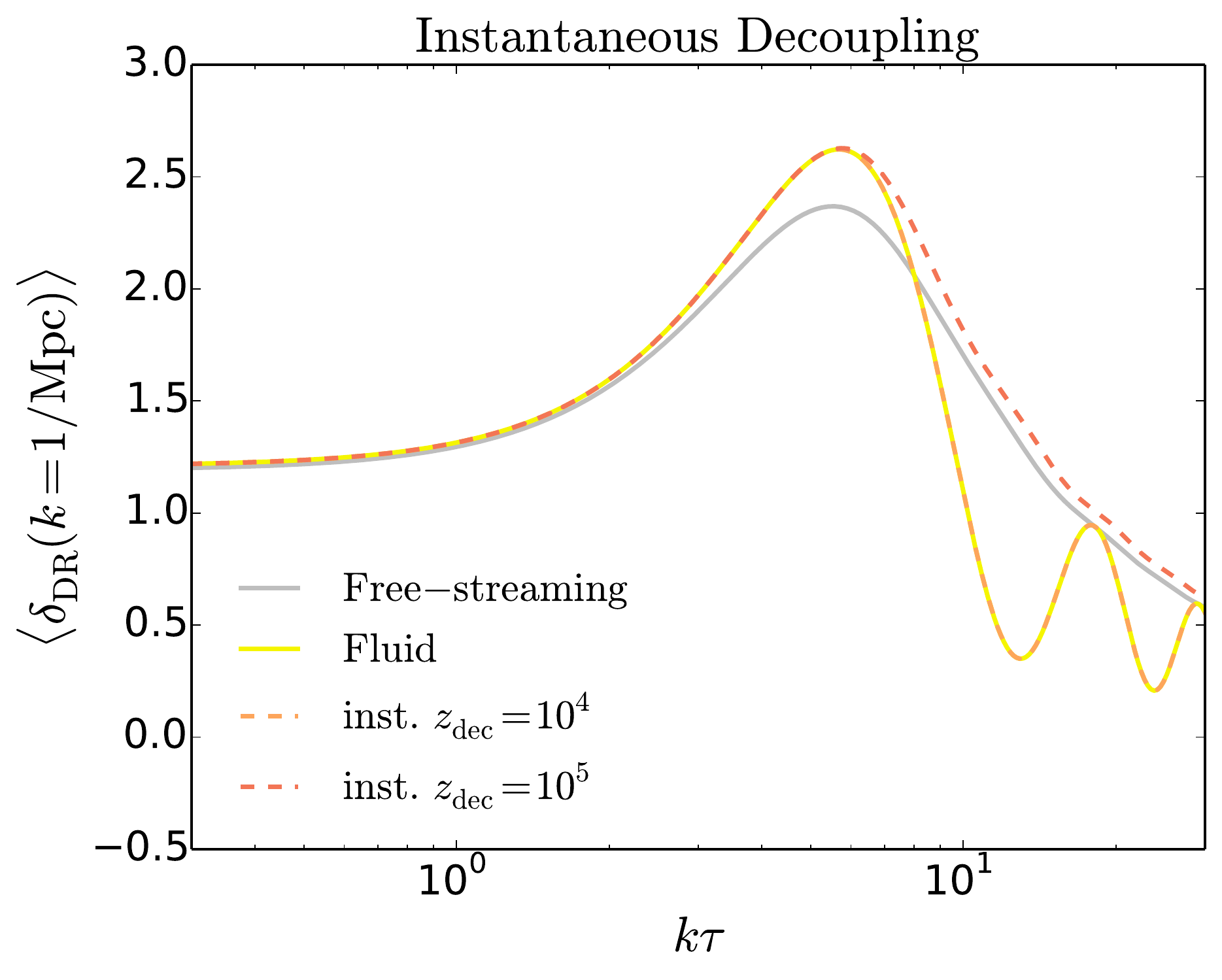} &\includegraphics[width = 5.8cm]{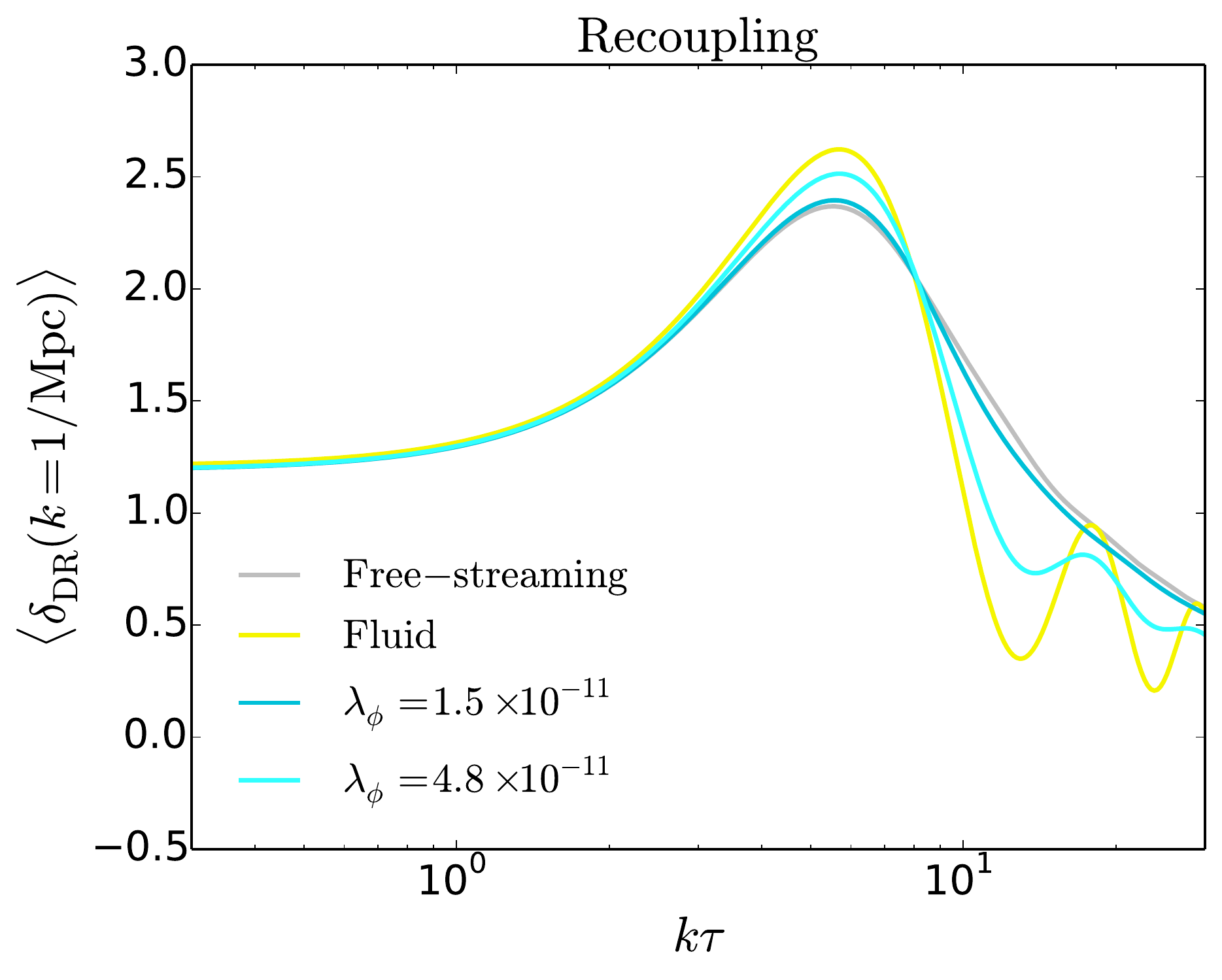}\\
		\mbox{(a)} & \mbox{(b)} & \mbox{(c)} 
	\end{array}$
 	\caption{The time average value of the Newtonian gauge dark radiation energy density, $\langle\delta_{\rm DR}\rangle$, plotted as a function of $k\tau$, with $k = 1/$Mpc. The left plot compares free-streaming and fluid-like radiation perturbations to decoupling models with values of $\Geff = 0.021/$(MeV)$^2$, and $6.1\times 10^{-4}/$(MeV)$^2$, corresponding to decoupling at $z_{\rm dec} \sim 10^4, 10^5$. The center plot illustrates perturbations that decouple instantaneously at $z_{\rm dec} \sim 10^4, 10^5$. The right plot compares free-streaming and fluid-like radiation perturbations to recoupling models with values of $\lambda_\phi =  1.5\times 10^{-11}$, and $4.8\times 10^{-11}$, corresponding to recoupling at $z_{\rm rec} \sim 10^5, 10^6$. }
	\label{fig:Deltaradavg}
	\end{figure}

\begin{figure}
	\centering
	$\begin{array}{ccc}
	\includegraphics[width = 5.8cm]{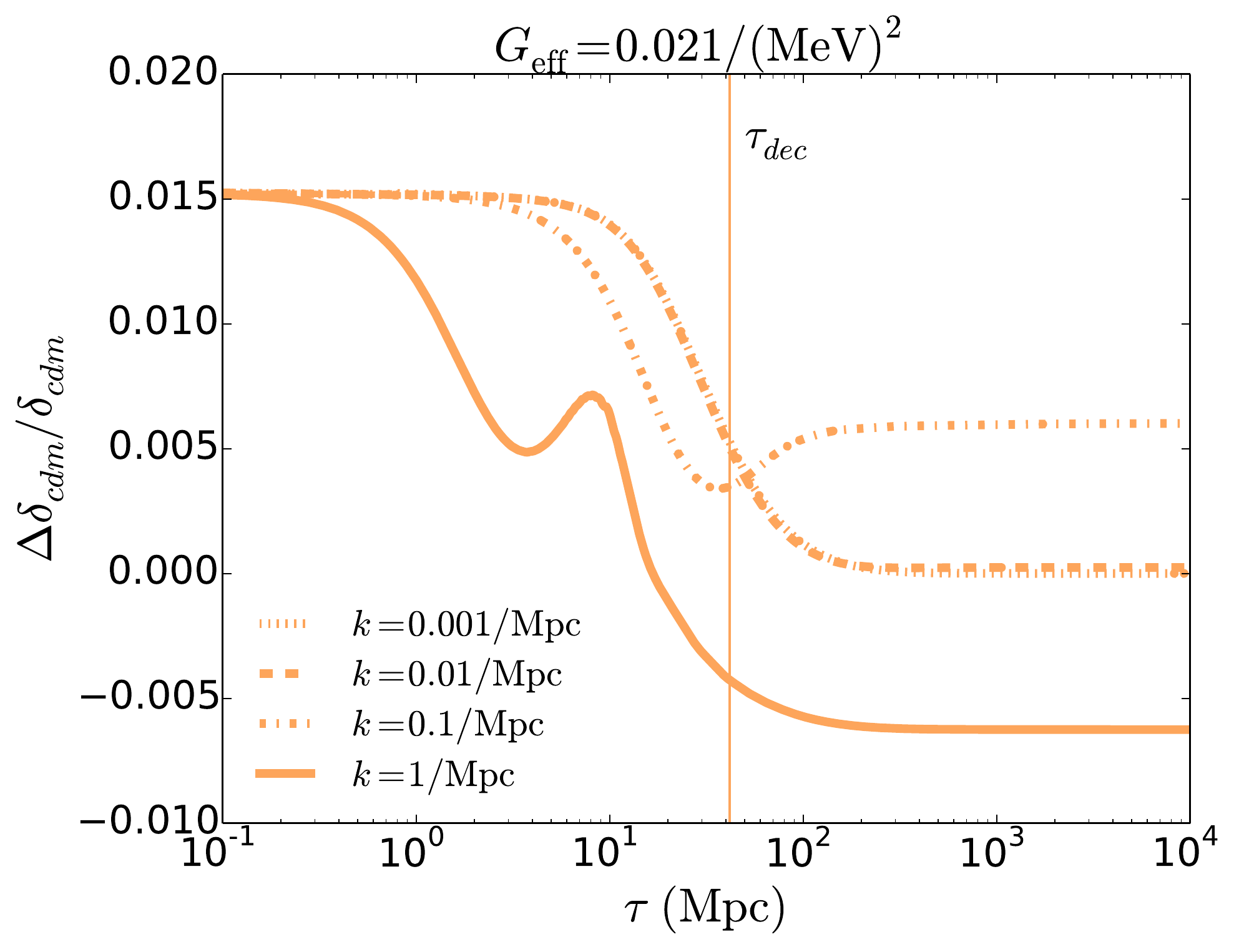} &\includegraphics[width =5.8cm]{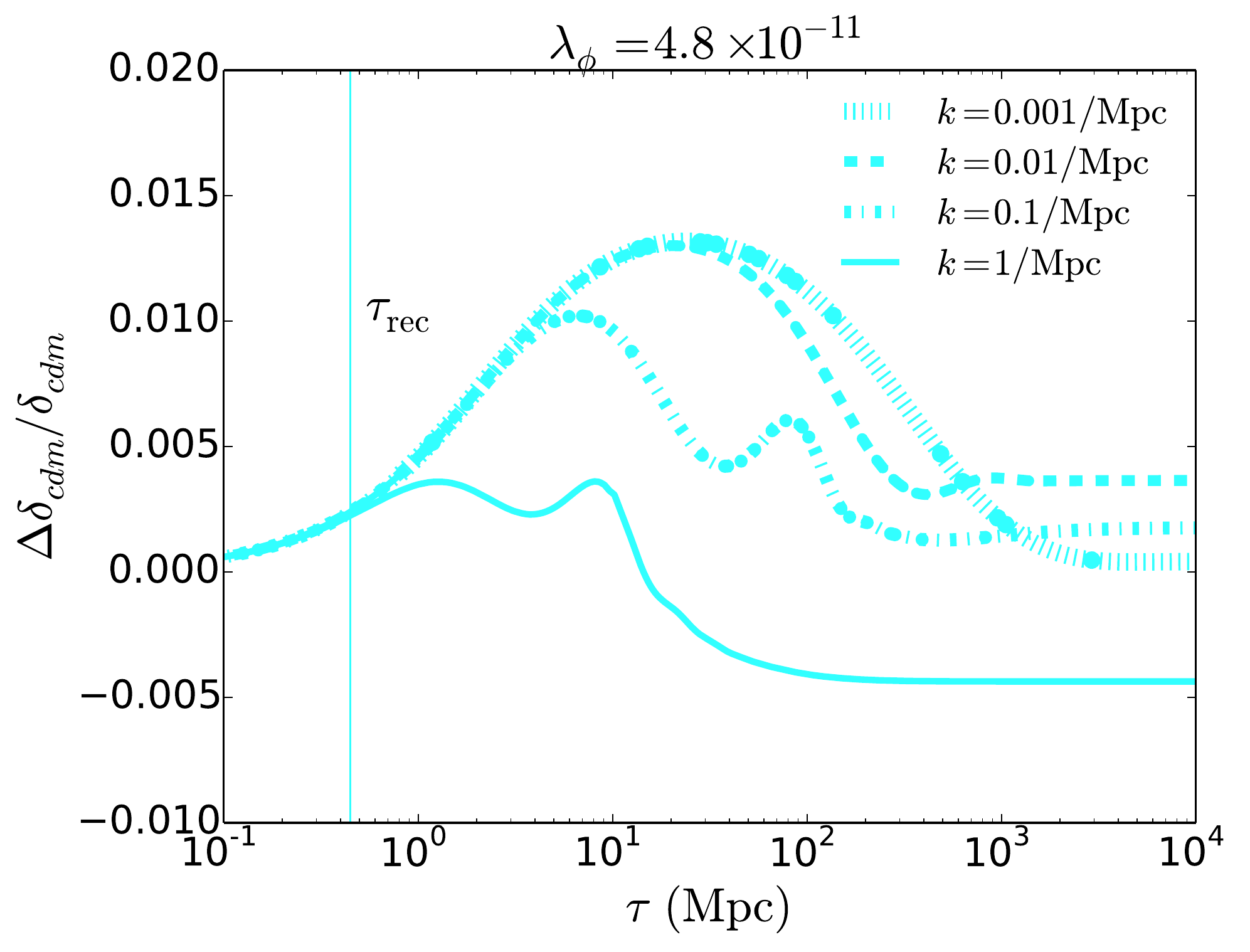}&\includegraphics[width = 5.8cm]{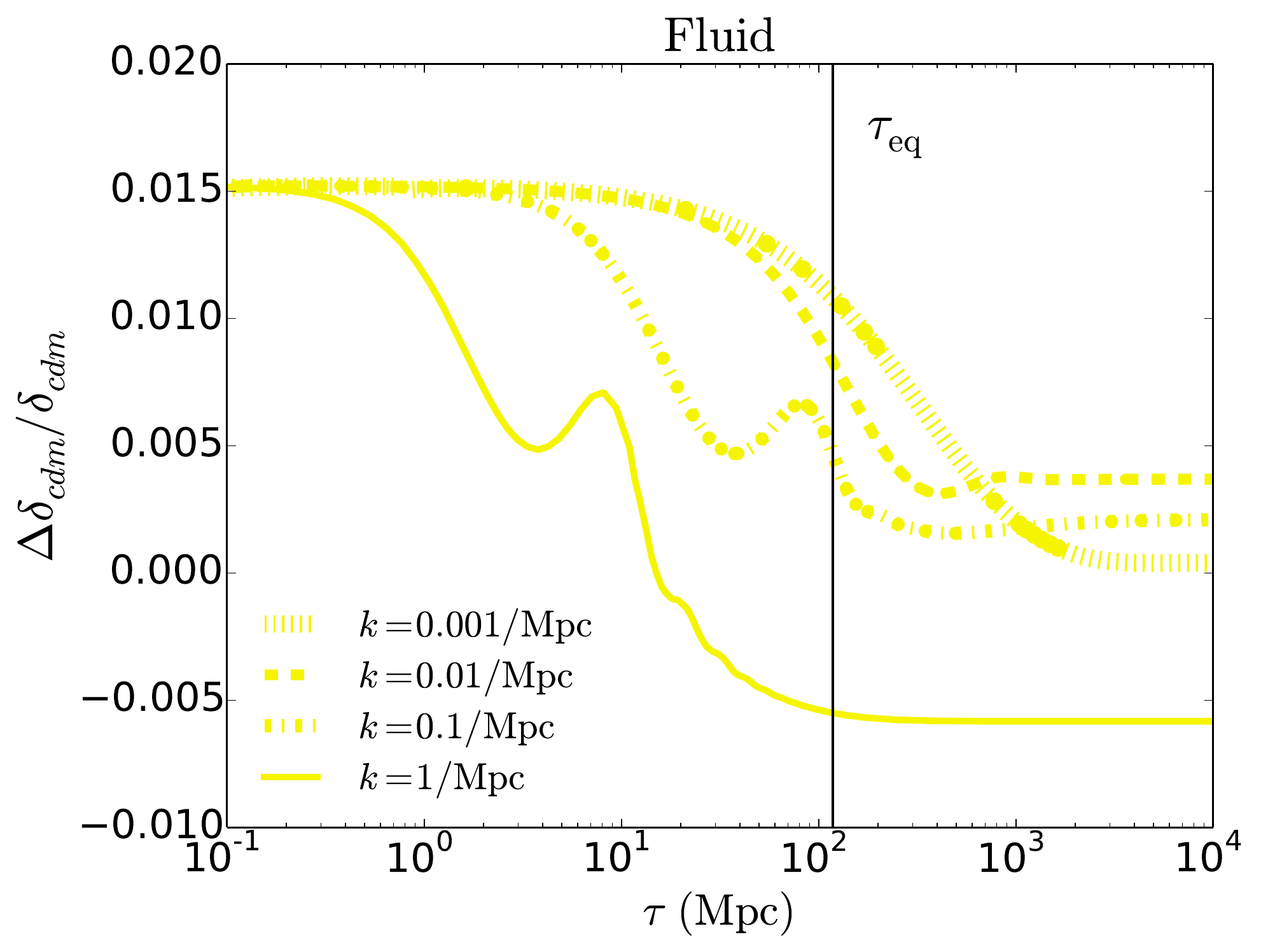}\\
	\mbox{(a)} & \mbox{(b)} & \mbox{(c)} 
	\end{array}$
 	\caption{Fractional change the evolution of dark matter density perturbations of different wavelengths when a component of the radiation density is in decoupling species (left), recoupling species (middle), or fluid-like species (right). The effect on the dark matter density perturbation depends on whether the modes reached $k\tau\sim1$ before or after the decoupling or recoupling transition, and for fluid-like radiation the matter-radiation equality time. These times are marked as vertical lines on each plot. }
	\label{fig:DeltaCDM}
	\end{figure}

During the radiation-dominated era, perturbations in radiation dominate the gravitational potential on large scales. The CDM perturbations are evolving in these potentials and are therefore also sensitive to the different behaviors of interacting DR. To understand how DR impacts CDM, it is helpful to look at the time average of the DR perturbations, which is a better estimate of the net source to the gravitational potential driving CDM. In Fig.~\ref{fig:Deltaradavg}, we show the time average ($\langle \delta_{DR}(\tau)\rangle \equiv \int^\tau_0 d\tau' \delta_{DR}/\tau$) of the radiation perturbations plotted in Fig.~\ref{fig:Deltarad}. It is clear that while the interacting cases have a larger amplitude around horizon crossing, at later times the free-streaming perturbations are larger and therefore source a larger gravitational potential. This means that we expect CDM perturbations entering the horizon during the radiation dominated era to be larger in cosmologies with free-streaming DR in comparison to those with fluid DR. The case where DR decouples around horizon crossing is special, the time-average of the DR is larger at horizon crossing, yet does not decrease subsequently (see Fig.~\ref{fig:Deltarad}) and at late times is slightly larger than the free-streaming case. We will see that the CDM perturbations that cross the horizon around decoupling indeed experience a slight boost. 

The effect of DR on the CDM perturbations is illustrated in Fig.~\ref{fig:DeltaCDM}. Plotted is the fractional change in the amplitude of the Newtonian gauge CDM perturbation, $\delta_{\rm cdm}$, when $\Delta \Neff = 0.5$ in DR is made interacting, as opposed to free-streaming.  Each panel shows a range of $k$-values corresponding to modes that cross the horizon at a variety of epochs relative to the de/re-coupling transitions and matter radiation equality. On super-horizon scales and during the radiation-dominated era, the matter perturbation in the Newtonian gauge is slightly larger in a cosmology with interacting DR (see e.g., \cite{Bashinsky:2003tk} and note that this difference disappears in the synchronous gauge). Interacting radiation experiences a larger boost at horizon crossing (Fig.~\ref{fig:Deltarad}) relative to free-streaming radiation that is transmitted to the CDM. Eventually, however, $\delta_{\rm cdm}$ modes that cross the horizon during the radiation-dominated era with interacting DR experience suppressed growth because, time-averaged over an oscillation, the source to the gravitational potential is smaller than for free-streaming radiation (see Fig.~\ref{fig:Deltaradavg}). Modes that cross the horizon near the decoupling transition experience a mixture of effects, a boost in amplitude at horizon crossing while the DR is interacting, but then a decreased subsequent decay as the radiation transitions to free-streaming before a full oscillation is complete. This is visible in the $k= 0.1/{\rm Mpc}$ mode in panel $(a)$. A similar feature occurs as the Universe transitions to matter domination. Modes that cross the horizon during this transition may experience an initial boost due to interactions, but the subsequent suppression in amplitude does not occur as the radiation perturbations cease to dominate the gravitational potential. This is visible for the $k= 0.1/{\rm Mpc}$ and $k= 0.01/{\rm Mpc}$ modes in the recoupling and fluid plots in panels $(b)$ and $(c)$. 

\begin{figure}
	\centering
	$\begin{array}{ccc}
	\includegraphics[width = 5.8cm]{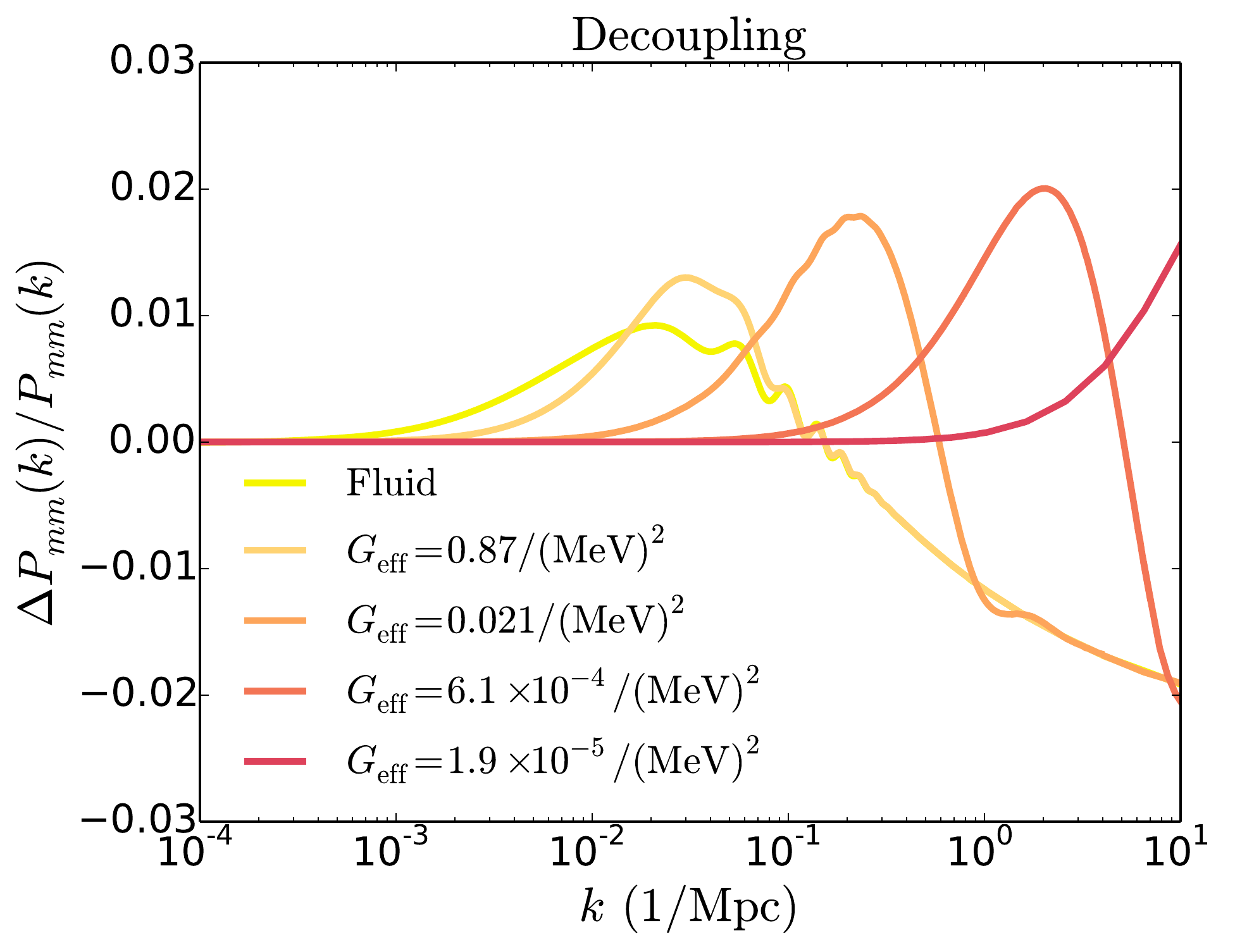} &\includegraphics[width = 5.8cm]{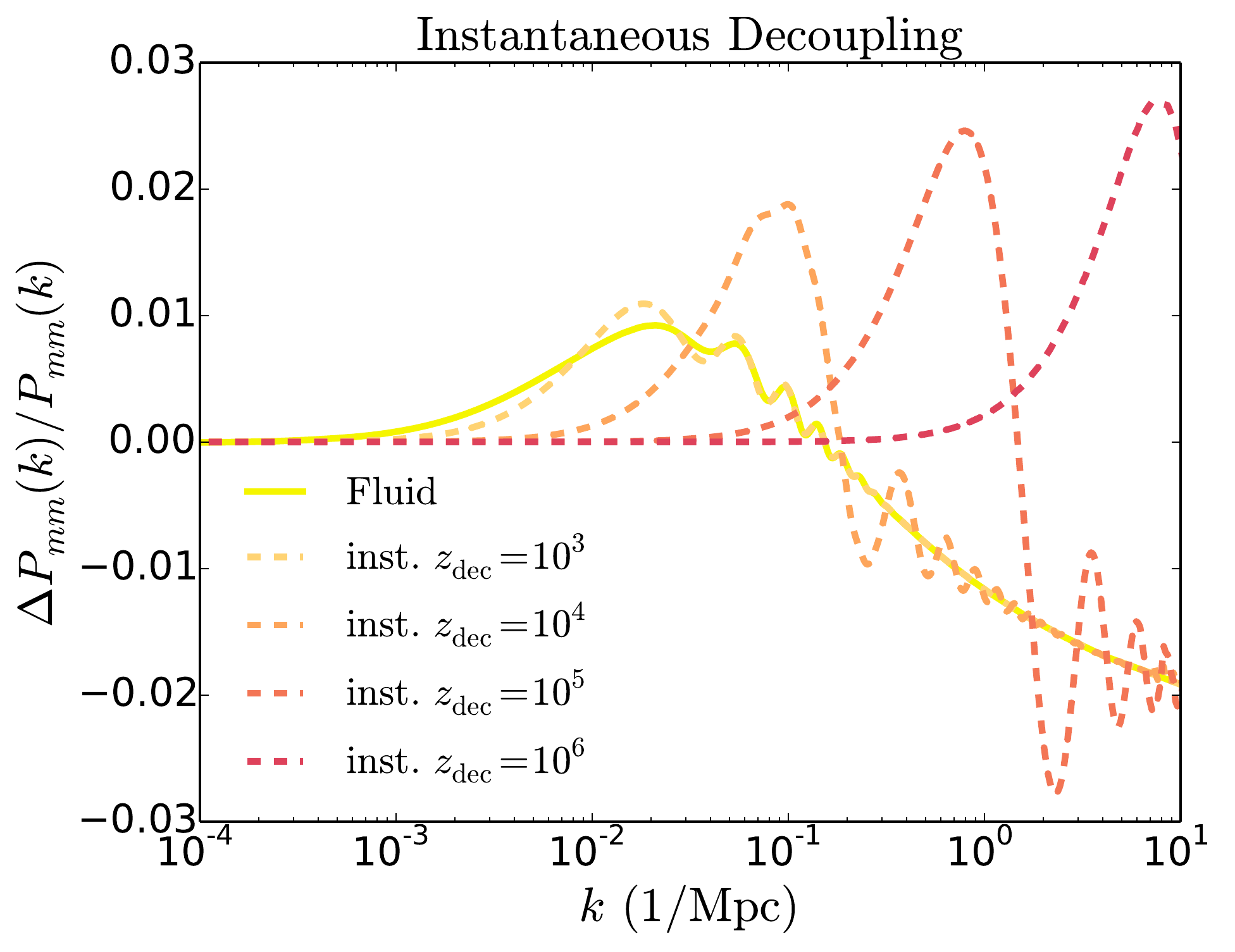} &\includegraphics[width = 5.8cm]{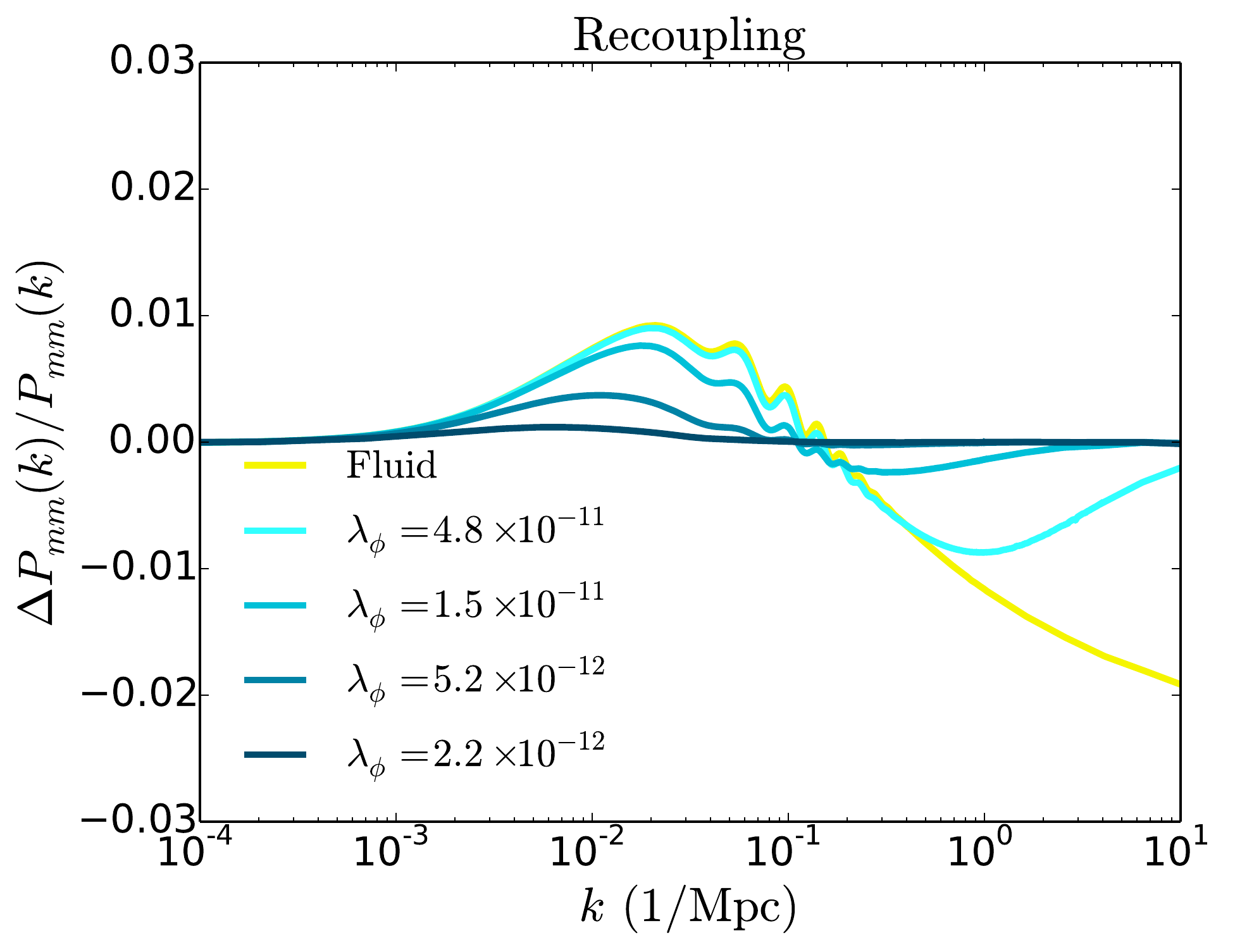} \\
	\mbox{(a)} & \mbox{(b)} & \mbox{(c)} 
	\end{array}$
 	\caption{Fractional changes in the matter power spectrum at $z=0$ when an additional DR contribution of $\Delta \Neff = 0.5$ is interacting instead of free-streaming. The left panel shows $\Delta P_{mm}(k)/P_{mm}(k)$ for the decoupling model, the center panel for the instantaneous decoupling model, and the right panel for the recoupling model. The values of $\Geff$ and $\lambda_\phi$ are chosen to give de/re-coupling redshifts of $z \sim 10^3$, $10^4$, $10^5$, $10^6$.}
	\label{fig:DeltaPmms}
\end{figure}

The scale-dependent time evolution in Fig.~\ref{fig:DeltaCDM} leads to scale-dependent features in the matter power spectrum. Figure~\ref{fig:DeltaPmms} illustrates the $k$-dependent changes to the matter power spectrum when an additional DR contribution of $\Delta \Neff = 0.5$ is interacting instead of free-streaming. For the decoupling cases, the matter power spectrum is suppressed at high $k$, corresponding modes that entered the horizon both during radiation domination and while the DR was interacting. In the cases where decoupling occurs during the radiation-dominated era, a peak appears around wavenumber $k \sim \pi/(c_s\tau_{\rm dec})$ corresponding to the scales that have just reached their first maximum after horizon crossing. As the Universe transitions from radiation to matter domination whether the relativistic species are fluid-like or free-streaming becomes less relevant to the evolution of the matter perturbations and the low-$k$ matter power spectra agree for all scenarios. The time-dependent change from radiation to matter domination leads to a scale-dependent feature near $k_{\rm eq}$ that is qualitatively similar to the feature caused by the decoupling transition even for a fluid-like species that never decouples. Finally, the change in the power spectra for the recoupling scenarios show a mix of behavior. The highest $k$-modes are identical to the free-streaming case. As $k$ decreases, they begin following the fluid curves below $k_{\rm rec} \sim \pi/(c_s\tau_{\rec})$, but then agree with the free-streaming case for $k \ll k_{\rm eq}$. Finally, the duration of the de-/recoupling transition is also important. If the transition from fluid-like to free-streaming is instantaneous, a ringing feature is visible in the matter power spectrum. This occurs because modes that have entered the horizon prior to decoupling oscillate, along with the photon-baryon fluid, and an instantaneous transition will imprint these oscillatory features on the final matter power spectrum. For the more gradual decoupling transition these features are washed out. The recoupling models we consider have very gradual transitions from free-streaming to fluid-like and overall a much smaller imprint on the matter power spectrum. We note that while we have implemented instantaneous decoupling as a transition of width $\Delta z_{\rm dec}/z_{\rm dec} = 0.01$ (see Eq.~\ref{eq:gdec}), the ringing features seen in Fig.~\ref{fig:DeltaPmms} are nearly unchanged for decoupling transitions as wide as $\Delta z_{\rm dec}/z_{\rm dec} = 0.1$. 

\begin{figure}
	\centering
	$\begin{array}{ccc}
	\includegraphics[width = 5.8cm]{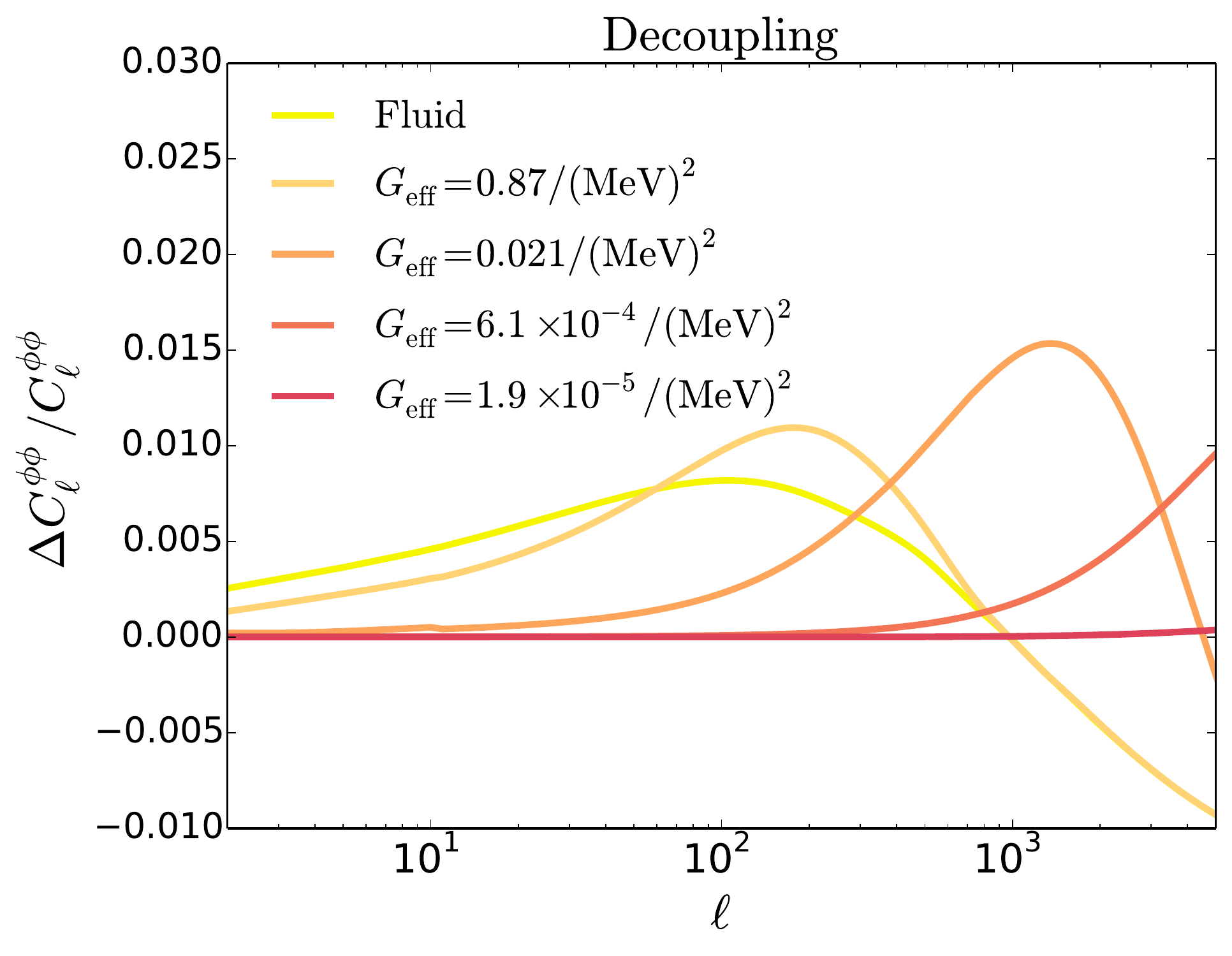} &\includegraphics[width = 5.8cm]{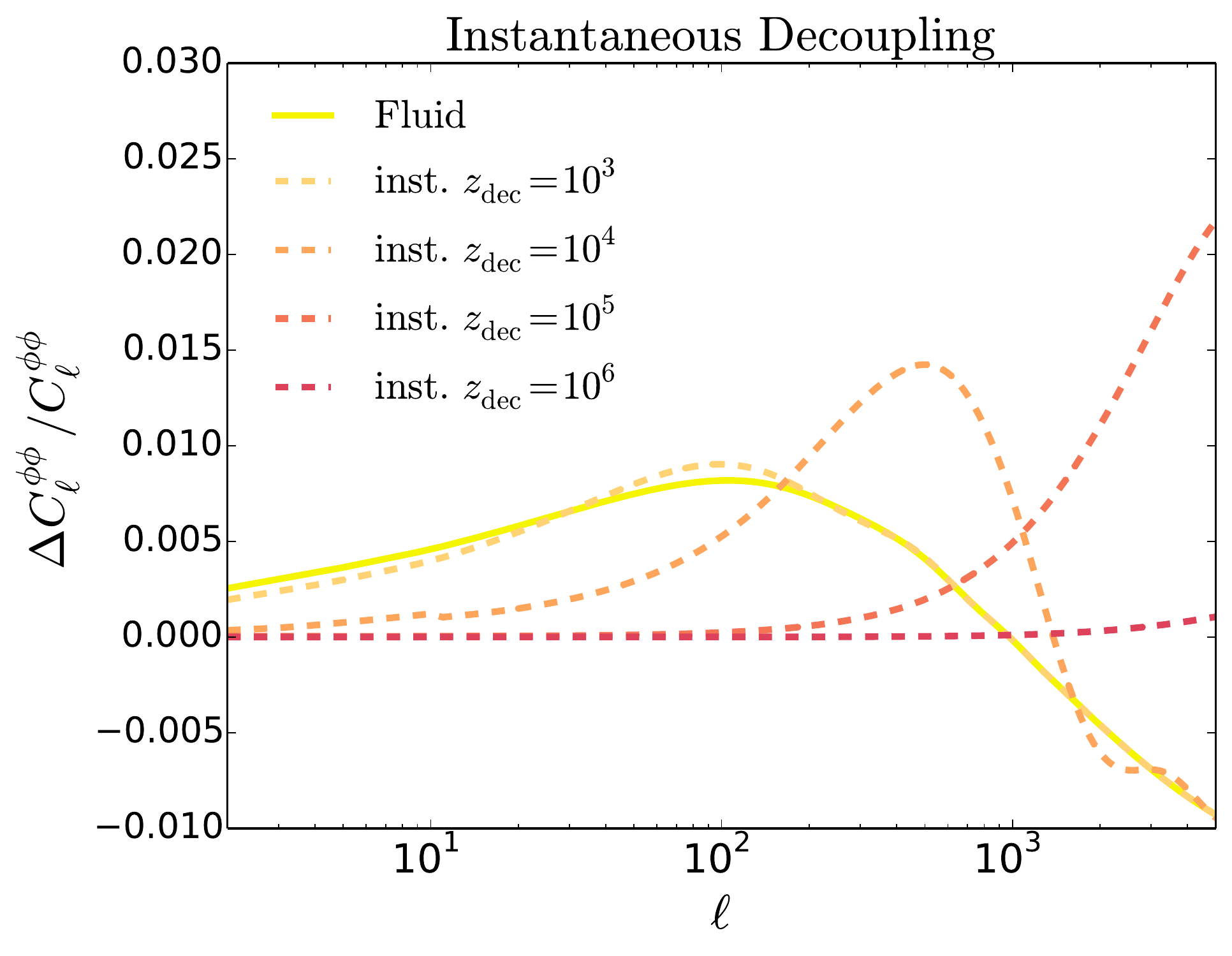}&\includegraphics[width = 5.8cm]{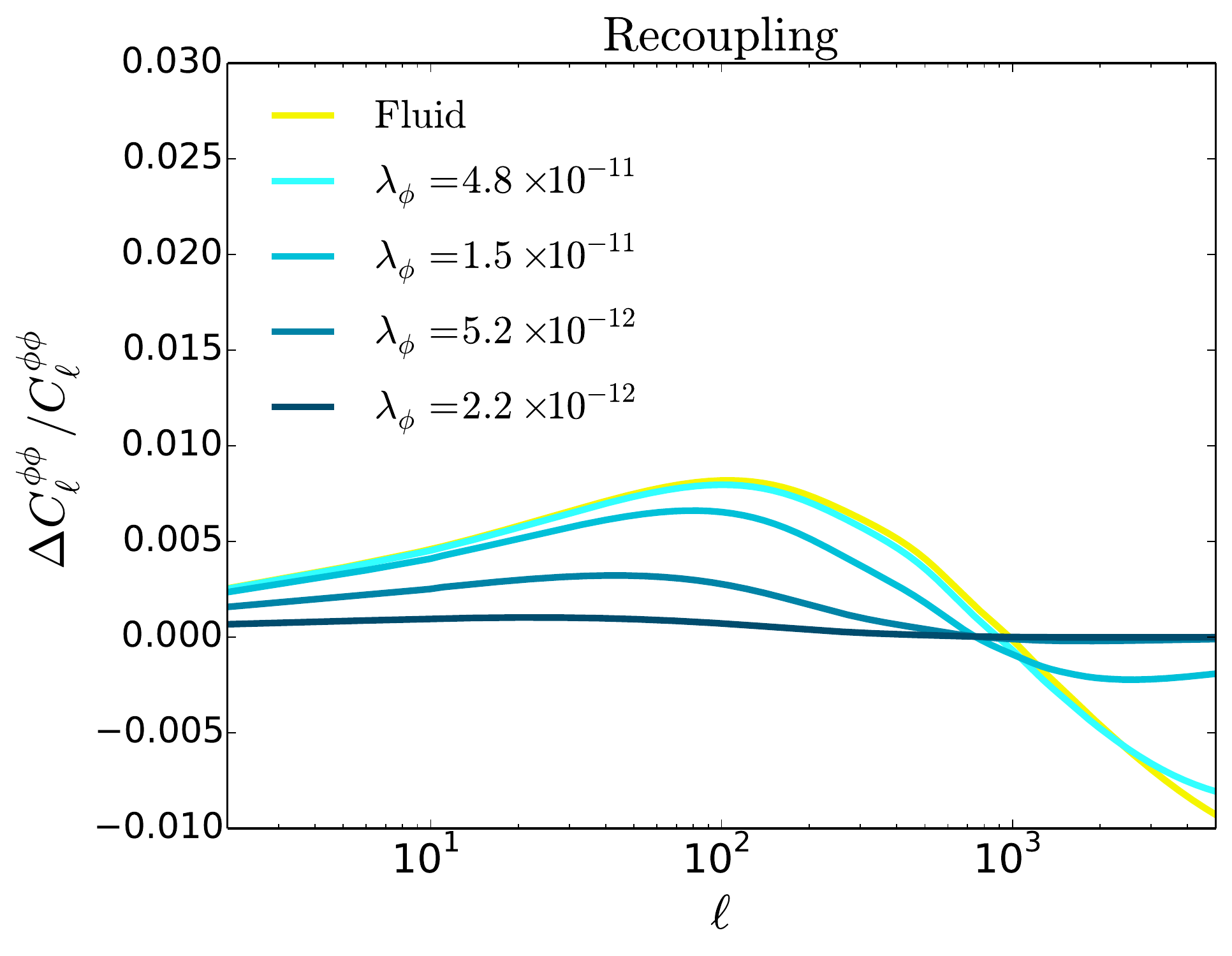} \\
	\mbox{(a)} & \mbox{(b)} & \mbox{(c)} 
		\end{array}$
 	\caption{Fractional changes in the CMB lensing power spectrum when an additional DR contribution of $\Delta \Neff = 0.5$ is interacting instead of free-streaming. The left panel shows $\Delta C_\ell^{\phi\phi}/C_{\ell}^{\phi\phi}(k)$ for the decoupling model, the center panel for the instantaneous decoupling model, and the right panel for the recoupling model. The values of $\Geff$ and $\lambda_\phi$ are chosen to give de/re-coupling redshifts of $z \sim 10^3$, $10^4$, $10^5$, $10^6$, the same values used for instantaneous decoupling in the center panel.}
	\label{fig:DeltaCphiphi}
	\end{figure}
	
The changes to the matter power spectrum in Fig.~\ref{fig:DeltaPmms} lead to similar, but less pronounced changes to the CMB lensing power spectrum. Figure~\ref{fig:DeltaCphiphi} shows the fractional change to the CMB lensing spectum when an additional DR contribution of $\Delta \Neff = 0.5$ is interacting instead of free-streaming. For each scenario the qualitative features are similar to the changes to the matter power spectrum, but suppressed in amplitude and sharp spectral features are washed out, due to the fact that CMB lensing at a given multipole moment depends on the matter power spectrum at a range of $k$. Nevertheless the suppression at high $\ell$ is apparent, as is the peak in $\Delta C_\ell^{\phi\phi}$ near $\ell_{\rm dec}$ and $\ell_{\rm eq}$. The instantaneous and standard decoupling cases, however, do not have distinguishable features.

\section{MCMC analysis}
\label{sec:method}
In this section, we present constraints on five interacting DR scenarios we have discussed. For the Markov chain Monte Carlo (MCMC) analysis we use the Metropolis-Hastings algorithm\footnote{As we will see in the following, the posterior distribution for the coupling parameters show local minima and maxima, which can be difficult for the Metropolis-Hastings to properly sample and care should be taken. However, in all our cases, no parts of the coupling parameter space is ruled out with any notable significance, so we know all possible values are allowed and have been sampled, with chains moving freely through different parts of parameter space, and we can be confident no modes have been missed by the sampling algorithm. We note that problems are known to arise when modes are clearly separated in parameter space and the MCMC chains are unable or unlikely to jump from one mode to another, in which case Metropolis-Hastings is not a suitable algorithm for exploring the posterior space and alternatives should be used, such as, e.g., nested sampling algorithms.}~\cite{Metropolis:1953am,Hastings:1970aa} of the cosmological sampling package \texttt{MontePython v3.2}\footnote{Find the new \texttt{MontePython v3.5} at https://github.com/brinckmann/montepython\_public}~\cite{Audren:2012wb,Brinckmann:2018cvx}, connected to an altered version of the Boltzmann Solver \texttt{CLASS v2.7}\footnote{Get the most recent \texttt{CLASS v3.2} at https://github.com/lesgourg/class\_public}~\cite{Blas:2011rf,Lesgourgues:2011re,Lesgourgues:2011rh}. We impose a Gelman-Rubin convergence criterion of $R-1 \lesssim 0.01$ for determining when the MCMC chains are converged.\\
\\
\noindent We use the following dataset combination:
\begin{itemize}
	\item CMB: Planck 2018 CMB temperature and polarization auto- and cross-correlation, both high-$\ell$ and low-$\ell$~\cite{Aghanim:2019ame}, and with the full set of nuisance parameters, as well as Planck 2018 CMB lensing\footnote{Although the data in our MCMC runs do include the changes to the CMB lensing spectra shown in Fig.~\ref{fig:DeltaCphiphi}, we expect those features are too small to be resolved with current data and do not contribute to our constraints.}~\cite{Aghanim:2018oex}.
	\item BAO\footnote{It is, in principle, possible that constraints using BAO data could be biased for models of self-interacting dark radiation, as these models introduce a phase shift in the acoustic peaks. However, Ref.~\cite{Bernal:2020vbb} studied this problem for a number of extensions to the $\Lambda$CDM model, including a fraction dark matter interacting with neutrinos, a model with a similar phase-shifting property. Therefore, for current BAO data combined with other probes, such as CMB in our case, models of self-interacting dark radiation is not likely to exhibit a bias in constraints derived using standard BAO likelihoods. For future data, this conclusion may be different, and it may be necessary to include BAO data, or full shape power spectrum information, in the analysis in a consistent way, or at least check that it is not a problem in a way similar to Ref.~\cite{Bernal:2020vbb}.}: The Six-degree Field Galaxy Survey (6dFGS, $z=0.106$)~\cite{Beutler:2011hx}, the Sloan Digital Sky Survey Data Release 7 Main Galaxy Sample (SDSS DR7 MGS, $z=0.15$)~\cite{Ross:2014qpa}, and the Baryon Oscillation Spectroscopic Survey Data Release 12 (BOSS DR12, $z=0.38, 0.51, 0.61$) three redshift bin sample~\cite{Alam:2016hwk} (containing the CMASS and LOWZ galaxy samples~\cite{Anderson:2013zyy}).
\end{itemize}

\begin{figure}[t!]
	\centering
	\includegraphics[width=8.4cm]{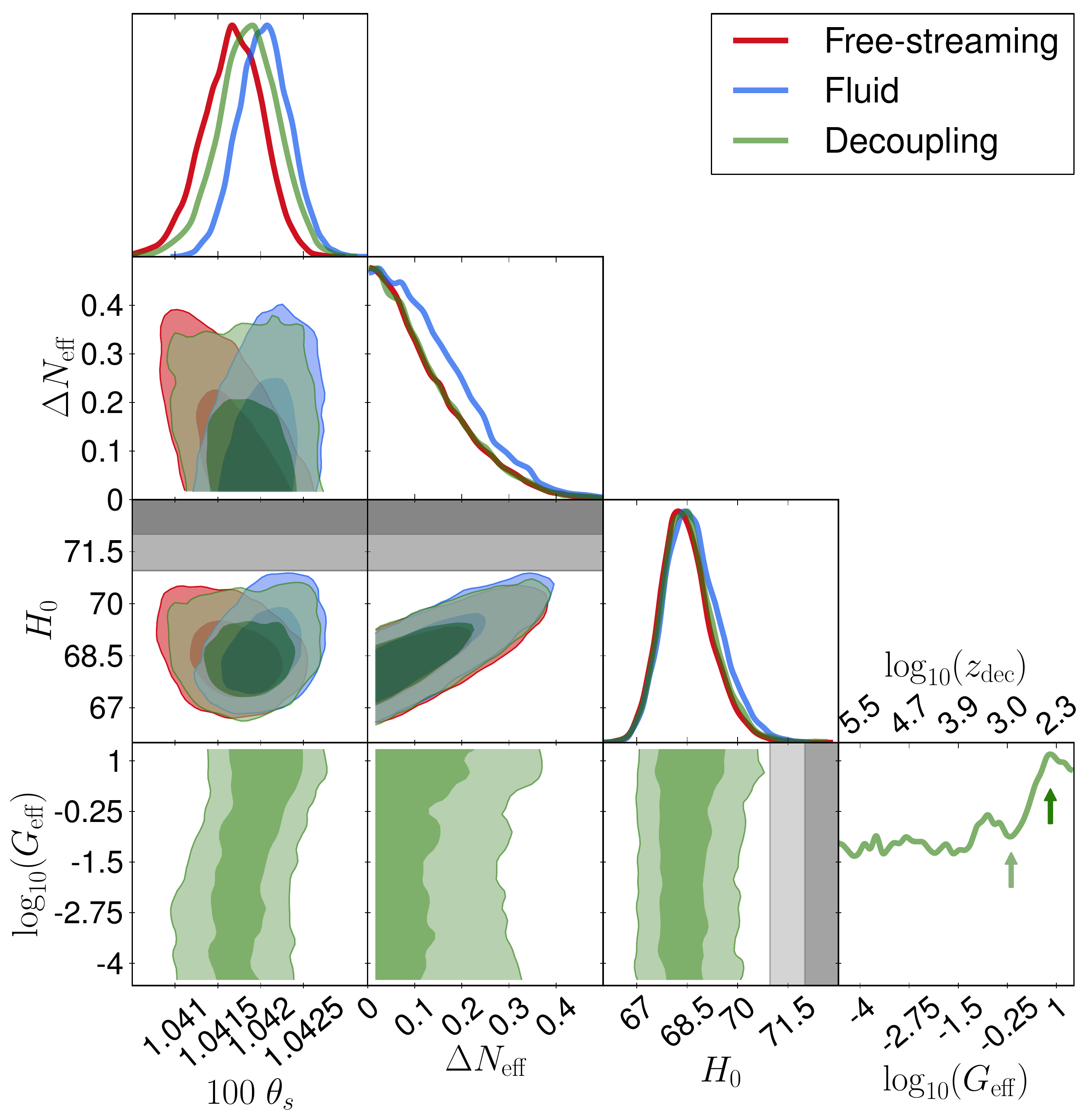}
	\hspace{0.1cm}\includegraphics[width=8.4cm]{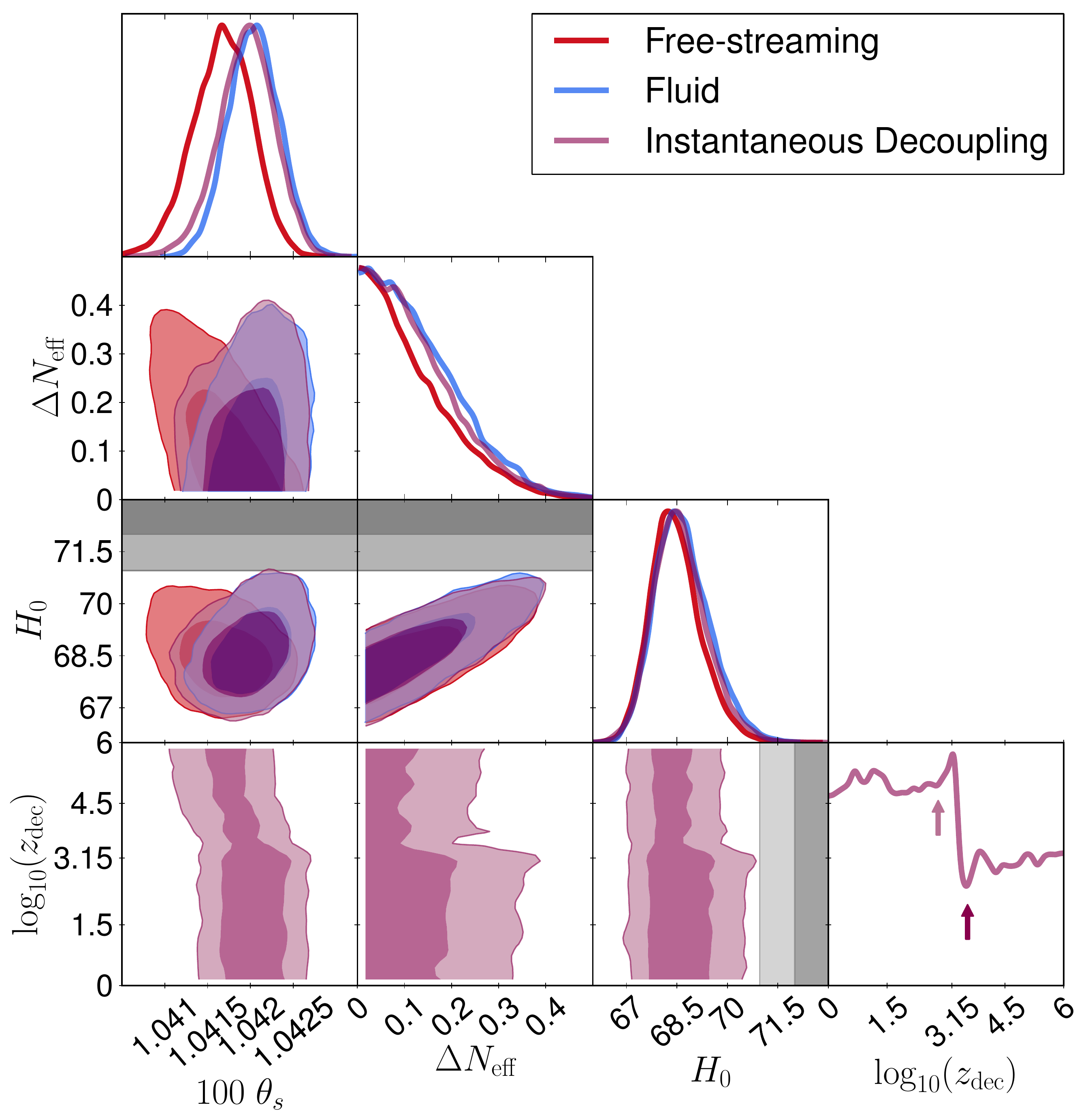}\\
	\includegraphics[width=8.4cm]{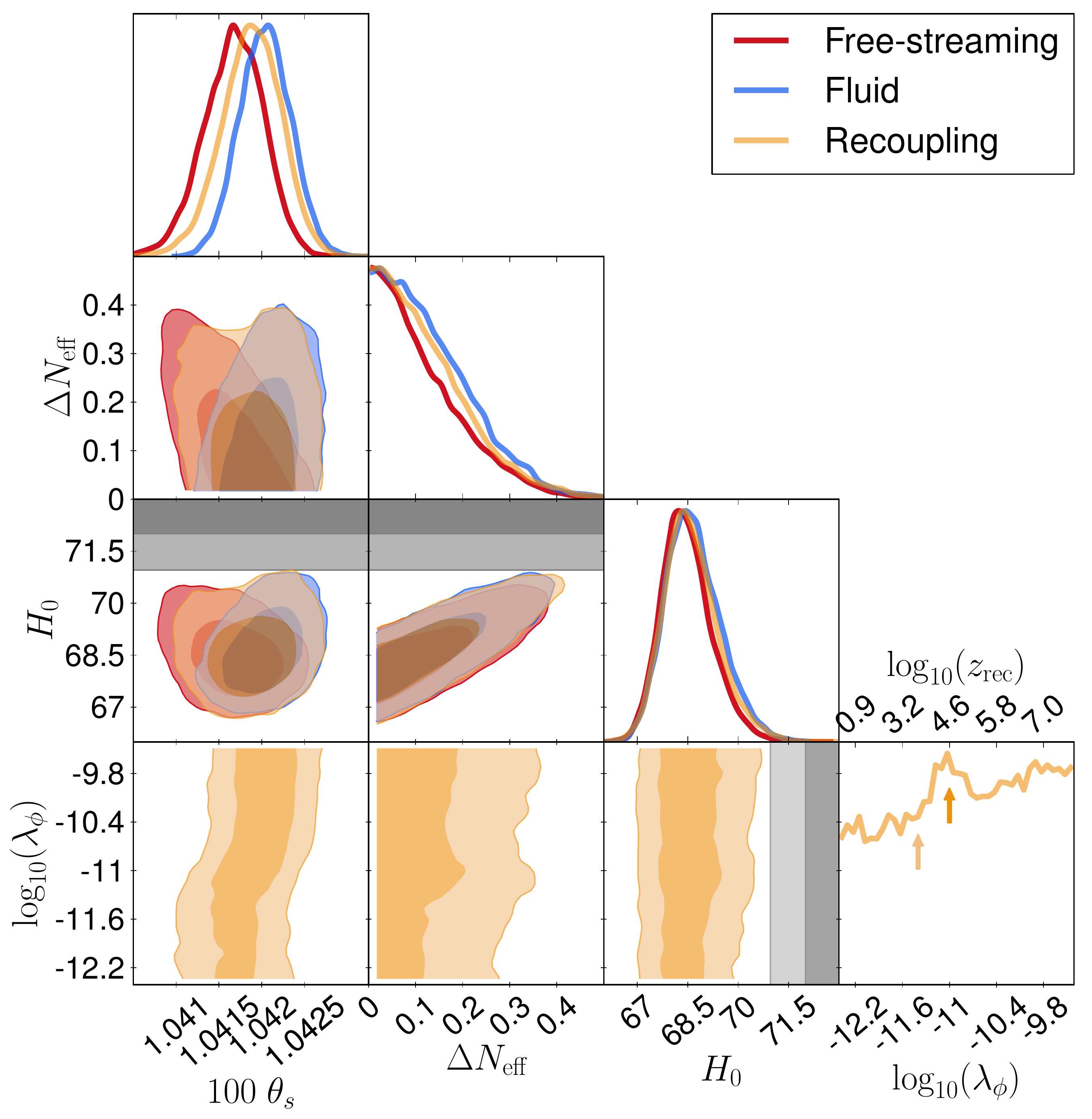}
	\caption{MCMC-derived posterior distributions for select parameters, with 1-$\sigma$ and 2-$\sigma$ intervals as dark and light shaded contours, respectively. \textbf{All:} Planck 2018 primary CMB anisotropies and lensing plus BAO. Free-streaming $\Delta\Neff$ in red and an always self-interacting Fluid $\Delta\Neff$ in blue. For comparison, we show the $H_0$ measurement from Ref.~\cite{Riess:2021jrx} as grey bounds (note only the edge of the 1- and 2-$\sigma$ range is visible). \textbf{Top left:} Decoupling $\Delta\Neff$ in green. Arrows on the bottom right sub-panel indicate the transition point at $\log_{10}(\Geff) \approx -0.15$ [corresponding to $\log_{10}(\zdec) \approx 3.05$] and the peak at $\log_{10}(\Geff) \approx 0.85$ [corresponding to $\log_{10}(\zdec) \approx 2.5$]. \textbf{Top right:} Instantaneous Decoupling $\Delta\Neff$ in purple. Arrows on the bottom right sub-panel indicate the first transition point at $\log_{10}(\zdec) \approx 2.8$ and the second transition point at $\log_{10}(\zdec) \approx 3.55$, while the peak is at $\log_{10}(\zdec) \approx 3.15$. \textbf{Bottom:} Recoupling $\Delta\Neff$ in yellow. Arrows on the bottom right sub-panel indicate the transition point at $\log_{10}(\lambda_\phi) \approx -11.4$ [corresponding to $\log_{10}(\zdec) \approx 3.7$] and the peak at $\log_{10}(\lambda_\phi) \approx -11.0$ [corresponding to $\log_{10}(\zdec) \approx 4.6$].}
	\label{fig:Pb_4param}
	\vspace{-0.2cm}
\end{figure}

\noindent We make the following notable modelling choices (unless otherwise specified): Three degenerate massive free-steaming neutrinos, as a good approximation to either the normal or inverted hierarchy~\cite{Lesgourgues:2006nd}, with a varying neutrino mass sum and a prior $\meff_\nu > 0$ eV. The neutrinos contribute 3.046 to $\Neff$. The helium fraction is inferred from Big Bang Nucleosynthesis and we make other standard $\Lambda$CDM assumptions, such as spatial flatness and a cosmological constant. All CMB power spectra (including lensing reconstruction) are modelled using linear theory, since we do not have an accurate non-linear prescription for the interacting DR models and all publicly available non-linear methods assume free-streaming DR. In any case, differences due to non-linear modelling are expected to be smaller than the errorbars for Planck on all scales, but this point needs to be carefully considered for future data. While we have seen in Sec.~\ref{sec:observables} that DR can also impact the amplitude and shape of the high-$k$ matter power spectrum, we do not include any datasets that probe this directly. We expect that the current bounds on $\Delta \Neff$ from CMB will be stronger than those available from, e.g. current Lyman-$\alpha$ forest data \cite{Chabanier:2018rga, Boera:2018vzq}, but these signatures would be interesting to explore in the future. \\
\\
The main results are shown in Table~\ref{tab:results} and Fig.~\ref{fig:Pb_4param}. In Table~\ref{tab:results}, we show mean values and 68\%CL intervals (unless otherwise noted) for each parameter. We also show the best-fit values for $\Delta \Neff$ and the interaction parameters as well as the minimum effective chi-squared and the chi-squared contribution from each likelihood.

In Fig.~\ref{fig:Pb_4param}, we show posterior distributions (1-$\sigma$ and 2-$\sigma$ intervals as dark and light shaded contours, respectively) for our Decoupling (top left, in green), Instantaneous Decoupling (top right, in purple) and Recoupling (bottom, in yellow) cases, compared to Free-streaming (in red) and Fluid (in blue) reference cases in each subplot. As the most interesting parameters for these models, we opted to show the angular scale of the CMB, $\theta_s$, the amount of DR beyond standard neutrinos, $\Delta \Neff$, and the Hubble parameter today, $H_0$, in addition to the relevant interaction parameters: $\Geff$ in units of (MeV)$^{-2}$ and an estimate of $\zdec$ for Decoupling (see Eq.~\ref{eq:z_dec_rec} and surrounding text), $\zdec$ for Instantaneous Decoupling (as a model parameter, in this case), and $\lambda_\phi$ and an estimate for $\zdec$ for Recoupling (see Eq.~\ref{eq:z_dec_rec}). For the set of full triangle plots see Appendix~\ref{app:full_MCMC}. In the bottom right panel of each subplot, we highlight interesting points in parameter space with arrows, e.g. transition points and peaks in the 1-d marginalized posterior distribution, discussed in the following.\\
\\
\noindent Decoupling (Fig.~\ref{fig:Pb_4param}, top left, green):
\begin{itemize}
	\item Looking at the $\Delta\Neff$ vs $\theta_s$ panel (first column, 2nd from the top), we see that the posterior for the Decoupling case neatly extends across the Free-streaming to the Fluid case. This comes from no strong preference by the datasets being considered for either model, so an MCMC exploration is allowed to freely vary across the full parameter space.
	\item This is further illustrated by the 1-d posterior panel for $\log_{10}(\Geff)$ (bottom right), where the posterior is flat towards smaller $\log_{10}(\Geff)$ (larger $\zdec$) values -- with the free-streaming regime to the far left of the plot and smaller values asymptoting to the Free-steaming case -- through intermediate decoupling redshifts, until there's a small (but not statistically significant) preference for a late decoupling beginning right around recombination, i.e. around the leftmost arrow at $\log_{10}(\Geff) \approx -0.15$ [corresponding to $\log_{10}(\zdec) \approx 3.05$] and peaking at $\log_{10}(\Geff) \approx 0.85$ [corresponding to $\log_{10}(\zdec) \approx 2.5$]. We cut the posterior exploration a bit after this point, since large values of $\log_{10}(\Geff)$ (small $\zdec$) asymptote to the Fluid case. This is because when decoupling happens sufficiently far after recombination, we are no longer sensitive to the decoupling behavior with the current datasets and the posterior would remain flat except for MCMC noise. Indeed, due to the notable decoupling width, we can understand why the posterior of the Decoupling case is not flat beyond recombination (as opposed to the Instantaneous Decoupling case), but instead peaks after recombination.

	\item Considering $\log_{10}(\Geff)$ vs $\theta_s$ (bottom row, 1st panel) and $\log_{10}(\Geff)$ vs $\Delta\Neff$ (bottom row, 2nd panel), we see a change in the preferred values for $\theta_s$ when the interaction strength gets larger [larger $\log_{10}(\Geff)$, smaller $\zdec$] and that $\Delta\Neff$ bounds are somewhat relaxed towards larger $\log_{10}(\Geff)$ values. This can be understood from the fact that interacting DR does not cause a phase-shift of the acoustic peaks like free-streaming DR, or DR with intermediate decoupling redshifts causes a reduced phase-shift. This feature brings the model into slightly better agreement with the data, allowing for a larger contribution of the species, and causes the preferred value for $\theta_s$ to shift towards larger values.
	
	\item Finally, there is no notable impact on the Hubble parameter (bottom row, 3rd panel), beyond a slight widening of the posterior for a late decoupling towards the Fluid asymptote, far from enough to be relevant in terms of resolving tensions.
\end{itemize}

\noindent Instantaneous Decoupling\footnote{The MCMC results in this section parameterize instantaneous decoupling as a transition of width $\Delta \zdec = 0.01 \zdec$, as discussed in Sec. \ref{sec:models}. We have, however, checked that our parameter constraints are virtually unchanged for a decoupling width of $\Delta \zdec = 0.1 \zdec$.} (Fig.~\ref{fig:Pb_4param}, top right, purple):
\begin{itemize}
	\item Considering the $\Delta\Neff$ vs $\theta_s$ panel (first column, 2nd from the top), we see the posterior for Instantaneous Decoupling mostly tracks that of the Fluid comparison case. This can be easily understood from the 1-d posterior panel for $\log_{10}(\zdec)$ (bottom right), where we see a preference for low decoupling redshift values, asymptoting to the Fluid case once we get beyond recombination. This preference dominates the other marginalized posterior distributions. Two transition points are indicated with arrows, one at $\log_{10}(\zdec) \approx 3.55$, from where the preference starts to increase for lower decoupling redshifts, and one at $\log_{10}(\zdec) \approx 2.8$, where we are fully beyond recombination and have reached the Fluid case asymptote and the posterior flattens to just MCMC noise. There appears to be a peak (albeit not statistically significant) at $\log_{10}(\zdec) \approx 3.15$, so the data seems to prefer a decoupling just before recombination.
	\item Considering $\log_{10}(\zdec)$ vs $\Delta\Neff$ (bottom row, 2nd panel) we see the $\Delta\Neff$ bound is significantly relaxed for late decoupling times (small $\zdec$) from near recombination onwards, leading to larger uncertainty for the Hubble parameter (bottom row, 3rd panel). Note that since the 1-d marginalized posterior distributions for $\Delta\Neff$ and $H_0$ include the high decoupling redshift part, the relaxed bounds are less apparent there.
\end{itemize}

\noindent Recoupling (Fig.~\ref{fig:Pb_4param}, bottom, yellow):
\begin{itemize}
	\item Looking at the $\Delta\Neff$ vs $\theta_s$ panel (first column, 2nd from the top), we see the recoupling case nearly spans the posterior space of the Free-streaming and Fluid comparison cases, like the Decoupling case. The reason is the same as before, because we have a mostly unconstrained parameter space, with a neat transition between the two asymptotes. Indeed, considering the 1-d posterior panel for $\log_{10}(\lambda_\phi$) (bottom right) we see a mostly flat posterior, with a small increase in probability starting from $\log_{10}(\lambda_\phi) \approx -11.4$ [corresponding to $\log_{10}(\zrec) \approx 3.7$], indicated by an arrow, towards larger values of the coupling strength, corresponding to an earlier recoupling. Interestingly, there appears to be a small peak (not statistically significant) at intermediate recoupling redshifts, around $\log_{10}(\lambda_\phi) \approx -11.0$ [corresponding to $\log_{10}(\zdec) \approx 4.6$], indicated by a second arrow, but we clearly need more sensitive datasets to constrain this model.
	\item Considering the other bottom row panels, we see a similar shift in $\theta_s$ (first panel) as we did for the Decoupling case. This is not surprising, larger values of the coupling parameters in both examples, corresponding to earlier recoupling or later decoupling, asymptote to the Fluid case whereas the late recoupling, or early decoupling, models asymptote to the Free-streaming case. In the second of the bottom panels, we see the bound on $\Delta\Neff$ is relaxed for models recoupling early or at intermediate redshifts, while late recoupling bounds are similar to the Free-streaming bounds. Finally, for the Hubble parameter (third panel), we see a slight relaxation of the confidence interval compared to the Free-streaming asymptote, but not enough to play a role in resolving cosmological tensions.
\end{itemize}

\begin{table}[htb]
	\scriptsize
	\centering
	\begin{tabular}{ l | c c | c c | c }
		& Free-streaming & Fluid & Decoupling & Instantaneous Dec. & Recoupling \\ \hline
		$\omega_{b}$ & $0.02250 \pm 0.00015$ & $0.02254 \pm 0.00017$ & $0.02250 \pm 0.00016$ & $0.02252 \pm 0.00016$ & $0.02252 \pm 0.00017$ \\
		$\omega_{cdm}$ & $0.1214 \pm 0.0017$ & $0.1219 \pm 0.0019$ & $0.1215 \pm 0.0018$ & $0.1217 \pm 0.0018$ & $0.1216 \pm 0.0018$ \\
		$100 \times \theta_s$ & $1.04167 \pm 0.00036$ & $1.04204 \pm 0.00031$ & $1.04185 \pm 0.00036$ & $1.04195 \pm 0.00033$ & $1.04188 \pm 0.00035$ \\
		$\ln({10}$$^{10} A_s)$ & $3.051 \pm 0.016$ & $3.046 \pm 0.015$ & $3.049 \pm 0.015$ & $3.047 \pm 0.015$ & $3.049 \pm 0.015$ \\
		$n_s$ & $0.9697 \pm 0.0050$ & $0.9674 \pm 0.0040$ & $0.9685 \pm 0.0046$ & $0.9676 \pm 0.0043$ & $0.9691 \pm 0.0049$ \\
		$z_{reio}$ & $7.85 \pm 0.76$ & $7.91 \pm 0.75$ & $7.86 \pm 0.76$ & $7.89 \pm 0.75$ & $7.89 \pm 0.75$ \\
		$\log_{10}(\Geff)$ & --- & --- & unbounded & --- & --- \\
		$\log_{10}(\zdec)$ & --- & --- & --- & unbounded & --- \\
		$\log_{10}(\lambda_\phi)$ & --- & --- & --- & --- & unbounded \\
		$\Delta\Neff$ & $< 0.304$ (95\%CL) & $< 0.324$ (95\%CL) & $< 0.311$ (95\%CL) & $< 0.314$ (95\%CL) & $< 0.320$ (95\%CL) \\
		$\meff_\nu$ & $< 0.120$ (95\%CL) & $< 0.125$ (95\%CL) & $< 0.122$ (95\%CL) & $< 0.127$ (95\%CL) & $< 0.123$ (95\%CL) \\
		$H_0$ $\left[\frac{\textrm{(km/s)}}{\textrm{Mpc}}\right]$ & $68.5 \pm 0.7$ & $68.7 \pm 0.8$ & $68.6 \pm 0.8$ & $68.6 \pm 0.8$ & $68.6 \pm 0.8$ \\
		$S_8$ & $0.831 \pm 0.012$ & $0.828 \pm 0.011$ & $0.829 \pm 0.012$ & $0.829 \pm 0.012$ & $0.829 \pm 0.012$ \\ [0.5mm] \hline
		\multicolumn{6}{ c }{Best fit} \\ [0.5mm] \hline
		$\Delta\Neff$ & 0.050 & 0.012 & 0.0046 & 0.0087 & 0.069 \\
		$\log_{10}(\Geff)$ & --- & --- & -1.62 & --- & --- \\
		$\log_{10}(\zdec)$ & --- & --- & --- & 2.61 & --- \\
		$\log_{10}(\lambda_\phi)$ & --- & --- & --- & --- & -10.1 \\
		P18 highTTTEEE & 2347.1 & 2346.5 & 2345.9 & 2345.3 & 2347.3 \\
		P18 lowTT & 23.3 & 23.6 & 23.6 & 23.8 & 23.1 \\
		P18 lowEE & 395.9 & 395.9 & 396.8 & 398.4 & 396.2 \\
		P18 lensing & 8.9 & 8.8 & 8.7 & 8.7 & 8.9 \\
		P18 total & 2775.3 & 2774.9 & 2775.1 & & 2775.4 \\
		BAO & 5.2 & 5.5 & 5.2 & 5.2 & 5.4 \\
		$\chi^2_{\rm{eff}}$ & 2780.4 & 2780.4 & 2780.2 & 2781.4 & 2780.8 \\
		$\Delta \chi^2_{\rm{eff}}$ & --- & 0.0 & -0.2 & +1.0 & +0.4 \\
	\end{tabular}
	\caption{Statistical information for all models. $\chi^2_{\rm{eff}} = -2 \ln \mathcal{L}$ is the minimum effective chi square, $\Delta \chi^2_{\rm{eff}}$ is with regards to the corresponding free-streaming case. All credibility intervals are 68\%CL centered around the mean unless otherwise noted. The clustering parameter $S_8$ is defined as $S_8 = \sigma_8 (\Omega_m/0.3)^{0.5}$.}
	\label{tab:results}
\end{table}

It is worth remembering that our choice for how to implement the neutrino mass sum, $\meff_\nu$, impacts the bounds on $\Delta\Neff$, as there's a direct correlation between neutrino mass sum and allowed values of $\Delta\Neff$. In many works, the neutrino mass sum is taken to be fixed (a common choice is around the minimal allowed value for a normal hierarchy neutrino configuration with two low mass and one more massive neutrino, $\meff_\nu = 0.06$ eV), but the absolute scale of the neutrino mass sum is not known, only the mass difference between states. To reflect this uncertainty, we allow the neutrino mass sum to vary in our main analysis, which impacts bounds on $\Delta\Neff$ compared to fixing the neutrino mass sum, as shown in Tables~\ref{tab:mnu_mass} (fixed neutrino mass sum) and~\ref{tab:mnu_mass2} (varying neutrino mass sum).

For the fixed neutrino mass sum comparison in Table~\ref{tab:mnu_mass}, we limit the scope to only the Free-streaming and Fluid reference cases, as they can be considered the boundary cases for self-interacting dark radiation. In this case, we consider $\meff_\nu = [0, 0.06, 0.11]$ eV to cover massless neutrinos, minimum normal hierarchy mass in two configurations (degenerate masses, denoted ``deg", and a normal hierarchy approximation with two massless and one massive neutrino, denoted ``2 massless"), and minimum inverted hierarchy mass (only degenerate masses).

In Table~\ref{tab:mnu_mass2}, we allow the neutrino mass sum to vary, with two common choices, one remaining agnostic about the mass of cosmological neutrinos freely varying the neutrino mass sum $\meff_\nu > 0$ (acknowledging that new physics could allow for different $\meff_\nu$ in cosmology compared to that from terrestrial measurements, see, e.g.,~\cite{Lorenz:2017fgo,Koksbang:2017rux,Lorenz:2018fzb,Chacko:2019nej,Chacko:2020hmh,Escudero:2020ped,Lorenz:2021alz,FrancoAbellan:2021hdb,Alvey:2021xmq}) and the other imposing a minimal allowed neutrino mass sum roughly corresponding to the minimal mass sum allowed in a normal hierarchy configuration with $\meff_\nu > 0.06$ eV. In both cases, we consider degenerate massive neutrinos, as the difference compared to more realistic configurations is very small~\cite{Lesgourgues:2006nd}.

\begin{table}[htb]
	\scriptsize
	\centering
	\begin{tabular}{ c | c c c c }
		$\meff_\nu$ prior & $0$ eV & $0.06$ eV (deg) & $0.06$ eV (2 massless) & $0.11$ eV \\ \hline
		Free-streaming $\Delta\Neff$ (95\%CL) & $< 0.285$ & $< 0.314$ & $< 0.314$ & $< 0.334$ \\ \hline
		Fluid $\Delta\Neff$ (95\%CL) & $< 0.298$ & $< 0.325$ & $<0.336$ & $< 0.363$ \\ \hline
	\end{tabular}
	\caption{Bounds on the effective number of extra relativistic species, $\Delta\Neff$, for different choices of a fixed neutrino mass sum $\meff_\nu$.}
	\label{tab:mnu_mass}
\end{table}

\begin{table}[htb]
	\scriptsize
	\centering
	\begin{tabular}{ c | c c c c c c }
		$\meff_\nu$ prior & & Free-streaming & Fluid & Decoupling & Instantaneous Decoupling & Recoupling \\ \hline
		$> 0$ eV & $\Delta\Neff$ (95\%CL) & $< 0.304$ & $< 0.324$ & $< 0.311$ &$< 0.314$ & $< 0.320$ \\ \hline
		$> 0.06$ eV & $\Delta\Neff$ (95\%CL) & $< 0.325$ & $< 0.339$ & $< 0.340$ & $< 0.346$ & $< 0.338$ \\ \hline
	\end{tabular}
	\caption{Bounds on the effective number of extra relativistic species, $\Delta\Neff$, for different choices of a varying neutrino mass sum $\meff_\nu$. In both cases the neutrino mass is implemented as three degenerate states with $m_{\nu i} = \meff_\nu/3$. }
	\label{tab:mnu_mass2}
\end{table}

We find differences of order 10\% in the bounds on $\Delta\Neff$ for common choices used in the literature for how to treat neutrino masses, a difference that increases for less common choices. For simplicity, throughout the paper we assumed the standard neutrinos behave as expected with a contribution to $\Neff$ of 3.046, but somewhat different results might be obtained if this assumption is relaxed and all of $\Neff$ is varied.

As a side note, if a very large neutrino mass sum were to be detected from cosmology, $\Delta\Neff$ bounds would dramatically change. This has been hinted at by some recent analyses, e.g. an analysis that factors out the Planck lensing anomaly\footnote{See, e.g., Refs.~\cite{Planck:2018vyg,Motloch:2018pjy} for discussions on the Planck lensing anomaly.} found $\meff_\nu = 0.51^{+0.21}_{-0.24}$ eV at 68\% CL from a combination of BOSS and Planck~\cite{Sgier:2021bzf} (see also the ACT DR4+WMAP results~\cite{ACT:2020gnv}, as well as Ref.~\cite{DiValentino:2021imh}, for similarly large neutrino mass sum bounds from cosmology when avoiding the Planck lensing anomaly). As a rudimentary check, fixing $\meff_\nu = 0.5$ we find a Free-streaming $\Delta\Neff < 0.738$ and a Fluid $\Delta\Neff < 0.727$ at 95\%CL, wildly different from conventional results. Of course, a proper analysis with the full lensing-anomaly treated likelihoods and varying neutrino mass sum should be conducted, but if a large neutrino mass were to be a reality, all $\Neff$ bounds would change accordingly.

\section{Discussion and conclusions}
\label{sec:discussion}

Dark radiation commonly arises in BSM scenarios and may contribute a detectable level of radiation density in the Universe in addition to photons and neutrinos. This contribution is typically parametrized with $\Delta\Neff$ and can be constrained by cosmological data. Yet, the signatures of DR can vary depending on the nature of DR and the parameter $\Delta\Neff$, which solely characterizes the energy density, is insufficient to capture the full range of scenarios. More generally, DR will have perturbations, along with the SM particles, and their energy density, pressure, and anisotropic stress evolve differently depending on whether the DR is interacting with itself or another constituent.  In general, interaction rates evolve differently from the Hubble parameter, and dark radiation can decouple or recouple from interactions during cosmic history. In this paper, we explore the cosmological signatures of and constraints on DR with interactions that modify the time-dependence of the sound speed and anisotropic stress of DR and thereby leave distinct signatures in cosmological observables. 

We classify DR into five interaction types: free-streaming, fluid-like, decoupling, instantaneous decoupling, and recoupling. Free-streaming and fluid-like DR are the usual scenarios in which radiation is either non-interacting or interacting for the whole observable cosmic history. The free-streaming case is part of standard cosmological analyses (e.g.~\cite{Planck:2018vyg}) and the fluid-like case have been studied by, e.g.~\cite{Baumann:2015rya,Blinov:2020hmc}. Decoupling DR is the case where its self-interaction is through a heavy mediator, so that interactions cease and radiation transitions from fluid-like to free-streaming, as with SM neutrinos. In this case, the decoupling transition happens smoothly as $\Gamma/H \propto T^3$. In contrast, instantaneous decoupling occurs suddenly due to the presence of a heavier or bound state that exponentially suppresses the scattering rate, similar to SM hydrogen recombination. Recoupling DR arises in scenarios with a light mediator or a renormalizable self-interaction term such as $\phi^4$ theory. In this case, the DR is initially free-streaming but interactions change the behavior of DR to behave more like a fluid~(for decoupling and recoupling types of models, see, e.g.,~\cite{Cyr-Racine:2013jua, Kreisch:2019yzn, Choi:2018gho, Archidiacono:2013dua}). These scenarios are described in Sec.~\ref{sec:models}.  We implemented the DR scenarios above in the Boltzmann equations, using the relaxation time approximation to include the collision terms in the Boltzmann hierarchy.  The interaction terms include the thermal averaged interaction rate and the relaxation time coefficient, which are carefully calculated in this work for different interaction benchmarks. Since the relaxation time coefficients depend on multipole $\ell$, the time at which moments in the Boltzmann hiearchy transition between fluid-like and free-streaming also depends on $\ell$. We define decoupling and recoupling with respect to the $\ell=2$ moment in the Boltzmann hierarchy, related to the anisotropic stress, and provide a precise relation between parameters in the DR Lagrangian and the decoupling or recoupling redshift. Our approach is described in Sec. \ref{sec:Boltzmanneq} and the details of calculations are given in Appendix \ref{app:relaxation_time_coeff}. 

We have shown that the DR scenarios lead to distinct signatures on CMB and matter power spectra. These signatures, along with a detailed discussion of the physical mechanisms, are presented in Sec. \ref{sec:observables}. A summary of the changes to CMB power spectra is in Fig. \ref{fig:DeltaCls}, on the matter power spectra in Fig.~\ref{fig:DeltaPmms}, and on CMB lensing in Fig.~\ref{fig:DeltaCphiphi}. Of particular interest is the sensitivity of the matter power spectrum to the duration of the decoupling transition, visible by comparing panels $(a)$ and $(b)$ in Fig.~\ref{fig:DeltaPmms}. We anticipate that for currently allowed values of $\Delta \Neff$, the signature in the matter power spectrum is too small to be detected. Yet, it would be interesting to explore whether these features can be exploited to detect or constrain DR with future large-scale structure datasets. We leave this to future work.

We derive MCMC constraints from Planck 2018 CMB data and BAO data on these DR scenarios, using the cosmological sampling package MontePython interfaced with a modified version of the Boltzmann solver CLASS, in which the Boltzmann hierarchy with the relaxation time approximation is implemented for each of our interacting DR cases. The 1- and 2-d marginalized posterior distributions for each of our cases are shown for select parameters in Fig.~\ref{fig:Pb_4param} and for the full cosmological parameter space in Appendix~\ref{app:full_MCMC}. In Table~\ref{tab:results}, we show bounds for each of our cosmological parameters along with information on the best-fitting model. We find no statistically significant bounds on the coupling constants of DR, although we find a slight preference for a late transition redshift for instantaneous decoupling DR at around recombination, and for the fluid-like limit of all the cases. The data exhibits interesting features at some specific times/redshifts in the early universe, but more constraining data is required to derive statistically significant bounds. The constraints on $\Delta N_{\rm eff}$ differ marginally for the different cases, but only at the order of about $5-10\%$. We note that the models of interacting DR considered here do not help resolve the $H_0$ and $\sigma_8$ tensions, as Planck CMB bounds on $\Delta\Neff$ do not allow for a signifcant deviation from the standard prediction, a situation that is unchanged by the presence of interactions, and therefore the models under consideration are unable to alleviate these cosmological tensions. Finally, in Tables~\ref{tab:mnu_mass} and~\ref{tab:mnu_mass2}, we consider how our prior choice for the neutrino mass sum affects bounds on $\Delta\Neff$ for our DR cases, finding a difference of order $10\%$, comparable or larger than the differences due to interactions, highlighting the importance of being clear on prior assumptions when reporting results.

Although current data are only able to provide hints towards preferred couplings of interacting DR, without placing statistically significant bounds, future experiments and surveys will be able to detect the small scale signatures of interacting DR discussed in Sec.~\ref{sec:observables}. Upcoming CMB experiment Simons Observatory~\cite{SimonsObservatory:2018koc}, the future CMB-S4 experiment~\cite{Abazajian:2016yjj}, and possibly CMB-HD \cite{CMB-HD:2022bsz} will delve into increasingly smaller scales of the CMB anisotropies, potentially bringing signatures of interacting radiation within reach (compare to signatures of interactions seen in primary CMB in Fig.~\ref{fig:DeltaCls}, increasing significantly towards smaller scales, as well as CMB lensing in Fig.~\ref{fig:DeltaCphiphi}). These future datasets will simultaneously aid in detecting novel signatures of interactions and constraining $\Neff$ to much better precision, with nearly an order of magnitude improvement expected by CMB-S4 over current constraints. Concurrently, data from current and near-term galaxy surveys such as DESI \cite{DESI:2016fyo}, Euclid~\cite{Amendola:2016saw}, the Roman Telescope \cite{WFIRST:2018mpe}, and the Rubin Observatory \cite{LSSTScience:2009jmu} will dramatically improve our measurements of the linear-theory matter power spectrum. The signatures of interacting radiation shown in Fig.~\ref{fig:DeltaPmms} present an interesting target worthy of further exploration. In the longer term, gravitational waves present a clean and complementary probe of the DR interaction history \cite{Loverde:2022wih}. In summary, the prospects of confirming or ruling out non-standard interactions in the neutrino and DR sector are excellent.

\section*{Acknowledgements}

We are grateful for helpful conversations with Manuel A. Buen-Abad, Soubhik Kumar, Murali Saravanan, and Zach Weiner. TB and ML acknowledge support from DOE Grant DE-SC0017848. TB is also supported through the INFN project ``GRANT73/Tec-Nu", and by the COSMOS network (www.cosmosnet.it) through the ASI (Italian Space Agency) Grants 2016-24-H.0 and 2016-24-H.1-2018. JHC is supported by NSF Grant PHY-2210361, the Maryland Center for Fundamental Physics (MCFP), and the Johns Hopkins University Joint Postdoc Fund. The work of PD~is supported by the US Department of Energy under grant DE-SC0010008.  ML gratefully acknowledges additional support from DOE grant DE-SC0023183, the Dr. Ann Nelson Endowed Professorship, and the Department of Physics and the College of Arts and Sciences at the University of Washington. Results in this paper were obtained using the high-performance computing system at the Institute for Advanced Computational Science at Stony Brook University.

\begin{appendix}

\section{Details of the collision terms in Boltzmann equations}
\label{app:relaxation_time_coeff}
In this appendix, we present more details of the collision terms $C[f]$ in the Boltzmann equation (Eq.~\ref{eq:Boltzmann_eq_general}) in Sec.~\ref{sec:Boltzmanneq}. As shown in  \cite{Oldengott:2014qra}, for the case of $2\leftrightarrow 2$ scattering process, where we label the momenta of initial (final) particles being $\mathbf q, \mathbf l\,( \mathbf q', \mathbf l')$,  the collision term $C[f]$ that is first order in $\Psi$ is defined as
\bea\label{eq:C[f]}
C[f]_{2\leftrightarrow2}(\mk, \mq,\tau)&=&\frac{g^3}{2 q (2\pi)^5}\int \frac{d^3\ml}{2l}\int \frac{d^3\mq'}{2q'}\int \frac{d^3\ml'}{2l'}\,\delta^4(q+l-q'-l')\nonumber\\
&~&\times \langle |\mathcal{M}|^2 \rangle\left[\bar f^{\rm eq}(q') \bar f^{\rm eq}(l')[\Psi(\mk,\mq',\tau)+\Psi(\mk,\ml',\tau)]-\bar f^{\rm eq}(q) \bar f^{\rm eq}(l)[\Psi(\mk,\mq,\tau)+\Psi(\mk,\ml,\tau)]\right],
\eea
where $g$ is the degree of freedom of the particle. 

To get Eq.~(\ref{eq:C[f]}) and analytic expressions for the collision terms after integration, we made several approximations. First, we assume all kinds of DR in our analysis have nearly thermal distributions: $\bar f(q,\tau)=\bar f^{\rm eq}(q)$.  Second, we assume $\bar f^{\rm eq}(q)$ has a Maxwell-Boltzmann distribution instead of a Bose-Einstein or Fermi-Dirac distribution. Third, we neglect Pauli blocking or Bose enhancement of final state particles.  The validity of the first assumption is discussed in Sec.~\ref{sec:Boltzmanneq} (see discussion below Eq.~(\ref{eq:f_expansion}). 
The last two assumptions are appropriate for all cases we study because we consider non-degenerate initial and final states.

Now we are ready to calculate the $C[f]$ for the different DR scenarios listed in Sec.~\ref{sec:models}. Our ultimate goal is to obtain the relevant $\Gamma$ and $\alpha_\ell$ in Eq.~(\ref{eq:Boltzmann_hierarchy_decoupling}). 

\subsection{Decoupling }
We first consider Majorana fermions as decoupling DR, which has self-interactions mediated by a heavy mediator (see discussion in Sec.~\ref{sec:models}). In this case, we can get $\left< |\mathcal{M}|^2 \right> = \frac{1}{16} G_\eff^2 (s^2+t^2+u^2)$, where $G_\eff$ is the effective Fermi constant (see Eq.~(\ref{eq:Heavy_mediator}))  and $s,t,u$ are Mandelstam varibles. Then Boltzmann hierarchy for $\Psi$ can be written as \cite{Oldengott:2014qra,Oldengott:2017fhy}:
\bea
\dot{\Psi}_0(q) &=& -k \Psi_1(q) + \frac{1}{6}\frac{\partial \ln \bar{f}}{\partial \ln q} \dot{h}-\frac{10}{3} \frac{N T_{D,0}^4 G_\eff^2}{a^4 (2\pi)^3}q \Psi_0(q)\nonumber\\
&& \qquad + \frac{N G_\eff^2}{2 a^4 (2\pi)^3} \int \dd q' \left[ K_0^m(q,q')-\frac{10}{9} q^2 q'^2 e^{-q/T_{D,0}} \right ] \frac{q' \bar{f}(q')}{q \bar{f}(q)} \Psi_0 (q')\label{eq:Boltzmann_hierarchy_decoupling_Psi0}\\
\dot{\Psi}_1(q) &=& -\frac{2}{3}k \Psi_2(q) + \frac{1}{3} k \Psi_0(q)-\frac{10}{3} \frac{N T_{D,0}^4 G_\eff^2}{a^4 (2\pi)^3}q \Psi_1(q)\nonumber\\
&& \qquad + \frac{N G_\eff^2}{2 a^4 (2\pi)^3} \int \dd q' \left[ K_1^m(q,q')-\frac{5}{9} q^2 q'^2 e^{-q/T_{D,0}} \right ] \frac{q' \bar{f}(q')}{q \bar{f}(q)} \Psi_1 (q')\\
\dot{\Psi}_2(q) &=& -\frac{3}{5}k \Psi_3(q) + \frac{2}{5} k \Psi_1(q) - \frac{\partial \ln \bar{f}}{\partial \ln q} \left( \frac{2}{5} \dot{\eta} +\frac{1}{15}\dot{h} \right)-\frac{10}{3} \frac{N T_{D,0}^4 G_\eff^2}{a^4 (2\pi)^3}q \Psi_2(q)\nonumber\\
&& \qquad + \frac{N G_\eff^2}{2 a^4 (2\pi)^3} \int \dd q' \left[ K_2^m(q,q')-\frac{1}{9} q^2 q'^2 e^{-q/T_{D,0}} \right ] \frac{q' \bar{f}(q')}{q \bar{f}(q)} \Psi_2 (q')\\
\dot{\Psi}_{\ell>2}(q) &=& -\frac{k}{2\ell+1}[\ell \Psi_{\ell-1}(q) -(\ell+1)\Psi_{\ell+1}(q)] - \frac{10}{3} \frac{N T_{D,0}^4 G_\eff^2}{a^4 (2\pi)^3}q \Psi_\ell(q)\nonumber\\
&& \qquad + \frac{N G_\eff^2}{2 a^4 (2\pi)^3} \int \dd q' K_\ell^m(q,q') \frac{q' \bar{f}(q')}{q \bar{f}(q)} \Psi_\ell (q') .\label{eq:Boltzmann_hierarchy_decoupling_Psi3}
\eea
Here, $T_{D,0}$ is the temperature of the DR today. The quantity $\bar f$ denotes the averaged phase space distribution and $\Psi_\ell$ is the Legendre decomposition of $\Psi$,
\beq
\Psi(\mk,\mq,\tau) = \sum_{\ell=0}^{\infty} (-i)^\ell (2\ell+1) \Psi_\ell(k,q,\tau) P_\ell(\cos \epsilon)
\eeq
where $\cos \epsilon = \mk \cdot \mq/(kq)$ and $P_\ell(\cos \epsilon)$ is a Legendre polynomial of order $\ell$. The function $K_\ell^m(q,q')$ is defined as 
\beq
K_\ell^m(q,q') = \int_{-1}^1 \dd \cos \theta K^m(q,q',\cos \theta) P_\ell (\cos \theta),
\eeq
where
\bea
K^m(q,q',\cos \theta) &=& \frac{T_{D,0}^4}{16 P^5} e^{-(Q_- + P)/2}(Q_-^2-P^2)^2 \left[P^2(3P^2-2P-4)+Q_+^2(P^2+6P+12) \right],
\eea
with $P=|\mq-\mq'|/T_{D,0}$ and $Q_{\pm}=(q\pm q')/T_{D,0}$. As mentioned above, to get the Boltzmann equations, we assume that the background density follows a Maxwell-Boltzmann  instead of Bose-Einstein/Fermi-Dirac distribution: 
\beq\label{eq:feq}
\bar f(q,\tau)\approx\bar{f}^{\rm eq}(q) = N e^{-q/T_{D,0}},
\eeq
where the normalization factor $N$ is chosen to match the \textit{energy} density from Maxwell-Boltzmann  and Bose-Einstein/Fermi-Dirac distribution: $\int dq\, q^3 \bar{f}^{\rm eq}(q)=\int dq\, q^3 f^{\rm BE/FD}(q)$. Therefore, we will get  $N= \pi^4/90$ for bosons and $N=7 \pi^4/720$ for fermions.
$F_\ell$ is defined by
\beq
F(k,\cos \epsilon,\eta)=\frac{\int q^3 \dd q \bar{f}(q) \Psi}{\int q^3 \dd q \bar{f}(q)} = \sum_{\ell=0}^{\infty} (-1)^\ell (2\ell+1) F_\ell(k,\eta) P_\ell(\cos \epsilon).
\eeq

To get the Boltzmann hierarchy for $F_\ell$, we further use the relaxation time approximation. This can be achieved by making the following approximation~\cite{Oldengott:2017fhy}:
\bea\label{eq:separable_ansatz}
\Psi_\ell(k,q,\tau)\approx -\frac{1}{4}\frac{d \ln \bar f}{d\ln q} F_\ell(k, \tau).
\eea
We then integrate the Boltzmann hierarchy for $\Psi_\ell$ in Eqs.~(\ref{eq:Boltzmann_hierarchy_decoupling_Psi0}-\ref{eq:Boltzmann_hierarchy_decoupling_Psi3}) to get the Boltzmann hierarchy for $F_\ell$ shown in Eq.~(\ref{eq:Boltzmann_hierarchy_decoupling}). With the definition of $\langle \Gamma\rangle$ in Eq.~(\ref{eq:ave_rate_decoupling}), we can find the coefficients $\alpha_\ell$. We get $\alpha_0 = \alpha_1=0$ which is consistent with energy and momentum conservation. For higher order $\alpha_\ell$, we get $\alpha_2=1.39, \alpha_3=1.48, \alpha_4=1.57, \alpha_5=1.62$.

\subsection{Recoupling}
For the recoupling case, we consider scalar DR with a $\phi^4$ interaction (see Sec.~\ref{sec:models}). In this case, the averaged matrix element can be calculated as $\left< |\mathcal{M}|^2 \right> =\frac{1}{2} \lambda_\phi^2$, and the Boltzmann equations for $\Psi$ read
\bea
\dot{\Psi}_0(q) &=& -k \Psi_1(q) + \frac{1}{6}\frac{\partial \ln \bar{f}}{\partial \ln q} \dot{h} - \frac{N \lambda_\varphi^2 T_{D,0}^2}{128 \pi^3 q} \Psi_0(q)\nonumber\\
&& \qquad + \frac{N \lambda_\varphi^2}{128 \pi^3} \int \dd q' \left[ K_0^0(q,q') - e^{-q{/T_{D,0}}} \right ] \frac{q' \bar{f}(q')}{q \bar{f}(q)} \Psi_0 (q')\label{eq:Boltzmann_hierarchy_recoupling_Psi0}\\
\dot{\Psi}_1(q) &=& -\frac{2}{3}k \Psi_2(q) + \frac{1}{3} k \Psi_0(q) - \frac{N \lambda_\varphi^2 T_{D,0}^2}{128 \pi^3 q} \Psi_1(q)\nonumber\\
&& \qquad + \frac{N \lambda_\varphi^2}{128 \pi^3} \int \dd q' K_1^0(q,q') \frac{q' \bar{f}(q')}{q \bar{f}(q)} \Psi_1 (q')\\
\dot{\Psi}_2(q) &=& -\frac{3}{5}k \Psi_3(q) + \frac{2}{5} k \Psi_1(q) - \frac{\partial \ln \bar{f}}{\partial \ln q} \left( \frac{2}{5} \dot{\eta} +\frac{1}{15}\dot{h} \right) - \frac{N \lambda_\varphi^2 T_{D,0}^2}{128 \pi^3 q} \Psi_2(q)\nonumber\\
&& \qquad + \frac{N \lambda_\varphi^2}{128 \pi^3} \int \dd q' K_2^0(q,q') \frac{q' \bar{f}(q')}{q \bar{f}(q)} \Psi_2 (q')\\
\dot{\Psi}_{\ell>2}(q) &=& -\frac{k}{2\ell+1}[\ell \Psi_{\ell-1}(q) -(\ell+1)\Psi_{\ell+1}(q)]  - \frac{N \lambda_\varphi^2 T_{D,0}^2}{128 \pi^3 q} \Psi_\ell(q)\nonumber\\
&& \qquad + \frac{N \lambda_\varphi^2}{128 \pi^3} \int \dd q' K_\ell^0(q,q') \frac{q' \bar{f}(q')}{q \bar{f}(q)} \Psi_\ell (q')\label{eq:Boltzmann_hierarchy_recoupling_Psil}
\eea
where
\bea
K^0(q,q',\cos \theta) &=& \frac{e^{-(Q_- + P)/2}}{P}.
\eea
To get the Boltzmann hierarchy for $F_\ell$ in Eq.~(\ref{eq:Boltzmann_hierarchy_decoupling}), we integrate Eqs.~(\ref{eq:Boltzmann_hierarchy_recoupling_Psi0}-\ref{eq:Boltzmann_hierarchy_recoupling_Psil}) with the approximation in Eq.~(\ref{eq:separable_ansatz}).  Based on the definition of $\langle \Gamma\rangle$ in Eq.~(\ref{eq:ave_rate_recoupling}), we can get the $\alpha_\ell$ coefficients for the recoupling case: $\alpha_2=0.188, \alpha_3=0.294, \alpha_4=0.356, \alpha_5=0.395$. Again, $\alpha_{0,1}=0$ due to energy and momentum conservation. Note that $\alpha_{2}$ for the recoupling case is much smaller than unity. This means naively setting $\alpha_\ell=1$ will generate a relatively large deviation in the estimation of $z_{\rm rec}$ as shown in Eq.~(\ref{eq:z_dec_rec}). 

\section{A model for self-interacting scalar dark radiation }
\label{app:scalar_DR}
Here we present a model for dark radiation that consists self-interacting light scalars. We consider axion-like particles (ALP) in the dark sector, denoted as $\phi$, which are Goldstone bosons from a global dark U(1) symmetry breaking. Below the spontaneously breaking scale $f$, we demand the leading coupling between $\phi$ and other fields be $\frac{\phi}{f}G_D\tilde{G}_D$, where $G_D$ is the field strength of a dark SU(N) gauge field. This coupling can be rewritten as a total derivative. Therefore, the ALP mass is protected by a shift symmetry at perturbative level and can be naturally light. After the dark SU(N) field confines, $\phi$ acquires a potential non-perturbatively as
\begin{eqnarray}\label{eq:L_ALP}
\mathcal L_\phi& \supset& \Lambda^4\cos\left(\frac{\phi}{f}\right)\nonumber\\
&\approx& \Lambda^4-\frac{ \Lambda^4}{2f^2}\phi^2+\frac{ \Lambda^4}{24f^4}\phi^4\nonumber\\
&\approx& \Lambda^4-\frac{ m_\phi^2}{2}\phi^2+\frac{m_\phi^2}{24f^2}\phi^4
\end{eqnarray}
where $\Lambda$ characterizes the energy scale of the confinement. In the second and third lines of the above equation, we neglect terms from the expansion that contain $\phi^n, (n>4)$, because their effects will be suppressed by $f^{n-4}$. The third line of Eq.~(\ref{eq:L_ALP}) shows the ALP model predicts a naturally light scalar field with $m_\phi\equiv\Lambda^2/f$ and a $\frac{\lambda_\phi}{24}\phi^4$ interaction with $\lambda_\phi\equiv m_\phi^2/f^2$.

To check whether the ALP model can be a model for the recoupling dark radiation, we need to demand i) $\phi$ is a thermal relic  (similar to SM neutrinos) during the cosmic history relevant for CMB measurements, and ii) it has the correct self-interaction strength for recoupling near recombination. Since we are interested in the case of low $f$ and $\Lambda$ (see below), $\phi$ and SU(N) gauge fields are in thermal equilibrium for $T_D> \Lambda$ . The first condition can be achieved when $T_{D}$ drops below $\Lambda$, where a confinement phase transition occurs in dark SU(N) theory. Since $\phi$ is the only light particle after the phase transition, all SU(N) gauge degrees of freedom will eventually convert into $\phi$. This conversion is fast given the low $f$ and $\Lambda$ we consider, so it is plausible to treat $\phi$ as a thermal relic when $T_{D}\lesssim \Lambda$. Moreover, $\Lambda$ needs to be high enough  that $\phi$ remains as a thermal relic during the history relevant for CMB measurements. This translates to a constraint of $\Lambda\gtrsim O(100)$ eV, where $O(100)$ eV roughly corresponds to the highest temperature scale that current CMB data can probe.

 For the second condition, we need the coupling strength of the order of $\lambda_\phi\sim 10^{-12}$ to achieve recoupling around $z\sim 10^3$ (see discussion in Sec.~\ref{sec:models}). Given that $m_\phi$ and $f$ are correlated for fixed $\lambda_\phi$ ($\lambda_\phi\equiv m_\phi^2/f^2$), an arbitrary small $m_\phi$ leads to too small $f$ and $\Lambda$, violating the first condition. Assuming  the temperature of the dark sector is roughly the temperature of SM neutrino bath  ($T_{D}\sim T_\nu $), we need $m_\phi\lesssim 0.1$ eV to ensure $\phi$ remains as radiation before recombination. Saturating $m_\phi\sim 0.1$ eV, we can infer that $f \sim 0.1$ MeV and $\Lambda \sim 100$ eV. Therefore, the ALP model satisfies both conditions and we use it as a  benchmark model for recoupling dark radiation.
 
\section{Full MCMC results}
\label{app:full_MCMC}
\begin{figure}[H]
	\centering
	\includegraphics[width=15cm]{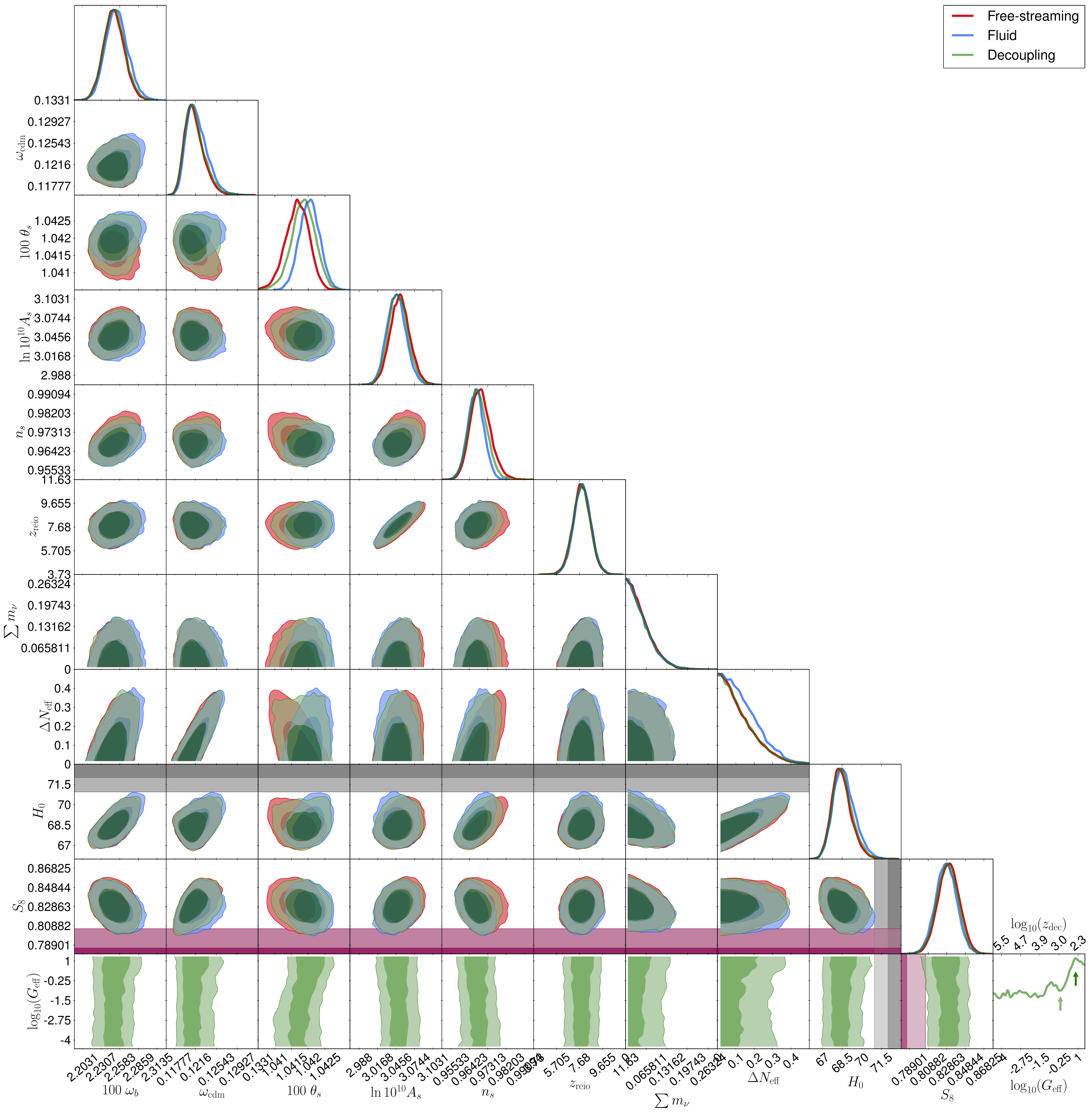}
	\caption{Full cosmological parameter space for the Free-streaming, Fluid, and Decoupling cases.}
	\label{fig:Pb_dec_full}
\end{figure}

\begin{figure}[H]
	\centering
	\includegraphics[width=15cm]{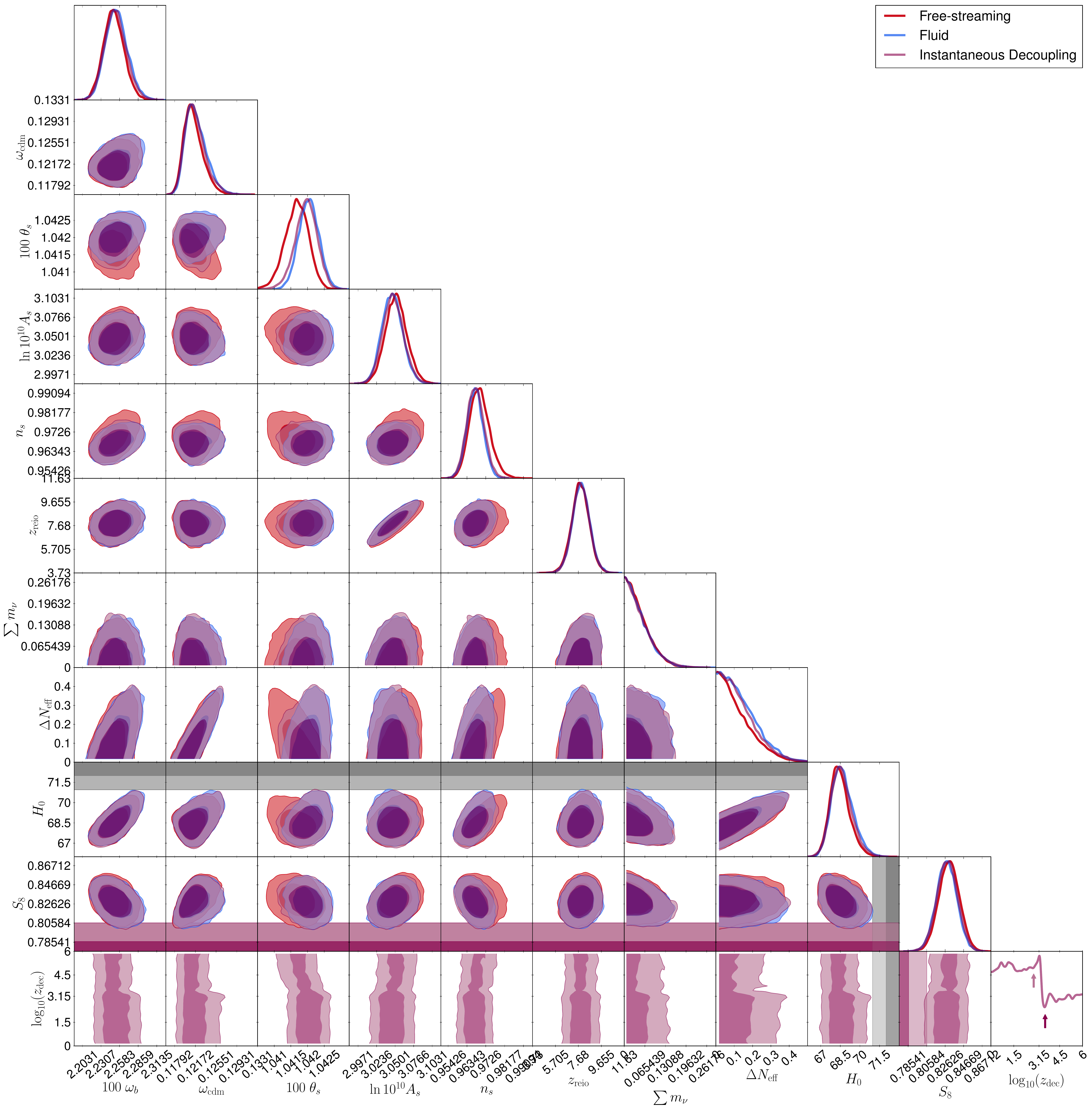}
	\caption{Full cosmological parameter space for the Free-streaming, Fluid, and Instantaneous Decoupling cases.}
	\label{fig:Pb_inst_full}
\end{figure}

\begin{figure}[H]
	\centering
	\includegraphics[width=15cm]{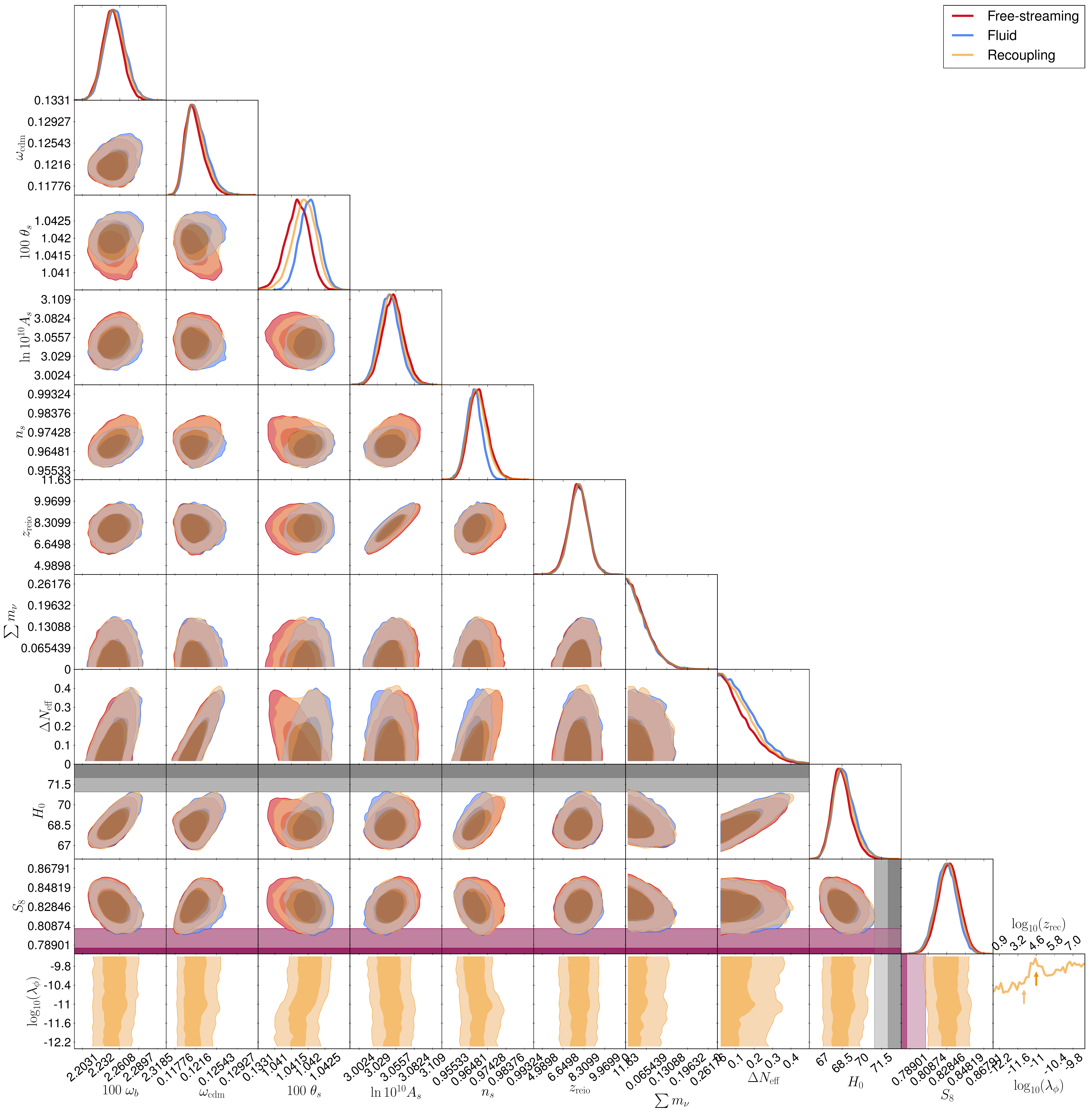}
	\caption{Full cosmological parameter space for the Free-streaming, Fluid, and Recoupling cases.}
	\label{fig:Pb_rec_full}
\end{figure}

Figures~\ref{fig:Pb_dec_full},~\ref{fig:Pb_inst_full}, and~\ref{fig:Pb_rec_full} show the full cosmological parameter space of the results in Fig.~\ref{fig:Pb_4param} top left, top right, and bottom, respectively. At the edge of the parameter space of the clustering parameter $S_8 = \sigma_8 (\Omega_m/0.3)^{0.5}$ we show a galaxy survey bound from KiDS-1000~\cite{Heymans:2020gsg} for comparison (in purple, note only the edge of the 1- and 2-$\sigma$ range is visible). We also show a bound on $H_0$ from Ref.~\cite{Riess:2021jrx} in grey at the edge of the $H_0$ parameter space (note only the edge of the 1- and 2-$\sigma$ range is visible).

\end{appendix}

\clearpage
\bibliography{references}
\bibliographystyle{apsrev4-1}
\end{document}